\documentclass[aps,prc,twocolumn,tightenlines,ece,amsmath,groupedaddress,showpacs,showkeys]{revtex4}
\usepackage{epsf}
\usepackage{graphicx}
\usepackage{graphics}
\usepackage{color}
\def\beg{\begin{equation}}
\def\ee{\end{equation}}

\usepackage{longtable}
\begin{document}
\def\nt{{\em Nature }}
\def\prl{{\em Phys. Rev. Lett. }}
\def\prc{{\em Phys. Rev. C }}
\def\jap{{\em J. Appl. Phys. }}
\def\ajp{{\em Am. J. Phys. }}
\def\nima{{\em Nucl. Instr. and Meth. Phys. A }}
\def\npa{{\em Nucl. Phys. A }}
\def\npb{{\em Nucl. Phys. B }}
\def\epjc{{\em Eur. Phys. J. C }}
\def\epja{{\em Eur. Phys. J. A }}
\def\plb{{\em Phys. Lett. B }}
\def\mpla{{\em Mod. Phys. Lett. A }}
\def\pr{{\em Phys. Rep. }}
\def\prv{{\em Phys. Rev. }}
\def\zpc{{\em Z. Phys. C }}
\def\zpa{{\em Z. Phys. A }}
\def\ppnp{{\em Prog. Part. Nucl. Phys. }}
\def\jpg{{\em J. Phys. G }}
\def\cpc{{\em Comput. Phys. Commun.}}
\def\app{{\em Acta Physica Pol. B }}
\def\aip{{\em AIP Conf. Proc. }}
\def\jhep{{\em J. High Energy Phy. }}
\def\ijmpa{{\em Int. J. Mod. Phys. A }}
\def\ijmpe{{\em Int. J. Mod. Phys. E }}
\def\ijp{{\em Indian J. of Phys.}}
\def\psc{{\em Prog. Sci. Culture }}
\def\snc{{\em Suppl. Nuovo Cimento }}
\def\phy{{\em Physics }}
\def\appb{{\em Acta  Phys.  Pol. B }}


\title{Charged Particle and Photon Multiplicity, and Transverse Energy Production in High-Energy Heavy-Ion Collisions}


\author{Raghunath Sahoo{$^1$}\footnote{Corresponding Author, Email:
   Raghunath.Sahoo@cern.ch}, Aditya Nath Mishra{$^1$}, Nirbhay K. Behera{$^2$}, and Basanta K. Nandi{$^2$}}

\affiliation{{$^1$}Indian Institute of Technology Indore, Indore, India-452017}
\affiliation{{$^2$}Indian Institute of Technology Bombay, Mumbai, India-400067}


\date{\today}

\begin{abstract}
We review the charged particle and photon multiplicity, and transverse energy production in heavy-ion collisions starting from few GeV to TeV energies. The experimental results of pseudorapidity distribution of charged particles and photons at different collision energies and centralities are discussed. We also discuss the hypothesis of limiting fragmentation and expansion dynamics using the Landau hydrodynamics and the underlying physics. Meanwhile, we present the estimation of initial energy density multiplied with formation time as a function of different collision energies and centralities. In the end, the transverse energy per charged particle in connection with the chemical freeze-out criteria is discussed.
 We invoke various models and phenomenological arguments to interpret and characterize the fireball created in heavy-ion collisions. This review overall provides a scope to understand the heavy-ion collision data and a possible formation of a deconfined phase of partons via the global observables like charged particles, photons and the transverse energy measurement. 

\end{abstract}


\maketitle


\section{INTRODUCTION} 
At extreme temperatures and energy density, hadronic matter
undergoes a phase transition to partonic phase called as Quark-Gluon
Plasma (QGP) \cite{x1,bjorken,lQCD}. The main goal of heavy-ion collision
experiments is to study the
QGP by creating such extreme conditions by colliding heavy nuclei at
relativistic energies. During the last decade, there are many
heavy-ion collision experiments carried out at SPS, RHIC and LHC to create and study QGP in the laboratory.  Global observables like transverse energy ($E_{\rm T}$), particle multiplicities ($N_{\gamma}, 
N_{\rm {ch}}$ etc.), $p_{\rm T}$-spectra of the produced particles and their pseudorapidity
distributions ($dE_{\rm T}/d\eta, dN/d\eta$), with different colliding species and beam energies provide
insight about the dynamics of the system and regarding the formation of QGP 
\cite{bjorken, kataja}. It is also proposed that the correlation of
mean transverse momentum, $\langle p_{\rm T} \rangle$ and the multiplicity of the produced particles may serve as
a probe for the Equation of State (EoS) of hot hadronic matter \cite{vanHove}. In a thermodynamic description of the produced system, the rapidity density ($dN/dy$) reflects the entropy and the mean transverse momentum ($\langle p_{\rm T} \rangle$), corresponds to the temperature of the system. Except at the
phase transition points, the rapidity density linearly scales with $\langle p_{\rm T} \rangle$. If the
phase transition is of first order, then the temperature remains constant at the
co-existence of the hadron gas and the QGP phase, thereby increasing the entropy density.
In such a scenario, $\langle p_{\rm T} \rangle$ shows a plateau with increase of entropy, thereby characterizing the phase transition associated with the time evolution of the system. Hence, the global observables
like, $dN/dy$ and $\langle p_{\rm T} \rangle$, give indication of a possible existence of a QGP phase and the order of phase 
transition. $dE_{\rm T}/d\eta$ gives the maximum energy density produced in the collision
process which is necessary to understand the reaction dynamics. The formation of QGP
may also change the shape of the pseudorapidity distribution \cite{sarkar, dumitru}.
The event multiplicity distribution gives information of the centrality and energy density
of the collision. The scaling of multiplicity with number of participant nucleons 
($N_{\rm {part}}$) reflects the particle production due to soft processes (low-$p_{\rm T}$). Whereas,
at high energy when hard processes (high-$p_{\rm T}$) dominate, it is expected that the 
multiplicity will scale with the number of nucleon-nucleon collisions ($N_{\rm {coll}}$).
There are models \cite{kharzeev} to explain the particle production taking a linear 
combination of $N_{\rm {part}}$  and $N_{\rm {coll}}$ (called two-component model).
The most viable way of studying QGP is via the
particles produced in the collision in their respective domain of
proposed methods. Then one of the most fundamental
questions arises about the mechanism of particle production and how they
are related with the initial energy density, gluon density in the
first stage of the collision evolution and
entropy of the system. Similarly, question can be put to figure out the role of soft and
hard process of particle productions. It is proposed that the charged particle multiplicity or
technically called as the pseudorapidity density distributions of charged
particles, $dN_{\rm{ch}}/d\eta$, can be used to address the above
questions \cite{a3,a4,a5,c3,c4,atlas,cmsNch}.  Here the pseudorapidity, $\eta =-ln ~tan~\theta/2$, where $\theta$ is the polar angle, the produced particles make with the detector. So $dN_{ch}/d\eta$
is called as one of the global variables to characterize the system
produced in the heavy-ion collisions.  Experimentally, it is more easy
to estimate this quantity as most of the detectors are capable of
detecting charged particles and it involves only kinematics of the
charged particles.  
\par
In this review, in Section-II, we discuss the method of experimental determination of collision centrality, which is followed by discussions on the midrapidity pseudorapidity density distributions of charged particles for different collision energies, collision species and centralities in Section-III. In this section, we discuss about the longitudinal scaling and factorization of charged particles.  The expansion dynamics of the system is discussed using the pseudorapidity density distributions of charged particles and the Landau-Carruthers hydrodynamics. In subsequent subsections, the scaling of total charged particles with collision centrality and its energy dependence are discussed. This follows with similar discussions on the photon pseudorapidity density at forward rapidities in Section-IV, which includes longitudinal scaling of photons.  Subsequently, in Section-V, discussions are made on the production of transverse energy and its use for centrality determination. Section-VI includes discussions on collision energy dependence of transverse energy, which is followed by discussions on the centrality dependence in Section-VII. Section-VIII includes discussions on estimation of initial energy density in Bjorken hydrodynamic scenario, and its energy and centrality dependences.  Further we correlate the energy and centrality dependence of transverse energy per charged particle with chemical freeze-out criteria in Section-IX. In Section-X, we summarize the review with conclusions. Appendix discusses on the important properties of Gamma and Negative Binomial Distributions.

\section{Centrality determination}
In heavy-ion collisions, the event centrality is of utmost importance to
understand the underlying physics of the collision. The event centrality
is related to the impact parameter, defined as the distance between the
centroids of the two colliding nuclei in a plane transverse to the beam
axis, of the collision. The impact parameter tells about the overlap volume
of the two nuclei. This overlap volume determines the geometrical
parameters, like number of participant nucleons ($N_{\rm{part}}$), number of spectator
nucleons ($N_{\rm{spec}}$) and the number of binary collisions ($N_{\rm{coll}}$). 

\par
The impact parameter can not be determined experimentally. However,
the global observables, like total charged particles ($N_{ch}$),
transverse energy ($E_T$) or energy deposited in ZDC ($E_{zdc}$) etc., are related to this
geometrical quantity. By combining the experimental observables with
simulation, one can estimate the impact parameter and hence, the centrality
of the event class. The centrality is expressed as the percentile ($c$) of
the total hadronic interaction cross section corresponding to the
charged particle multiplicity above certain threshold and is given by,
\begin{equation}
c = \frac{1}{\sigma_{AA}}\int_{0}^{b}
\frac{d\sigma}{db^{\prime}}db^{\prime} ~~.
\end{equation}
In Eq (1), $\sigma_{AA}$ is the total nuclear interaction cross section of
A+A collision. Assuming constant luminosity, the cross section can be
replaced by the number of observed events after the trigger efficiency
correction. But at very high energy, when these two nuclei pass by each
other, there is a large QED cross section because of the electromagnetic
field \cite{cent1,cent2}. This QED cross section is much larger than the hadronic cross
section and this contaminates the most peripheral events. That is why the
centrality determination is restricted to some percentile where the
QED contribution is negligible. The fraction of hadronic events excluded
by such cut as well as the trigger efficiency can be estimated by using a
Glauber model simulation.

For a given impact parameter,  the $N_{\rm{part}}$ and $N_{\rm{coll}}$ can be
estimated by  Glauber Monte Carlo method. The parametrized  Negative
Binomial Distribution (NBD) can be used to describe the nucleon-nucleon collisions. For heavy-ion collisions, $N_{\rm{part}}$ and $N_{\rm{coll}}$ are used to generate the number of charged particles
by incorporating two-component model in the following way:
\begin{equation}
N_{ancestors} = f \times N_{part} + (1-f) \times N_{coll}.
\end{equation}
This $N_{\rm{ancestors}}$ refers to the ``independent emitting
source''. The two-component model given in Eq (2) incorporates the
soft and hard interactions. Soft process is related to the $N_{\rm{part}}$
and hard process is related to $N_{\rm{coll}}$. 

\par
\begin{figure}
\begin{center}
\includegraphics[width=3.6in]{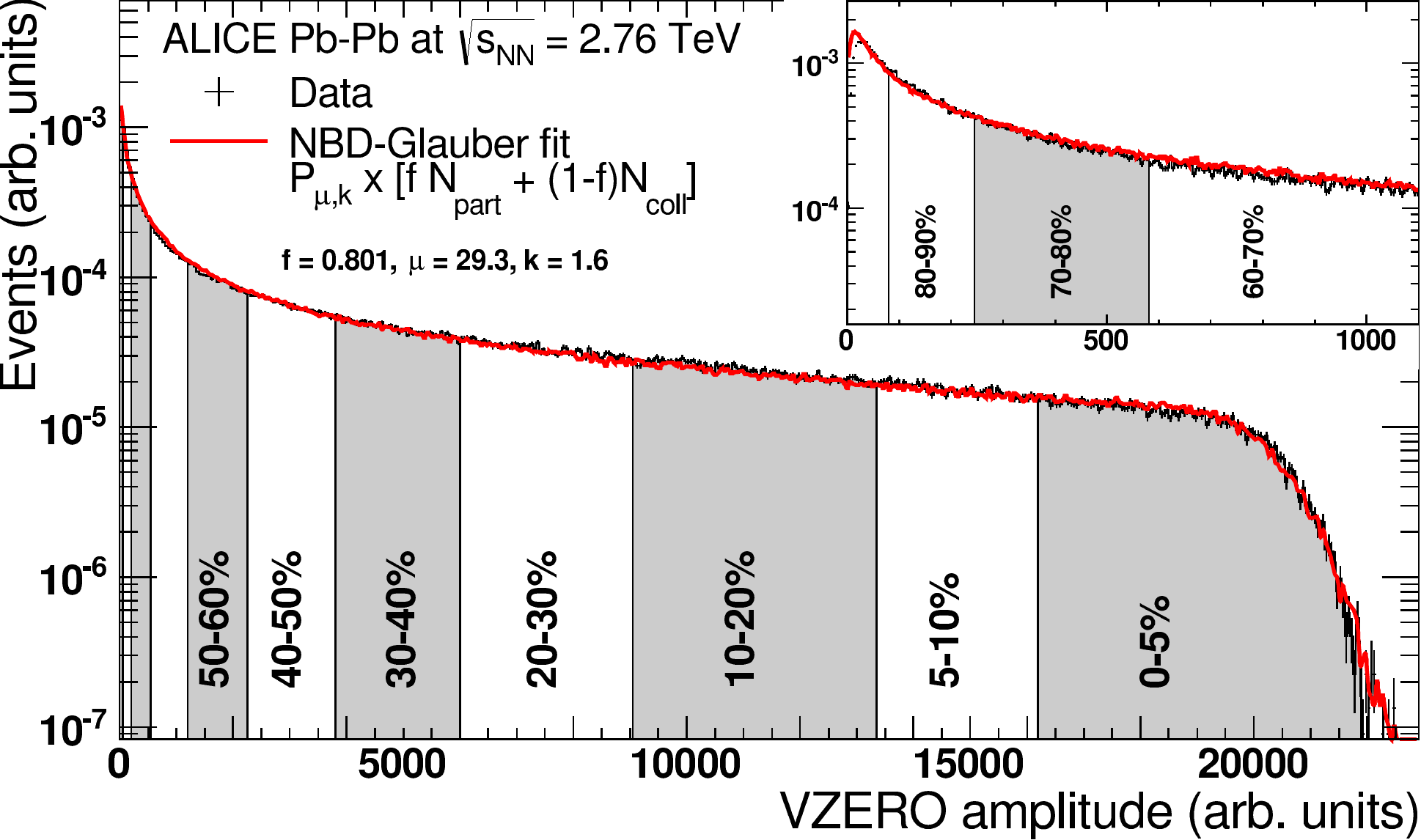}
\caption{Distribution of the summed amplitudes in the VZERO scintillator tiles (histogram); inset shows the low amplitude part of the distribution. The curve shows the result of the Glauber model fit to the measurement. The vertical lines separate the centrality classes used in the analysis, which in total correspond to the most central $80\%$ of hadronic collisions. Figure taken from Ref. \cite{alice2.76}} 
\label{centrality}
\end{center}
\end{figure}

The functional form of NBD distribution is given by,
\begin{equation}
P(\mu, k, n) = \frac{\Gamma(n+k)}{\Gamma(n+1)\Gamma(k)} . \frac{(\mu/k)^{n}}{(\mu/k+1)^{n+k}}.
\end{equation}
Eq (3) represents the probability of measuring $n$ hits per
ancestor. Here, $\mu$ represents the mean multiplicity per ancestor and
$k$ controls the width of the distribution. 
 In $p+p (\bar{p}$) collision, a Negative
Binomial Distribution (NBD) with a fixed value of $\mu$ and $k$ well
describes the charged particle multiplicity data. 
 The charged particle multiplicity for nucleus-nucleus collisions with
 a given impact parameter is
 generated by sampling $N_{\rm{ancestors}}$ times the $p+p$ multiplicity,
 which is generated by using NBD. Finally, a $\chi^2$ minimisation is done by fitting the Glauber
 Monte Carlo generated multiplicity and the charged particle
 multiplicity obtained from the collision data. The $\chi^2$ minimization
 will give us the value of $f$, $\mu$ and $k$. This gives a connection between
an experimental observable and a Glauber Monte
Carlo. From this one can have
access to $N_{\rm{part}}$ and $N_{\rm{coll}}$ for a given class of
centrality by NBD-Glauber fit. For example, the centrality determination in ALICE using
V0 amplitude is given in Figure \ref{centrality}. The two component model is fitted with
the V0 amplitude in Figure \ref{centrality} to find out the $N_{\rm{part}}$ and $N_{\rm{coll}}$ values for a
corresponding centrality \cite{cent1}.

\section{PSEUDORAPIDITY DENSITY DISTRIBUTION OF CHARGED PARTICLES ($dN_{ch}/d\eta$)}

\subsection{Energy dependence of $dN_{ch}/d\eta$ for different collision species}

\begin{figure}
\begin{center}
\includegraphics[width=3.6in]{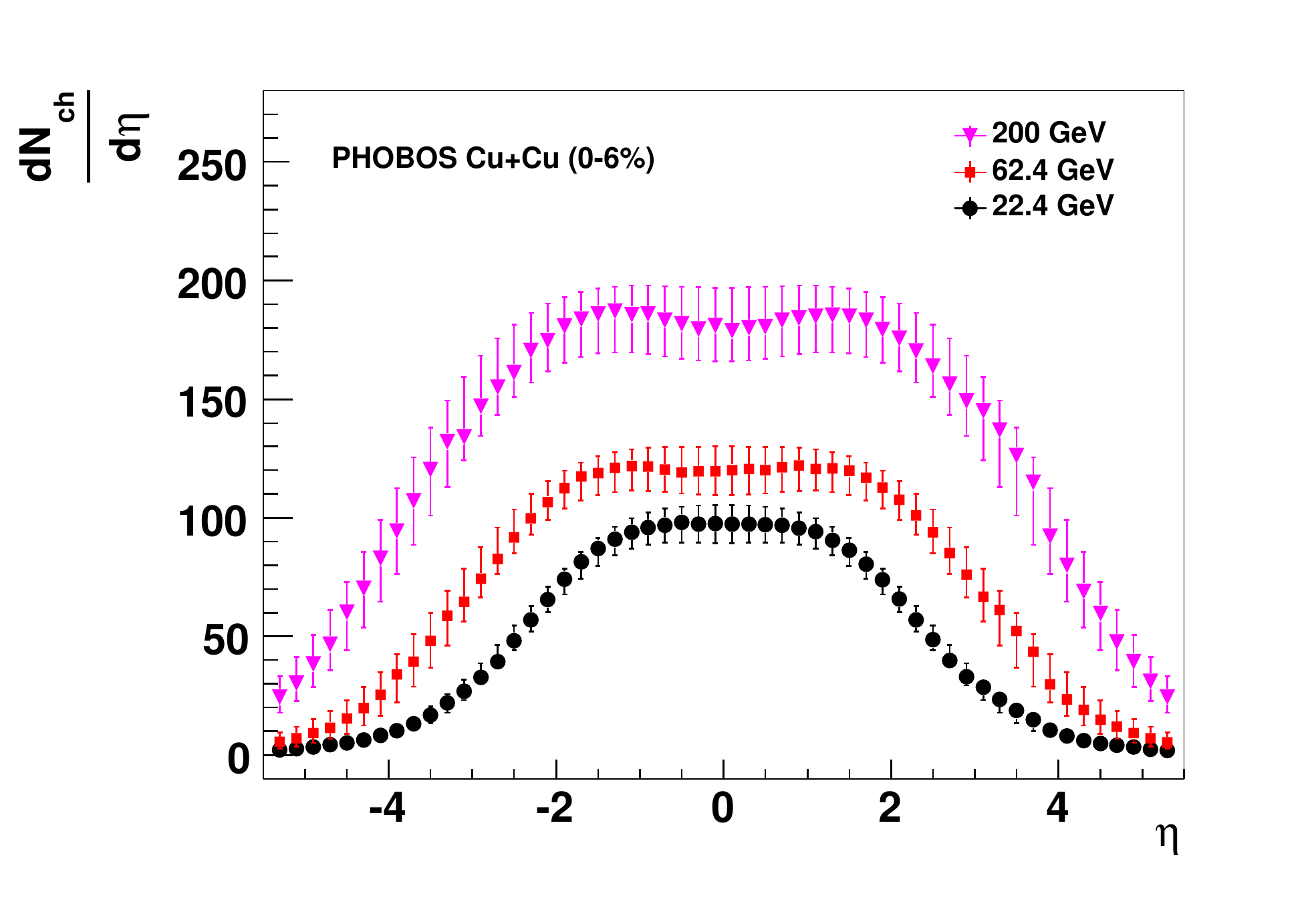}
\caption{Charged particle pseudorapidity distributions of Cu+Cu
  collision systems for the most central events for different collision
  energies.}
\label{dNchdEtaCuCu}
\end{center}
\end{figure}

\begin{figure}
\begin{center}
\includegraphics[width=3.6in]{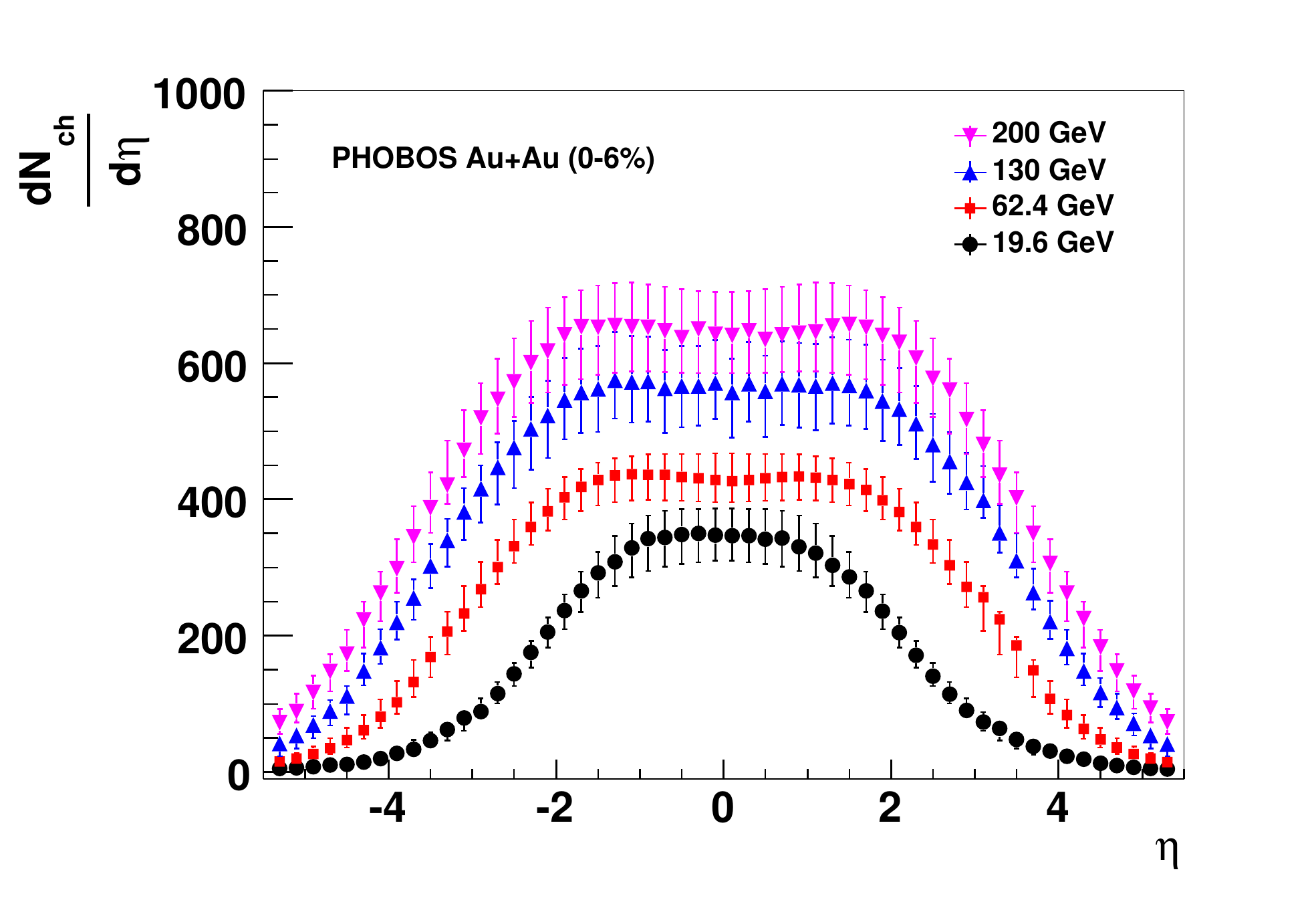}
\caption{Charged particle pseudorapidity distributions of Au+Au
  collision system for the most central events for different 
  collision energies.}
\label{dNchdEtaAuAu}
\end{center}
\end{figure}

\begin{figure}
\begin{center}
\includegraphics[width=3.6in]{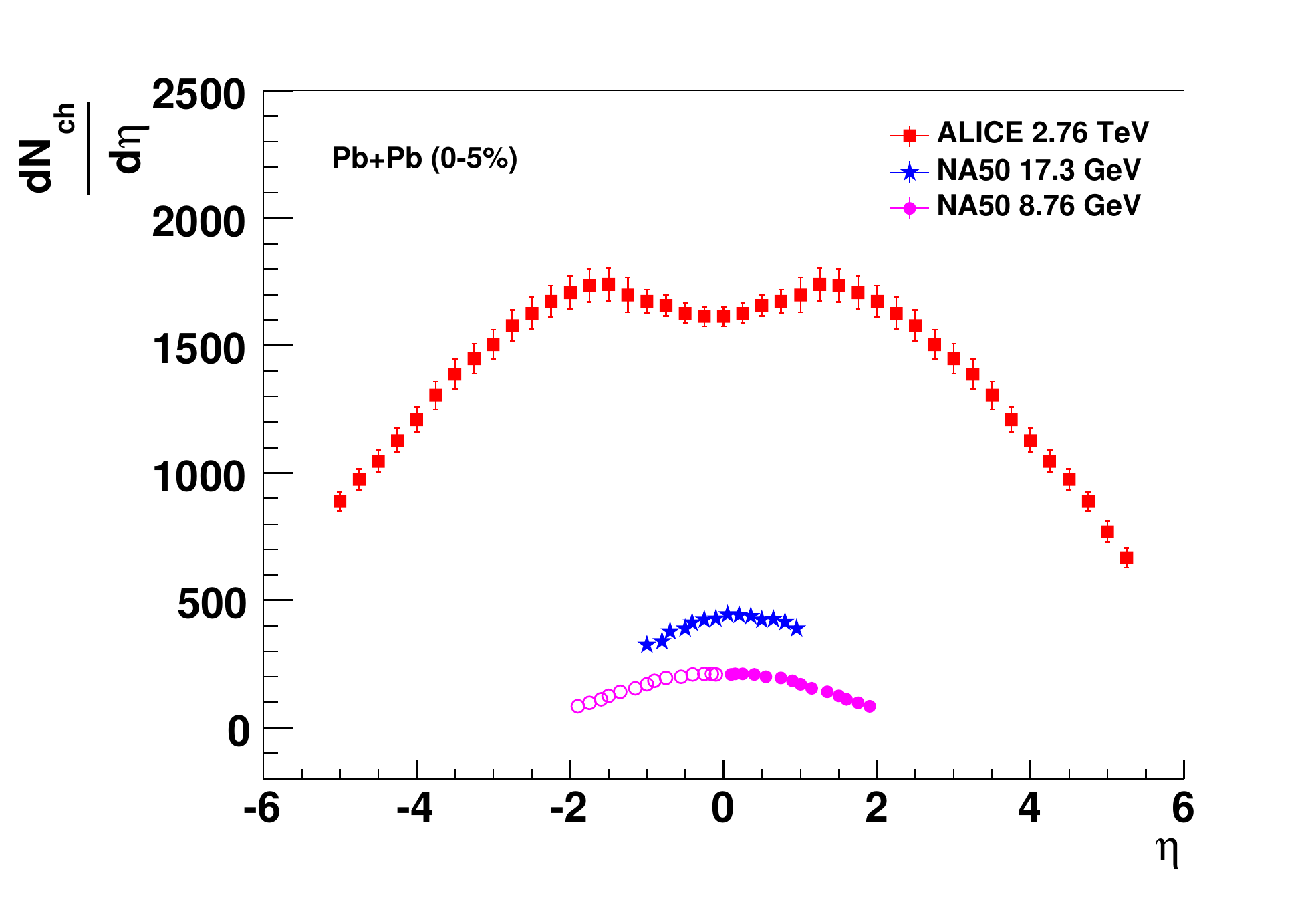}
\caption{Charged particle pseudorapidity distributions of Pb+Pb
  collision systems for the most central events
  at different collision energies.}
\label{dNchdEtaPbPb}
\end{center}
\end{figure}

The $dN_{\rm{ch}}/d\eta$ distributions as a function of
pseudorapidity of most
central events for Cu+Cu collisions at $\sqrt{s_{\rm{NN}}}$ = 22.4 GeV, 62.4 GeV
 and 200 GeV are given in Figure
\ref{dNchdEtaCuCu} \cite{a1}. 
Similarly, the $dN_{\rm{ch}}/d\eta$
distributions for Au+Au system at $\sqrt{s_{\rm{NN}}}$ = 19.6 GeV, 62.4 GeV, 130 GeV
and 200 GeV are given
in Figure \ref{dNchdEtaAuAu} \cite{a2,a3}.
Both the collision systems, i.e. Cu+Cu and Au+Au, data are from PHOBOS experiment
which has maximum pseudorapidity coverage of $|\eta| < 5.3$ at RHIC.
In Figure \ref{dNchdEtaPbPb}, the charged particle pseudorapidity
distributions of Pb+Pb collisions at different energies are
presented. The filled circles and star markers correspond to the fixed
target experiment for beam energies 40 AGeV and 158 AGeV,
respectively. For the fixed target experiment, the $x$-axis is $\eta-\eta_{\rm{peak}}$. 
Here, $\eta_{\rm{peak}}$ corresponds to the peak position of the $dN_{\rm{ch}}/d\eta$ distribution in the fixed target experiment. Theoretically, for fixed target environment, the $\eta_{\rm{peak}}= \eta_{\rm{mid}} = y_{\rm{beam}}/2$ = 2.24 at 40 AGeV and 2.91 at 158 AGeV for Pb+Pb collisions, respectively. In experiment, $\eta_{\rm{peak}}$  is obtained by fitting a Gaussian function to the $dN_{\rm{ch}}/d\eta$ distribution. From the fitting, the $\eta_{\rm{peak}}$ comes out to be 2.43 and 3.12 for 40 AGeV and 158 AGeV, respectively. 
The hollow points at 40 AGeV correspond to the mirror
reflection around the $\eta_{peak}$. The Pb+Pb data at 40 AGeV and 158
AGeV taken from NA50 have pseudorapidity coverage $(\eta- \eta_{peak})
< 1$ and $(\eta - \eta_{\rm{peak}}) \leq 2$, respectively \cite{a4}. In Figure
\ref{dNchdEtaPbPb}, the circles represent the beam energy of 40 AGeV,
the star markers represents the beam energy of 158 AGeV. The $dN_{\rm {ch}}/d\eta$
values of Pb+Pb collisions at $\sqrt{s_{\rm{NN}}}$=2.76 TeV which are
represented by squares and taken from ALICE
experiment \cite{a5}. ALICE has more wider pseudorapidity coverage ($|\eta| <
5.25$). The data shown in Figure
\ref{dNchdEtaCuCu}, \ref{dNchdEtaAuAu} and \ref{dNchdEtaPbPb}
correspond to the most central events in the
midrapidity. It is observed from Figure \ref{dNchdEtaCuCu},
\ref{dNchdEtaAuAu} and \ref{dNchdEtaPbPb} that the distribution is
symmetric around $\eta$ = 0. It is also found that with the increase of 
collision energy, the width and amplitude of the $\frac{dN_{\rm{ch}}}{d\eta}$
increase. Similarly, width of the central plateau region also increases with increase of
energy. Moreover, the plateau region converts into a dip for Pb+Pb
collisions at $\sqrt{s_{\rm{NN}}}$=2.76 TeV as shown in Figure \ref{dNchdEtaPbPb}. This can
be addressed by the particles compositions which is directly related
to the chemistry of the QGP. The pseudorapidity distributions of
kaon has more dip than pion and proton has more dip than pion and
kaons at $\eta$ = 0. This is because of the mass of the particles.
In other way, heavier is the particles, the more is the dip in its
pseudorapidity distribution.  In the mean time, the transverse
momentum spectra of identified particles show that the total
proton+anti-proton production cross section is higher at LHC than at
RHIC \cite{cent1}. That explains the dip observed for the Pb+Pb
collisions at LHC energy.
\par
The multiplicity distribution at LHC can be well described by the double
Gaussian function as follows \cite{a5},
\begin{equation}
A_{1}e^{\frac{-\eta^2}{\sigma_{1}^{2}}} - A_{2}e^{\frac{-\eta^2}{\sigma_{2}^{2}}}
\end{equation}
It is reported in Ref \cite{a5} that the values of $A_{1}/A_{2}$, $\sigma_1$, $\sigma_2$
are same within the errors for each measured centrality bin. To
test this whether it is valid for other systems and energies, we
tried to fit this double Gaussian function to other multiplicity
distributions of Au+Au and Cu+Cu systems measured at $\sqrt{s_{\rm{NN}}}$= 200 and 130 GeV. To
check the consistency, we considered $dN_{\rm{ch}}/d\eta$
distributions of three centralities: 0-6$\%$, 6-15$\%$
and 15-25$\%$. The $\chi^2$ of the fitting, the fitting parameters
$A_{1}, A_{2}$, the ratio of
$A_{1}/A_{2}$, $\sigma_1$, $\sigma_2$ are given in Table
\ref{fitvalue1}, \ref{fitvalue2} and \ref{fitvalue3}. It can be seen
from the tabulated values that the values of $A_{1}/A_{2}$,
$\sigma_1$, $\sigma_2$ are same within the errors for different
centralities at a particular energy. Hence, this
observation for RHIC energies agree with the observation made at LHC
energy. It can be seen from Figure
\ref{dNchdEtaCuCu}, \ref{dNchdEtaAuAu} and \ref{dNchdEtaPbPb} that
with increase of energy, the width of pseudorapidity distribution
increases. This can be related to the longitudinal flow 
and velocity of sound of the system ($c_{s}$) using Landau hydrodynamic model. It is observed that with increase
of energy, the velocity of sound increases and can be understood in
the rapidity space as follows \cite{a7,mohantyB}.

\begin{equation}
\sigma^{2}_{y} = \frac{8}{3} \frac{c_{s}^{2}}{1-c_{s}^{4}}
ln(\sqrt{s_{NN}}/2m_{p}) .
\end{equation}

where $m_{p}$ is mass of proton and $\sigma_{y}$ is the width of
rapidity distribution of charged particles and $c_s^2$ is the square of the velocity of sound, which equals to $1/3$ for an ideal gas. 

\begin{table*}[!h]
\centering
\begin{tabular}{|c | c | c | c | c | c | c |}
\hline
Centrality($\%$) & $\chi^{2}/ndf$ & $A_1$ & $A_2$ & $A_{1}/A_{2}$ &
$\sigma_1$ & $\sigma_2$\\\hline 

 0-6& 2.787/48 & 1130$\pm$60.52 & 951.2$\pm$56.7 & 1.19 & 2.94$\pm$0.06&
 2.62$\pm$0.08\\
\hline
 
6-15&1.238/48  & 821.7$\pm$36.66& 682.5$\pm$36.67 & 1.20 & 3.0$\pm$0.08&
 2.65$\pm$0.09 \\
\hline

15-25 & 0.913/48 &789.9$\pm$26.18 & 670.7$\pm$525 &
1.18 & 3.02$\pm$0.113 & 2.77$\pm$0.12  \\
\hline

\end{tabular}
\caption{PHOBOS Cu+Cu 200 GeV}
\label{fitvalue1}
\end{table*}

\begin{table*}[!h]
\centering
\begin{tabular}{|c | c | c | c | c | c | c |}
\hline
Centrality($\%$) & $\chi^{2}/ndf$ & $A_1$ & $A_2$ & $A_{1}/A_{2}$ &
$\sigma_1$ & $\sigma_2$\\\hline 

 0-6& 2.574/48 & 1987$\pm$106 & 1461$\pm$86.48 & 1.36 & 2.96$\pm$0.04&
 2.28$\pm$0.06 \\
\hline
 
6-15& 1.591/48  & 1831$\pm$183.9 & 1344$\pm$186.9 & 1.36 & 2.99$\pm$0.08&
 2.42$\pm$0.08 \\
\hline

15-25 & 1.427/48 & 1488$\pm$116.1 & 1125$\pm$78.8 &
1.32 & 3.01$\pm$0.50 & 2.53$\pm$0.06  \\
\hline

\end{tabular}
\caption{PHOBOS Au+Au 200 GeV}
\label{fitvalue2}
\end{table*}

\begin{table*}[!h]
\centering
\begin{tabular}{|c | c | c | c | c | c | c |}
\hline
Centrality($\%$) & $\chi^{2}/ndf$ & $A_1$ & $A_2$ & $A_{1}/A_{2}$ &
$\sigma_1$ & $\sigma_2$\\\hline 

 0-6& 4.987/48 & 1451$\pm$132.1 & 904.8$\pm$143.1 & 1.61 & 2.89$\pm$0.06&
 2.04$\pm$0.10 \\
\hline
 
6-15& 3.47/48  & 1128$\pm$24.3 & 699.3$\pm$88.9 & 1.61 & 2.97$\pm$0.08&
 2.13$\pm$0.08 \\
\hline

15-25 & 1.674/48 & 898$\pm$9.9 & 600$\pm$60.7 &
1.51 & 2.99$\pm$0.03 & 2.3$\pm$0.08  \\
\hline

\end{tabular}
\caption{PHOBOS Au+Au 130 GeV}
\label{fitvalue3}
\end{table*}


\subsection{Longitudinal Scaling}
Charged particle production in the higher
rapidity region are subject of interest in terms of hypothesis of
limiting fragmentation \cite{b1}. According to this hypothesis,
the observed pseudorapidity density of particle as a function of $\eta^{\prime} =\eta
- y_{\rm{beam}}$ approaches a limiting value in the fragmentation
region even if the colliding energy is increased. Here $y_{\rm{beam}} = ln \left( \sqrt{s_{\rm{NN}}}/m_{\rm{p}} \right)$. This can be explained by
considering the whole heavy-ion collision process in laboratory
frame of one of the nuclei. The hypothesis can be represented as follows. In the laboratory
frame, out of the
produced particles, some of them will have velocity increasing with
the increase of collision energy. But some of them will have fixed
velocity (or pseudorapidity) as collision energy increases which is postulated as they have
approached a limiting distribution. This can be explained as
follows. In the frame of the target nucleus, the projectile is Lorentz
contracted and appears like a disk, collides and produces particles. As
colliding energy is increased, the target will observe that a more
contracted disk is colliding with it. However, the momentum transfer process between the projectile and
target does not change with respect to the contraction rate. This
leads to the limiting distribution of produced particles in the
fragmentation region even if the collision energy is increased. One of
the advantages of this observation is that it can be seen both in
rapidity as well as in pseudorapidity distributions of the particles
because at large forward rapidity region, $\eta \sim y -
ln(p_{\rm T}/m_{\rm T})$. 

\begin{figure}
\begin{center}
\includegraphics[width=3.6in]{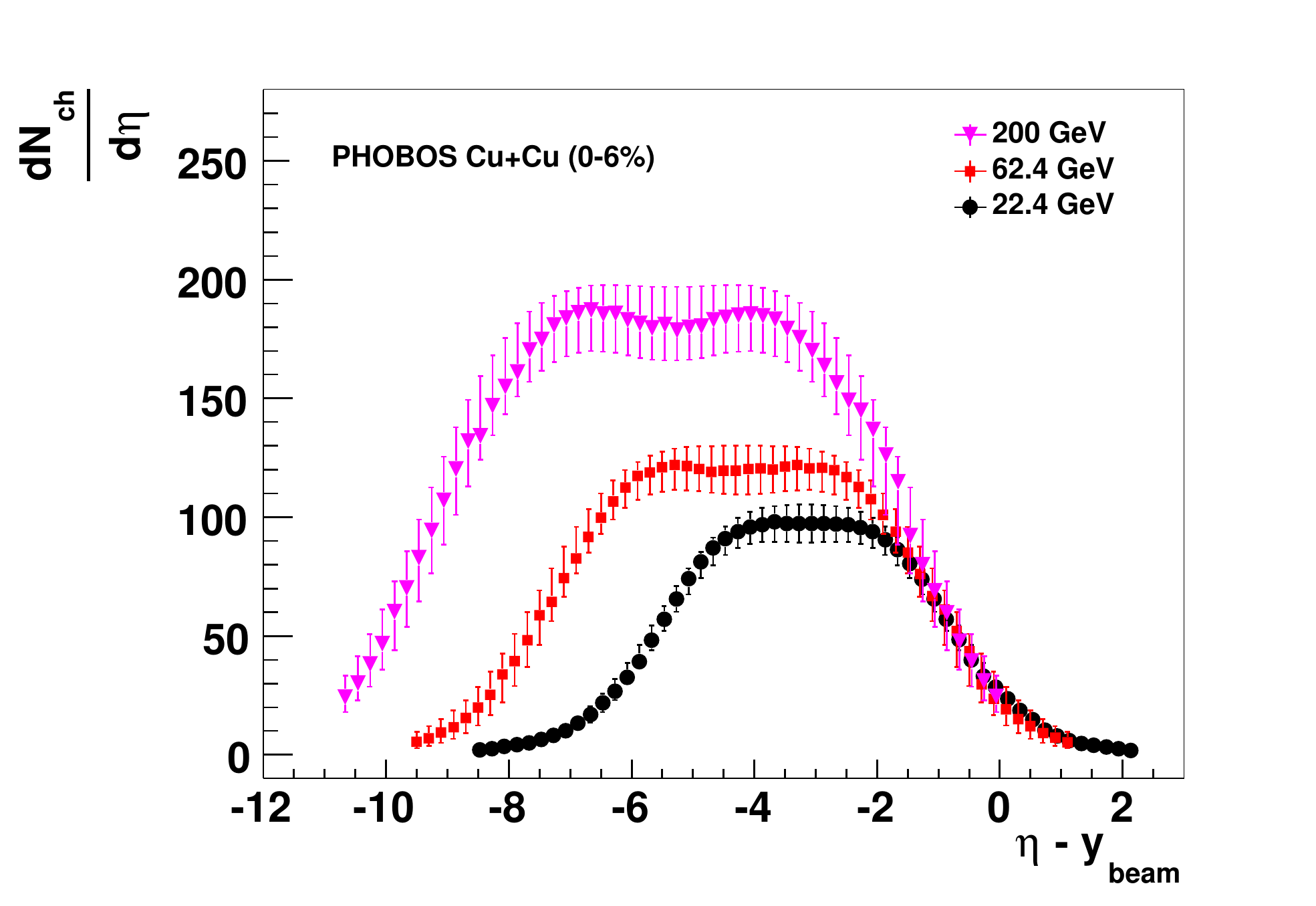}
\caption{ Charged particle multiplicity density normalized by
  participant pairs for Cu+Cu collisions at different energies, shown in
  the projectile rest frame by using $\eta^{\prime} = \eta - y_{\rm{beam}}$.}
\label{LongScalCuCu}
\end{center}
\end{figure}

\begin{figure}
\begin{center}
\includegraphics[width=3.6in]{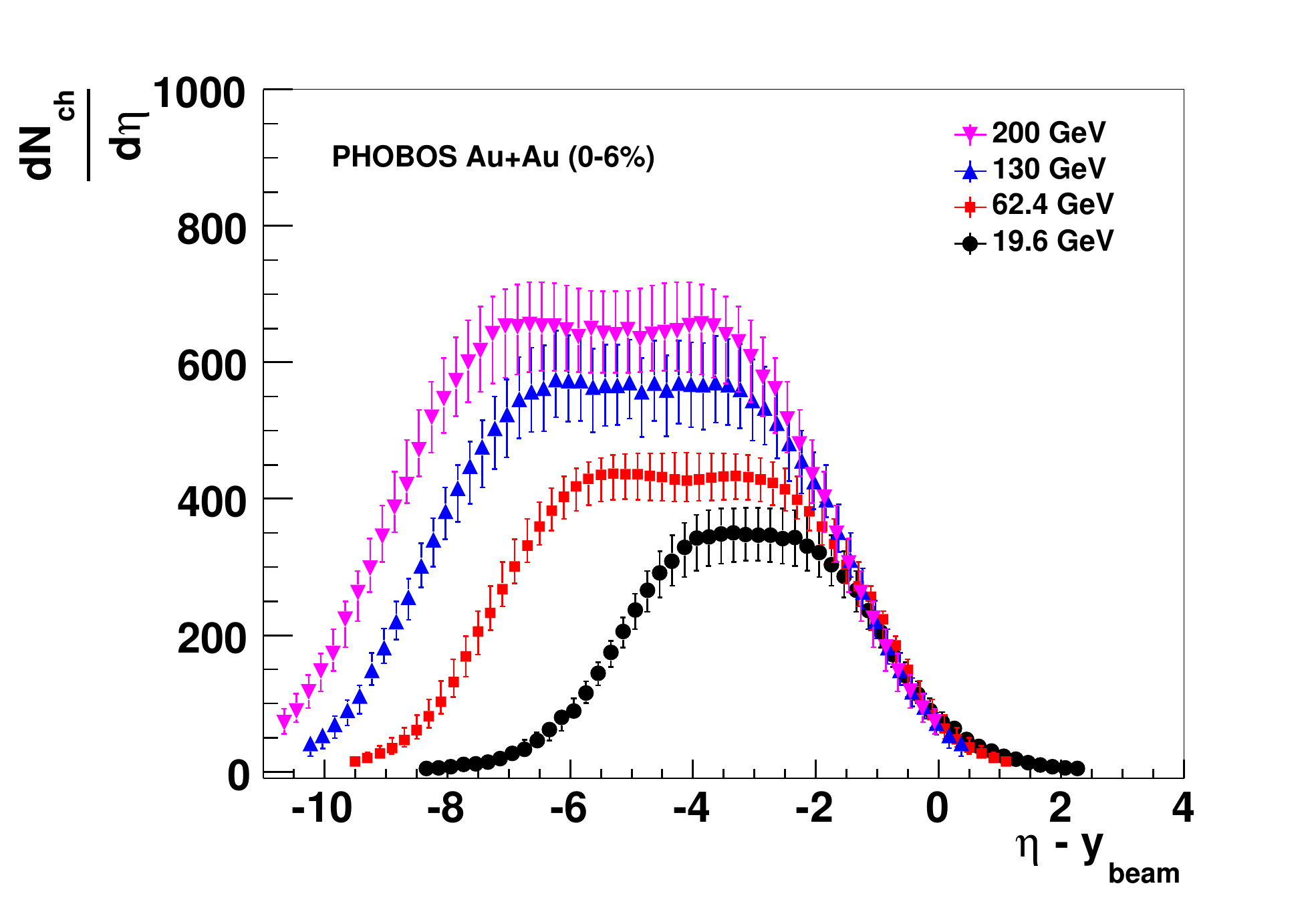}
\caption{ Charged particle multiplicity density normalized by
  participant pairs for Au+Au collisions at different energies, shown
  in the rest frame of projectile by using $\eta^{\prime} = \eta - y_{\rm{beam}}$.}
\label{LongScalAuAu}
\end{center}
\end{figure}

\begin{figure}
\begin{center}
\includegraphics[width=3.6in]{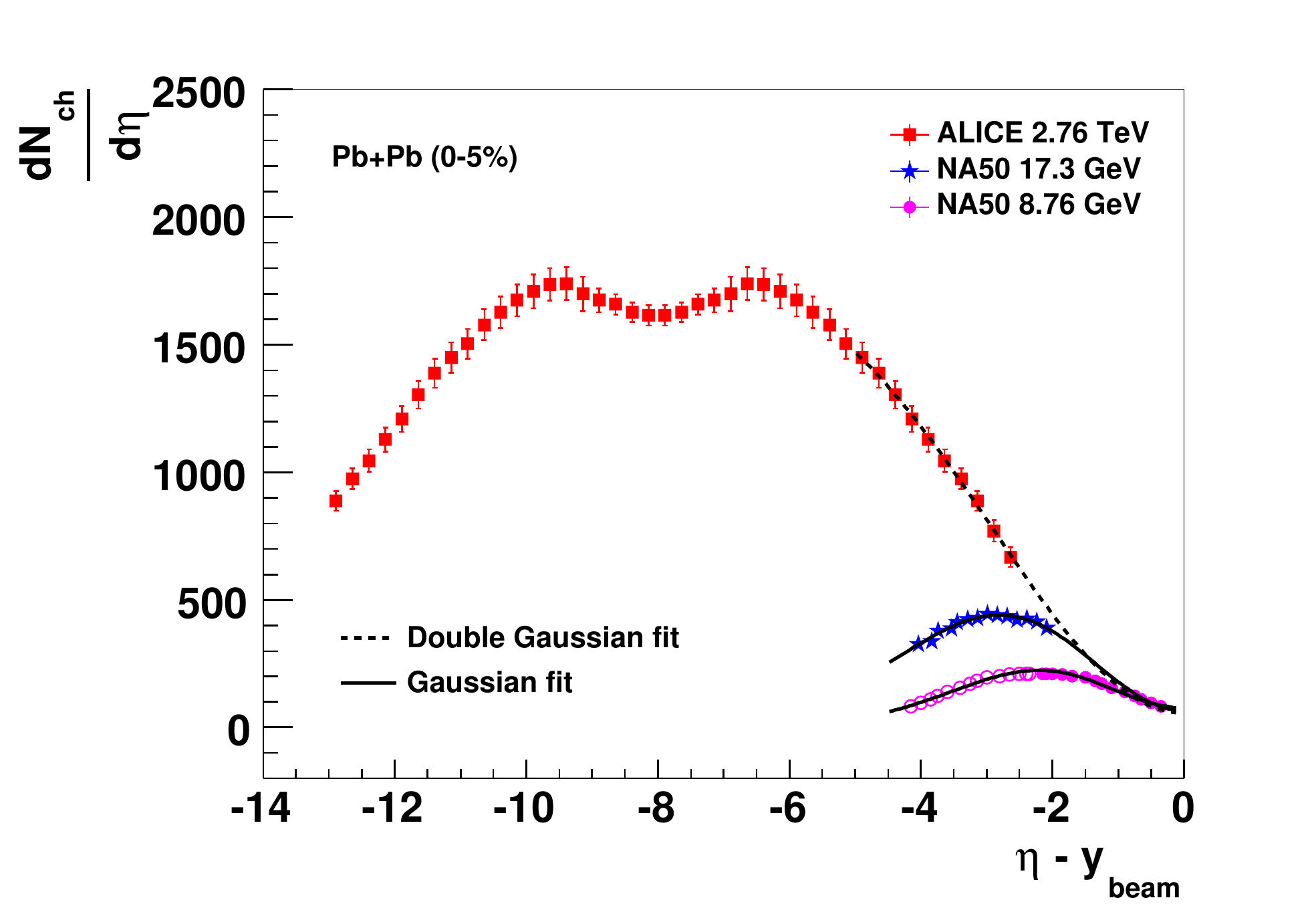}
\caption{ Charged particle multiplicity density normalized by
  participant pairs for Pb+Pb collisions at different energies, shown in the frame one of
  the rest frame of projectile by using $\eta^{\prime} = \eta - y_{\rm{beam}}$.}
\label{LongScalPbPb}
\end{center}
\end{figure}

The normalized charged particle
multiplicity density per participant pair as a function of $\eta^{\prime} = \eta - y_{\rm{beam}}$ for different collision systems and different energies are
shown in Figure \ref{LongScalCuCu}, \ref{LongScalAuAu} and
\ref{LongScalPbPb}.  In Figure \ref{LongScalCuCu}, the data are shown
for Cu+Cu collisions at $\sqrt{s_{NN}}$ = 22.4 GeV, 62.4 GeV and 200
GeV \cite{a1}. In Figure \ref{LongScalAuAu}, the data are shown
for Au+Au collisions at $\sqrt{s_{NN}}$ = 19.6 GeV, 62.4 GeV, 130 GeV
and 200 GeV \cite{a2,a3}. Similarly, in Figure \ref{LongScalPbPb}, the data are shown for
Pb+Pb collisions at beam energies of 40 AGeV, 158  AGeV and at
$\sqrt{s_{NN}}$ = 2.76 GeV \cite{a4,a5}. The charged particle numbers for
Pb+Pb collisions at $\sqrt{s_{NN}}$ = 2.76 TeV at
forward rapidity are estimated by extrapolating the double Gaussian function
used to explain the charged particle distribution \cite{a5}. Figure
\ref{LongScalCuCu}, \ref{LongScalAuAu} and 
\ref{LongScalPbPb} show the saturation or limiting nature of charged
particle density at very high value of $\eta - y_{beam}$ even if the energy of the projectile is
increased. It is also observed in high energy $e^{+}+e^{-}$, $p+p$ and $d+$Au
collisions \cite{b1a, b1b}. The
hypothesis of limiting fragmentation assumes that the hadronic cross
section approaches an asymptotic value at high
energy \cite{totem}. That means the hadronic excitation and breakup probability are
almost independent of projectile energy. But later it is found that
the hadronic cross section increases with increase of center of mass
energy. The most spectacular fact of this hypothesis is that still
this phenomenon is observed for a wide range of collision
energies. Later this limiting fragmentation was tried to explain through
Color Glass Condensate (CGC) model \cite{b2,b3}. The gluon saturation picture at very
small $x$ is used to understand this phenomenon. The charged particle
multiplicity density normalized to participant pair obtained from CGC
model is compared with the RHIC data at different energies \cite{b2,b3}. This CGC based model calculations provides reasonable
description of the data at the fragmentation region for $p+p$ and $A+A$
collisions systems by considering different scale parameters and
initial conditions. However, more precise
modelling of the impact parameter dependence of the ``unintegrated'' gluon
distribution functions is demanded in these models. In addition to this,
the precise estimation of final state effects and inclusion of quark distributions into this
frameworks are needed to explain the whole spectrum of data.
\par
In the framework of statistical thermal
model, the extended longitudinal scaling can also be explained upto RHIC
energies \cite{b4}. It is also predicted that the LHC data will not
show the longitudinal scaling. In Ref \cite{b4}, the string percolation model
predictions were
also used to support their predictions \cite{b5}. However, the recent LHC data violate
the predictions from thermal model and follow the universal
longitudinal scaling. It indicates that at LHC, some
non-equilibrium phenomenon may be playing a role, which needs to be understood \cite{b6}. 

\begin{figure}
\begin{center}
\includegraphics[width=3.6in]{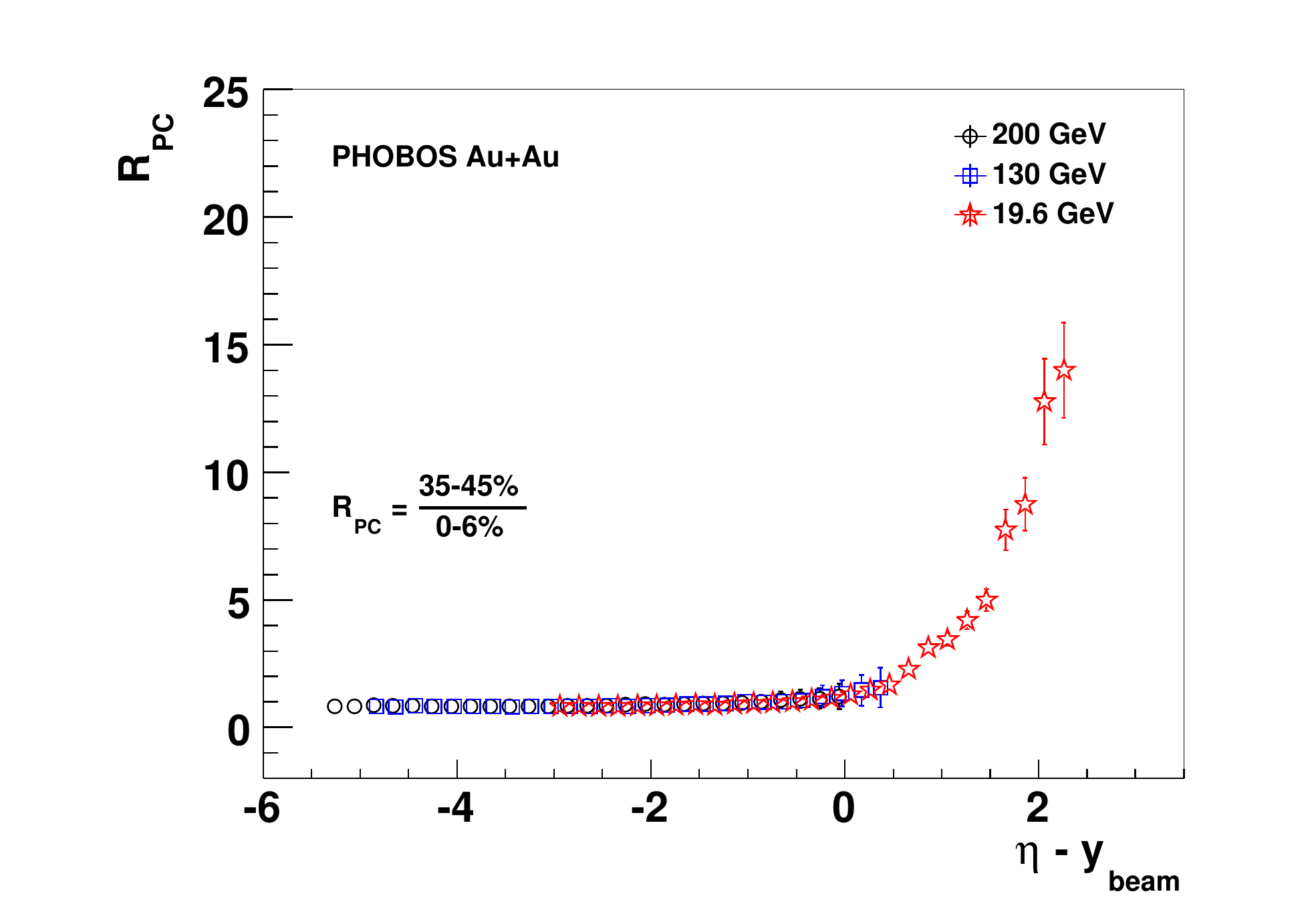}
\caption{$R_{PC}$ as a function of $\eta^{\prime} = \eta - y_{\rm{beam}}$ for Au+Au
  collisions at different energies.}
\label{RcpAu}
\end{center}
\end{figure}

\begin{figure}
\begin{center}
\includegraphics[width=3.6in]{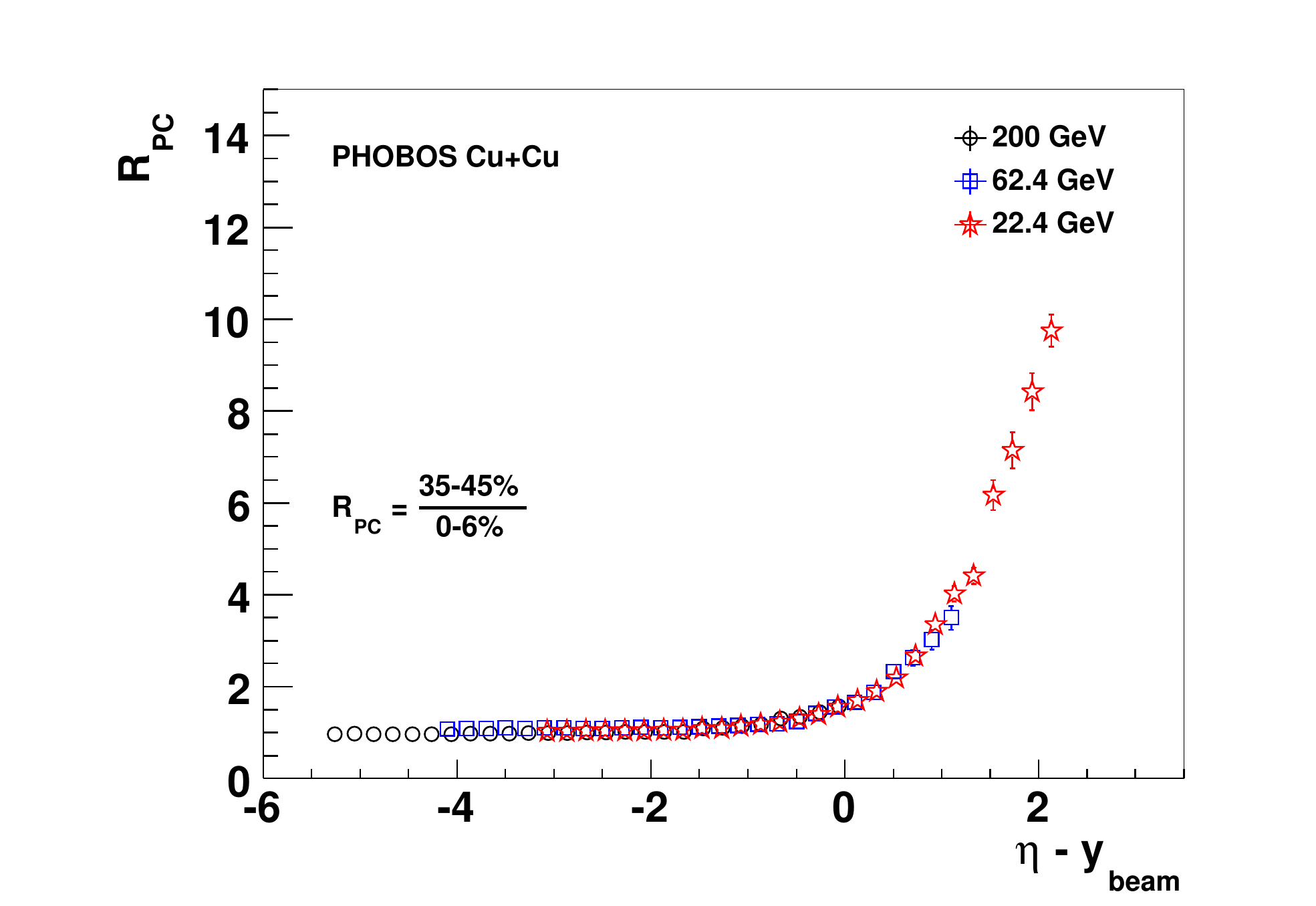}
\caption{$R_{PC}$ as a function of $\eta^{\prime} = \eta - y_{\rm{beam}}$ for Cu+Cu
  collisions at different energies.}
\label{RcpCu}
\end{center}
\end{figure}

\begin{figure}
\begin{center}
\includegraphics[width=3.6in]{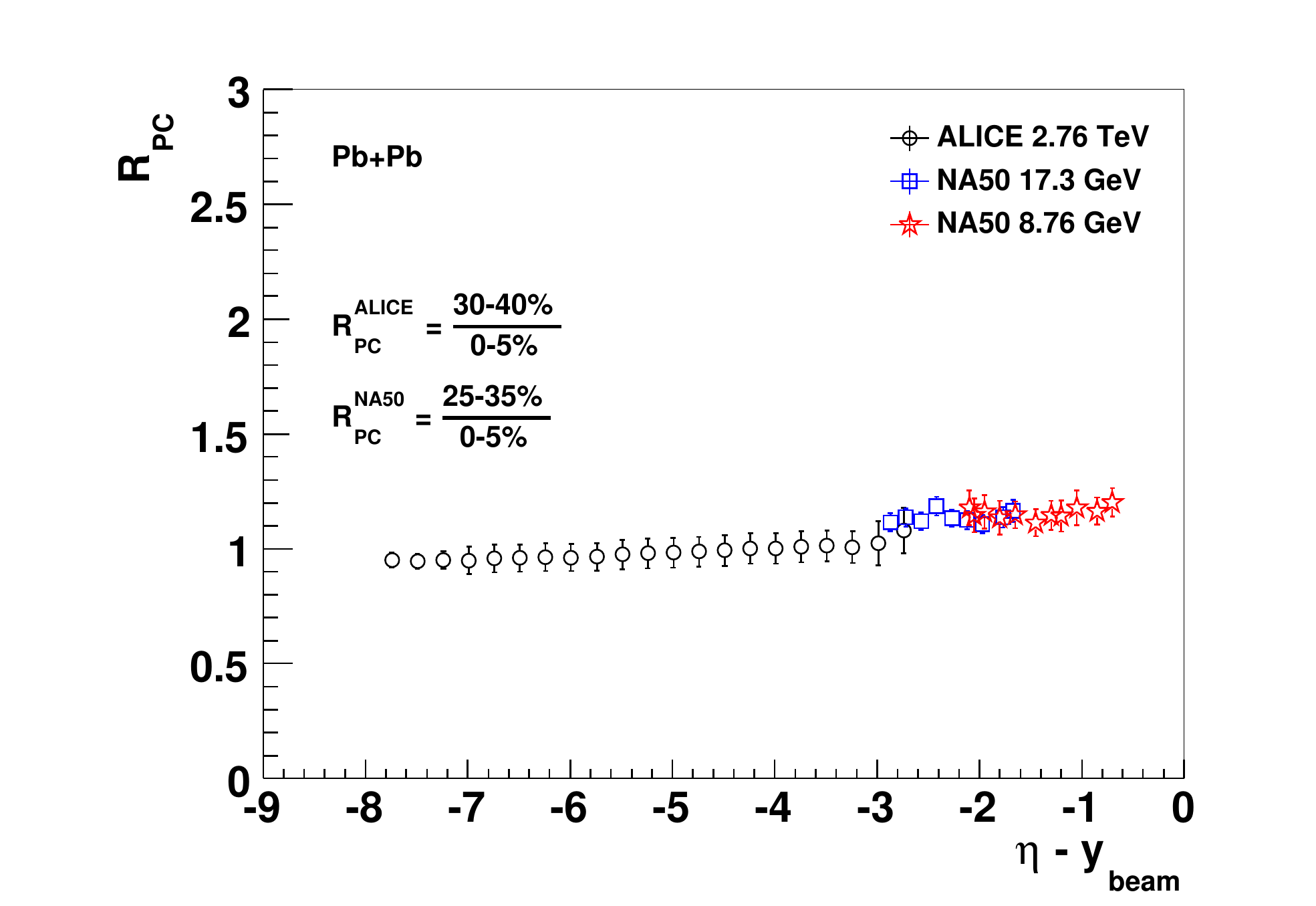}
\caption{$R_{PC}$ as a function of $\eta^{\prime} = \eta - y_{\rm{beam}}$ for Pb+Pb
  collisions at three different energies.}
\label{RcpPb}
\end{center}
\end{figure}
It is reported in Ref \cite{a2} that the shape of the scaled pseudorapidity
density in the rest frame of the projectiles is independent of the beam
energy. However, this shape differs when it is studied as a function
of different centralities. This centrality dependence mainly because
of an excess of particles at high $\eta$ and narrowing of the width of
the pseudorapidity distribution in peripheral $A+A$ collisions. The
excess particles basically originate from nuclear remnant in the
peripheral collisions. So it is realized that  the shape is mainly
a function of collision geometry. To cancel out the geometry effect, it is argued in Ref
\cite{a2,a3} that ratio of $dN_{\rm{ch}}/d\eta$ normalized to $N_{\rm{part}}$ of
central to peripheral events ($R_{PC}$) can be used to ensure the observations on
the energy-independence of the shape called as longitudinal scaling in
the forward rapidities. In Ref \cite{a2}, variable $R_{PC}$ is
defined as follows

\begin{equation}
R_{PC}\left(\eta^{\prime}, 35-45\%\right) =
\frac{\left ( dN_{ch}/d\eta \right )^{35-45\%}/N_{part}^{35-45\%}}{\left(
    dN_{ch}/d\eta\right )^{0-6\%} / N_{part}^{0-6\%}} 
\end{equation}
was introduced which shows the energy independence
behaviour for Au+Au collisions at $\sqrt{s_{\rm{NN}}}$ = 19.6, 130
and 200 GeV. This is shown in Figure \ref{RcpAu} \cite{a2}. The
similar observation is tried for other collision systems.
The $R_{\rm{PC}}$ as a function of $\eta - y_{\rm{beam}}$ for Cu+Cu collisions at
$\sqrt{s_{\rm{NN}}}$= 22.4, 62.4 and 200 GeV are shown in Figure
\ref{RcpCu} \cite{a1}. Similarly, in Figure \ref{RcpPb}, values of $R_{\rm{PC}}$
of Pb+Pb collisions at beam energies of 40 AGeV, 158 AGeV and
$\sqrt{s_{\rm{NN}}}$ = 2.76TeV are shown.
Very interestingly, we observe both for Au+Au ( Figure \ref{RcpAu})
and Cu+Cu (Figure \ref{RcpCu}) collision data that $R_{\rm{PC}}$ is independent of collision
energy. For Pb+Pb collisions at 2.76 TeV, the peripheral events
corresponding to 20-30$\%$ centrality and central events of 0-5$\%$ centrality \cite{a5}. 
For 158 AGeV and 40 AGeV Pb+Pb collisions, peripheral events correspond to 25-35$\%$ centrality and central events correspond to 0-5$\%$ centrality \cite{a4}. 
From Figure \ref{RcpPb}, it is difficult to conclude about the Pb+Pb collision data for the
three energies as the data are not available for the whole
pseudorapidity range as far as this discussion is concerned. However,
the trend of the $R_{PC}$ values as a function of $\eta - y_{beam}$ in
Figure \ref{RcpPb} goes in line with the observations at RHIC.

\subsection{Factorization}
In a typical heavy-ion collision process, the nucleons in the overlap
zones are called as participant nucleons which must have suffered at
least one inelastic collision. Hence, the charged particles
produced in the collision may have some relation with the number of
participant nucleons in the reaction zone as well as the number of
binary collisions. A nucleus-nucleus collision can be thought of
superposition of many individual $p+p$ collisions. So the final charged
particle density should have some empirical relationship with the
$\langle N_{\rm{part}} \rangle$ and number of binary collisions
($N_{\rm{coll}}$). In the framework of ``wounded nucleon model'', it is
observed that the $\frac{dN_{ch}}{d\eta}$ scales
with some power of $N_{\rm{part}}$ upto
the SPS energy \cite{c1}. That is called as power law fit and is given by,
\begin{equation}
\frac{dN_{ch}}{d\eta} \propto N_{part}^{\alpha}
\end{equation}
where $\alpha$ is found to be $\sim 1$ for SPS energies. This linear
relation with $N_{\rm{part}}$ is interpreted as that the particle production
upto SPS energies is mainly from the soft processes. However, the
particle multiplicity at RHIC energies could not be explained by the
above relationship. Then a two-component model was adopted which
incorporate both the contribution of soft and hard processes by considering
the $\langle N_{\rm{part}} \rangle$  and $\langle N_{\rm{coll}} \rangle$ to describe the final
state hadron multiplicity \cite{c1,c2}. The two-component model is given as,
\begin{equation}
\frac{dN_{ch}}{d\eta} = (1-x) n_{pp} \frac{\langle N_{part}
  \rangle}{2} + x n_{pp} \langle N_{coll} \rangle .
\end{equation}
 where $n_{pp}$ is the measured multiplicity density in $p+p$
 collisions due to $x$ fraction of hard processes and ($1-x$) fraction
 represents the soft process.
\par
 Number of binary collisions is proportional to nucleon-nucleon
inelastic cross section ($\sigma^{\rm{NN}}_{\rm{inel}}$). With
increase of collision energy, the $\sigma^{\rm{NN}}_{\rm{inel}}$ also
increases \cite{totem}. This results in dramatic increase of $N_{\rm{coll}}$ with
the increase of collision energy and therefore, the contribution of hard process will
be dominant for particles production. So it is expected
that there will be a strong centrality dependence of pseudorapidity
distributions at higher energies. This can be tested by
taking the ratio of scaled yield at the respective centralities at
different energies. It is reported in
Ref \cite{c3,c4} that the centrality dependence of particle production in the midrapidity
exhibits factorization of beam energy and collision centrality as follows,

\begin{equation}
\frac{2}{\langle N_{part} \rangle} \frac{dN_{ch}}{d\eta} = f(s) ~g(N_{part})
\label{eqn:factorization}
\end{equation}

\begin{figure}
\begin{center}
\includegraphics[width=3.6in]{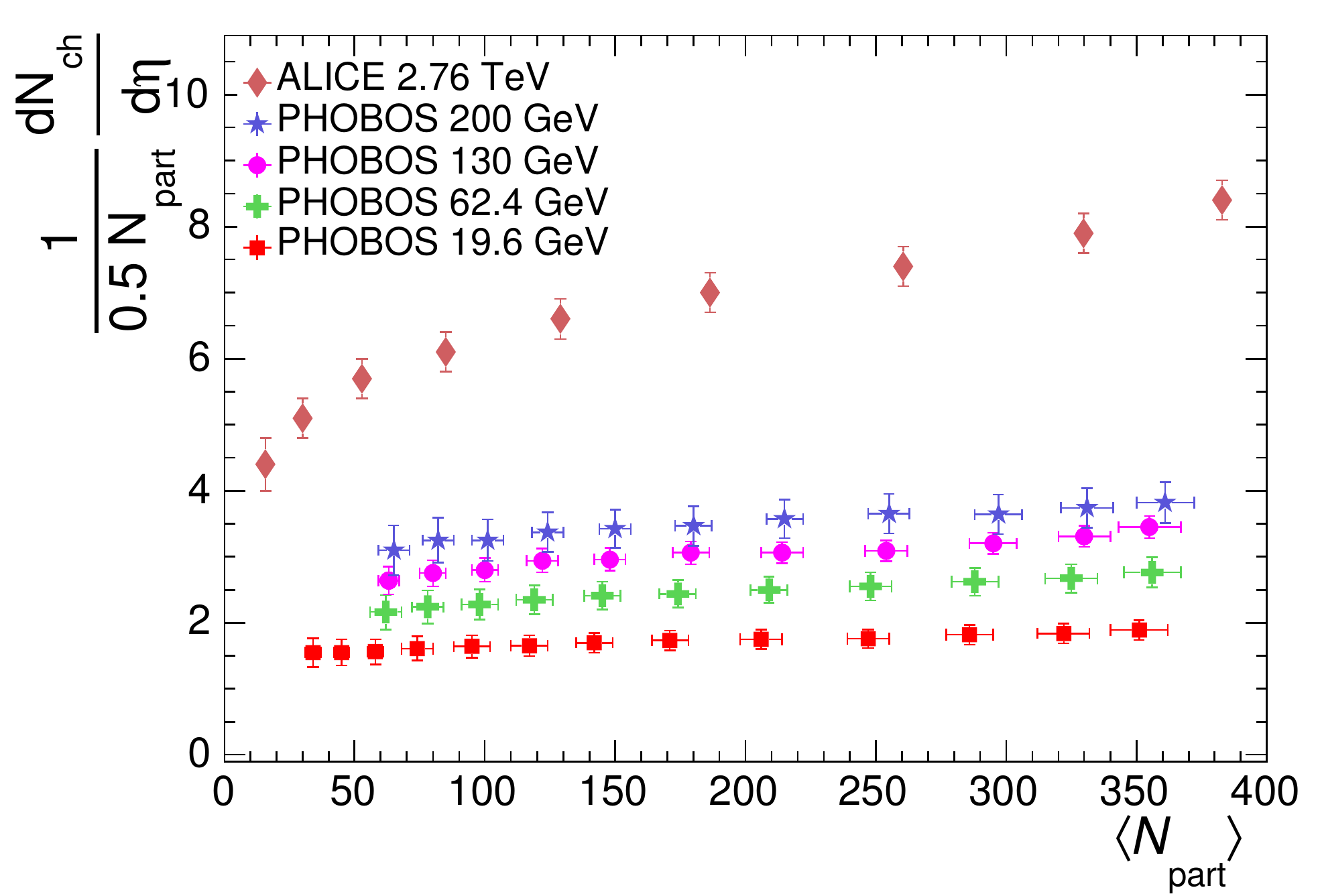}
\caption{ $N_{\rm{part}}$ normalized charged particle density for different
  collision energies.}
\label{fact1}
\end{center}
\end{figure}

Eq (9) basically illustrates the energy-centrality factorization. In
the right hand side of Eq (9), the first term, i.e. $f(s)$, depends
on the energy and the second term, i.e. $g(N_{\rm{part}})$, depends on the $\langle N_{\rm{part}} \rangle$. In the
midrapidity, the charged particle multiplicity density normalized to
the participant pair, ($\langle N_{\rm{part}} \rangle/2$) at different
energies is shown in Figure \ref{fact1}. The collision data are fitted
with the parametrized form of right hand side of Eq (9). For Au+Au
collision, the parametrized forms of $f(s)$ and $g(N_{\rm{part}})$ found
from Ref \cite{c4} are as follows,

\begin{equation}
f(s) =  0.0147 [ln(s)]^{2} + 0.6
\end{equation}
\begin{equation}
g(N_{part})  =  1 + 0.095 N_{part}^{1/3}
\end{equation}

\begin{figure}
\begin{center}
\includegraphics[width=3.6in]{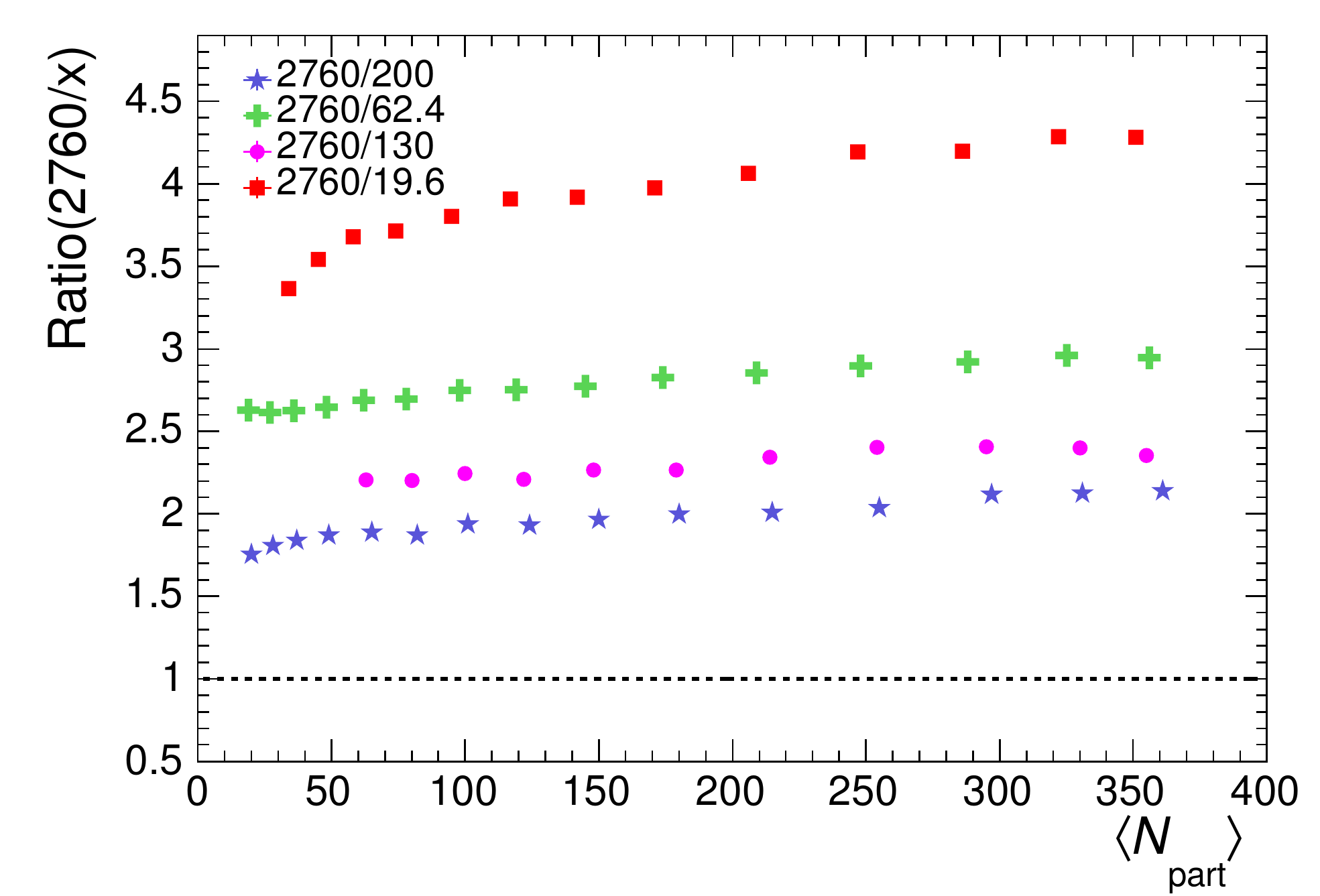}
\caption{ Ratio of charged particle density for different energies normalized
  per participant pair as a function of collision centrality.}
\label{fact2}
\end{center}
\end{figure}

\begin{figure}
\begin{center}
\includegraphics[width=3.6in]{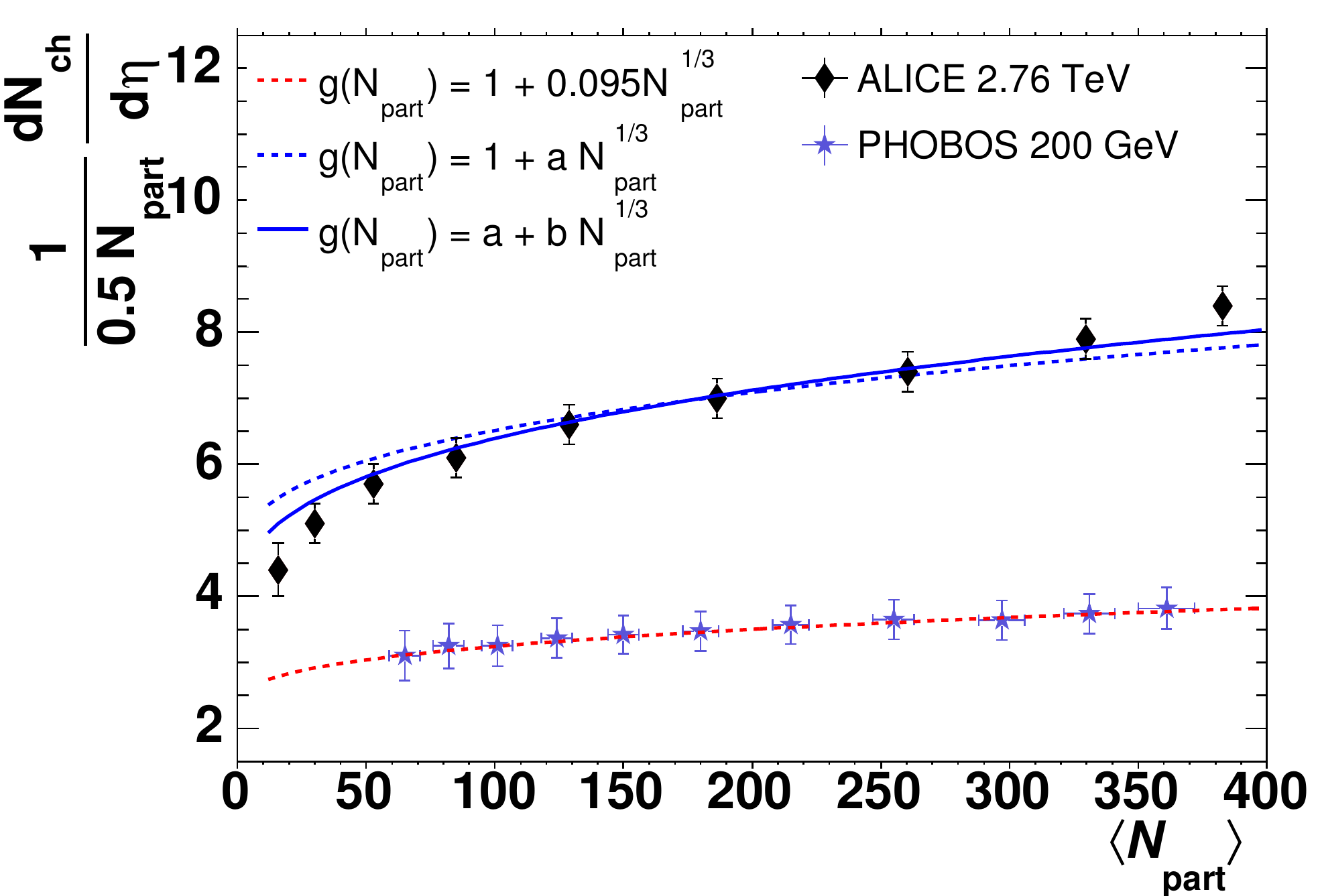}
\caption{ Factorization of RHIC and LHC data. The data points are fitted with various parametrized forms of Eq. \ref{eqn:factorization}.}
\label{fact3}
\end{center}
\end{figure}

Similarly, for Cu+Cu collisions, the co-efficients of $f(s)$ does not
change. However, the co-efficient of $N_{\rm{part}}^{1/3}$ in $g(N_{\rm{part}})$
changes, which is given by,
\begin{equation}
g(N_{part}) = 1 + 0.129 N_{part}^{1/3}
\end{equation}
In Figure \ref{fact2}, ratios of the charged particle multiplicity density normalized to
the participant pair of Pb+Pb collisions at $\sqrt{s_{NN}} = 2.76 $ TeV
and Au+Au collisions data at different energies are shown as a function of $\langle
N_{\rm{part}} \rangle$. This observation implies that the pseudorapidity density of particles
in the midrapidity normalized per participant pair can be
factorized. However, when the collision system changes, the $N_{\rm{part}}^{1/3}$
dependence comes into picture. We tried to fit the parametrized form
of Eq (11) with the LHC data. We keep the form of $f(s)$
same and set one parameter free of $g(N_{\rm{part}})$. However, it doesn't
fit the data. This is shown in Figure \ref{fact3}. Contrary, when both the
parameters of Eq (11) are set free, then it fits well to the data. This
observation contradicts the observation at RHIC. The RHIC data show
that only the co-efficient of $N_{\rm{part}}^{1/3}$ changes when collision system
changes at the same collision energy. However, at LHC energy, the
energy as well as the system size changes. After $\chi^2$
minimization, for better fit, we get the following form of $g(N_{\rm{part}})$ for LHC
data. It can be inferred that some other factor is playing a role for
the particle production at LHC energy in addition to the RHIC energy.
\begin{equation}
g(N_{part}) = 0.833 + 0.142  N_{part}^{1/3}
\end{equation}

\subsection{Expansion dynamics}
The space-time evolution of the fireball created in the heavy-ion
collisions can be explained by relativistic hydrodynamical approach
which assumes that the medium is continuously flowing. The elliptic
flow measurements, the two-particle correlations and transverse
momentum spectra results at RHIC have given ample evidence of a
strongly interacting medium created in the laboratory.  There are many
proposed statistical as well as hydrodynamical models in
the past to explain the multiplicity and expansion dynamics of the
systems. Landau hydrodynamics model is one of them, which is widely
used to explain the expansion of the system produced in
the collision, like $e^{+}+e^{-}$, $p+p$ and $A+A$ \cite{d1}. It has successfully
explained the low energy collision data including the charged pion
data at RHIC \cite{d2}. The form of Landau hydrodynamics has been
evolved with time to explain the global particle multiplicity and the differential rapidity
distribution \cite{d3,d4}. The width of the charged particle density distribution in the
midrapidity can shed some light on the longitudinal expansion
dynamics of the system, velocity of sound and initial and final state
rescattering. A detailed analysis about these are given in
Ref. \cite{d5}. It can also be used to define the degree of stopping or
transparency in the heavy-ion collision reactions. 
\par
The form of Landau-Carruthers rapidity distributions is given as \cite{d3}, 
\begin{equation}
\frac{1}{N} \frac{dN}{d\lambda} = \frac{ exp( -\lambda^{2}/2L) } { ( 2\pi L)^{1/2}} 
\end{equation}
where $\lambda = \eta = -ln~tan( \theta/2 )$ and $L = ln~\gamma = \frac{1}{2}
ln (s/4m^2)$ and $\gamma$, which is equal to $\sqrt{s_{NN}}/2m_{p} $,
is the Lorentz contraction factor. Here $m_{p}$ is the mass of the
proton. The Gaussian form of it is given as,
\begin{equation}
\frac{dN}{dy} \propto exp(- y^{2}/2L) .
\end{equation}
Later in Ref \cite{d4}, the pseudorapidity variable is substituted by
rapidity to describe the distribution appropriately (the rapidity
distribution of charged particles differs from pseudorapidity distribution at the smaller
rapidity region). Then the rapidity distribution is given as \cite{d4},
\begin{equation}
\frac{dN}{dy} \propto exp( \sqrt{y_{b}^{2} - y^2} ) ,
\end{equation}

where the beam rapidity, $y_{b}$, in the center of mass frame is
$\text{cosh}^{-1} (\sqrt{s_{NN}}/2m_{p}) = ln(\sqrt{s_{NN}}/m_{p})$. Then Ref
\cite{d4} connects the total entropy of the system with the
number density such that their ratio is constant for a thermally equilibrated system. 

\begin{figure}
\begin{center}
\includegraphics[width=3.6in]{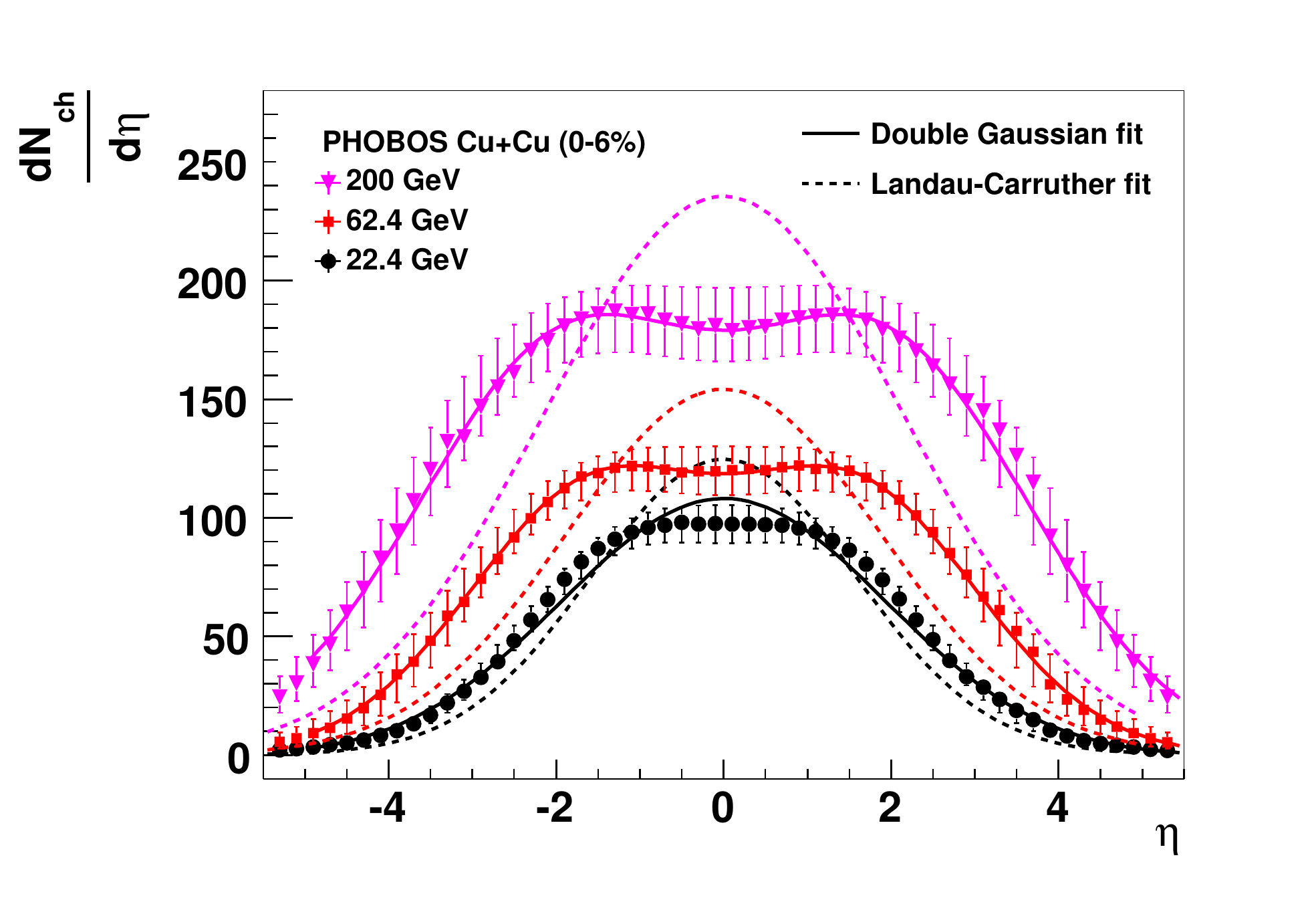}
\caption{The charged particle multiplicity density distributions of Cu+Cu collisions
  at three different energies, fitted with
  double Gaussian function and Landau-Carruthers functions.}
\label{HydroCu}
\end{center}
\end{figure}

\begin{figure}
\begin{center}
\includegraphics[width=3.6in]{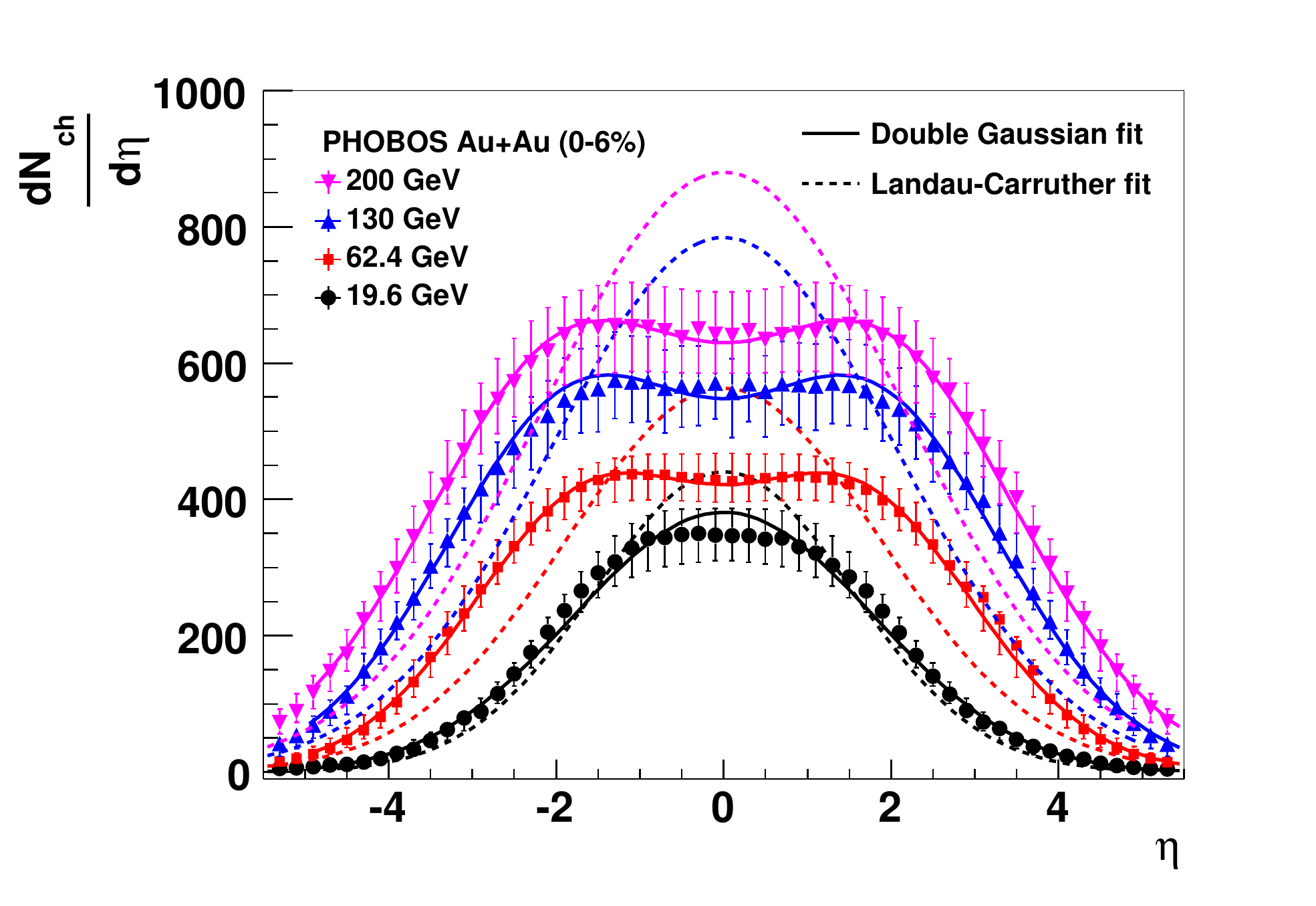}
\caption{The charged particle multiplicity density distributions of Au+Au collisions
  at three different energies, fitted with
  double Gaussian function and  Landau-Carruthers functions.}
\label{HydroAu}
\end{center}
\end{figure}

\begin{figure}
\begin{center}
\includegraphics[width=3.6in]{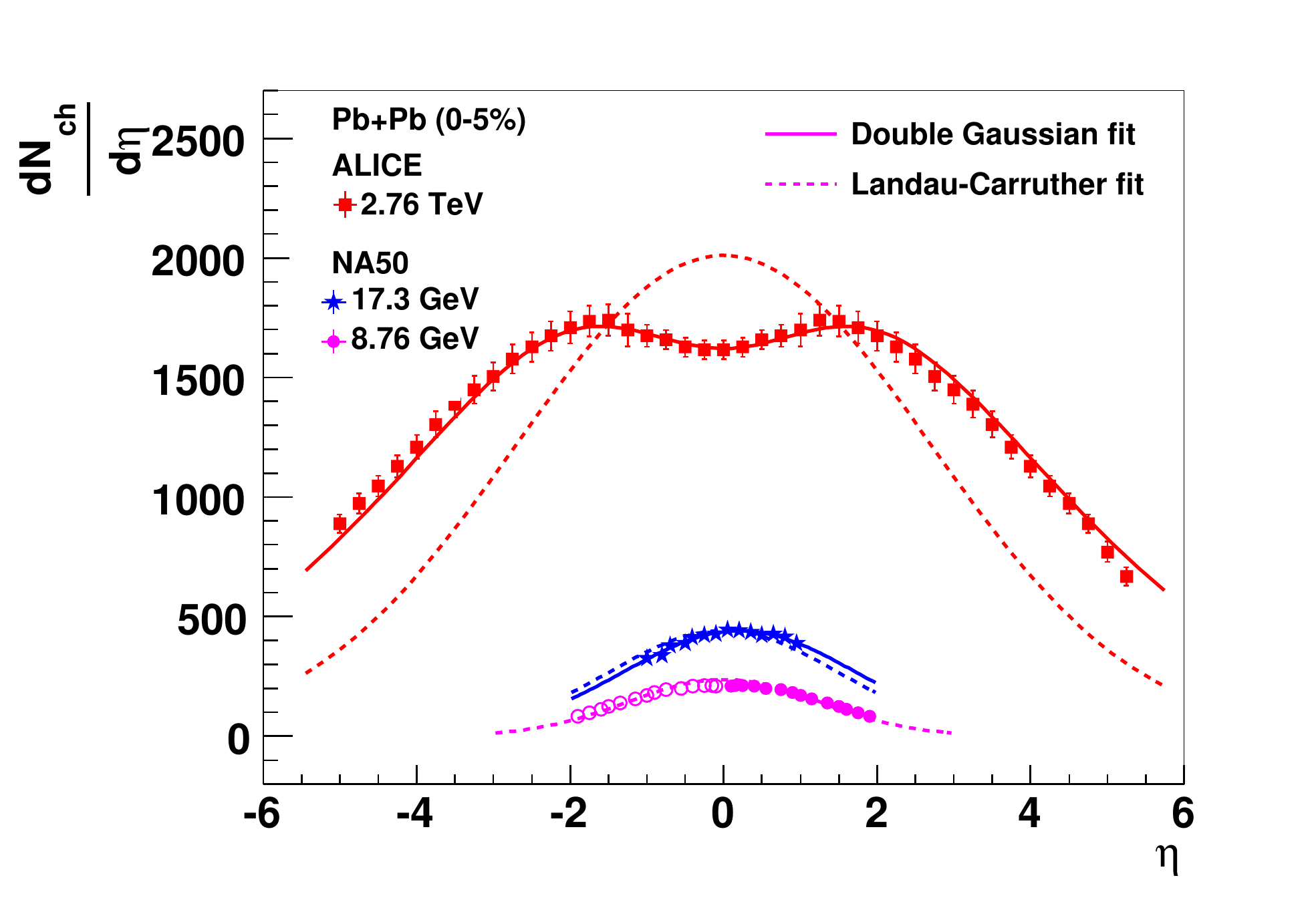}
\caption{The charged particle multiplicity density distributions of Pb+Pb collisions
  at three different energies, fitted with
  double Gaussian function and Landau-Carruthers functions.}
\label{HydroPb}
\end{center}
\end{figure}

\begin{figure}
\begin{center}
\includegraphics[width=3.6in]{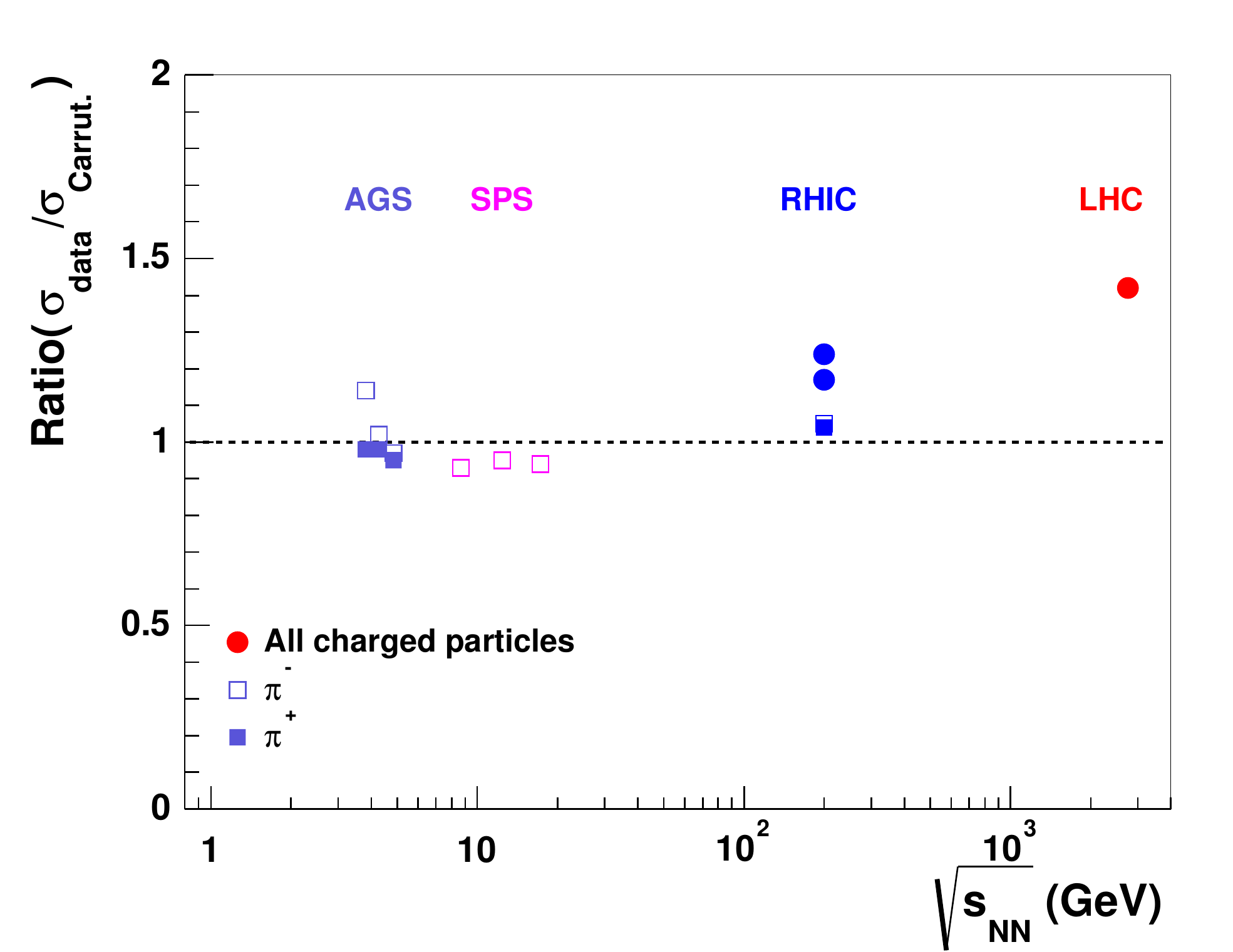}
\caption{The ratio of the widths of the data to the width obtained by
  fitting Landau-Carruthers function to the charged particle
  multiplicity density, as a function of collision energy.}
\label{WidthRatio}
\end{center}
\end{figure}
\par
It is found that when the transformation of the distribution is made
to rest frame of one of the colliding nuclei, the Gaussian form as
given in Eq (15) shows the limiting fragmentation behaviour. And
surprisingly, by setting some parameters, it also matches multiplicity
distributions with the CGC calculations \cite{d2}.

\par
In this review, we have tried to see the agreement of
pseudorapidity distributions of charged particles by
Landau-Carruthers function. The advantage of fitting
Landau-Carruthers form to the data is that the $\lambda$ variable
used in the function has similar form as the pseudorapidity. The multiplicity distribution of Cu+Cu
collision data as a function of rapidity for different energies are shown in Figure
\ref{HydroCu}. The Cu+Cu collision data are fitted with the
Landau-Carruthers functions. The multiplicity distribution of Au+Au
collisions as a function of rapidity for different energies are shown
in Figure \ref{HydroAu}. Similarly, the rapidity distributions of
charged particles of Pb+Pb collisions at different energies are shown in Figure
\ref{HydroPb}. The $dN_{\rm{ch}}/d\eta$ distribution of charged particles are
also fitted with double-Gaussian functions. The width of the
distributions obtained from the data and the models are divided and
shown as a function of collision energy in Figure \ref{WidthRatio}. It is
observed from Figure \ref{WidthRatio} that Landau-Carruthers
hydrodynamics form explains the data starting from AGS, SPS to RHIC as
the ratio is closed to one. However, the LHC data is far from one
which implies that Landau hydrodynamics does not explain the expansion
dynamics at LHC energy. 

 \subsection{Energy dependence of multiplicity density}
\begin{figure}
\begin{center}
\includegraphics[width=3.6in]{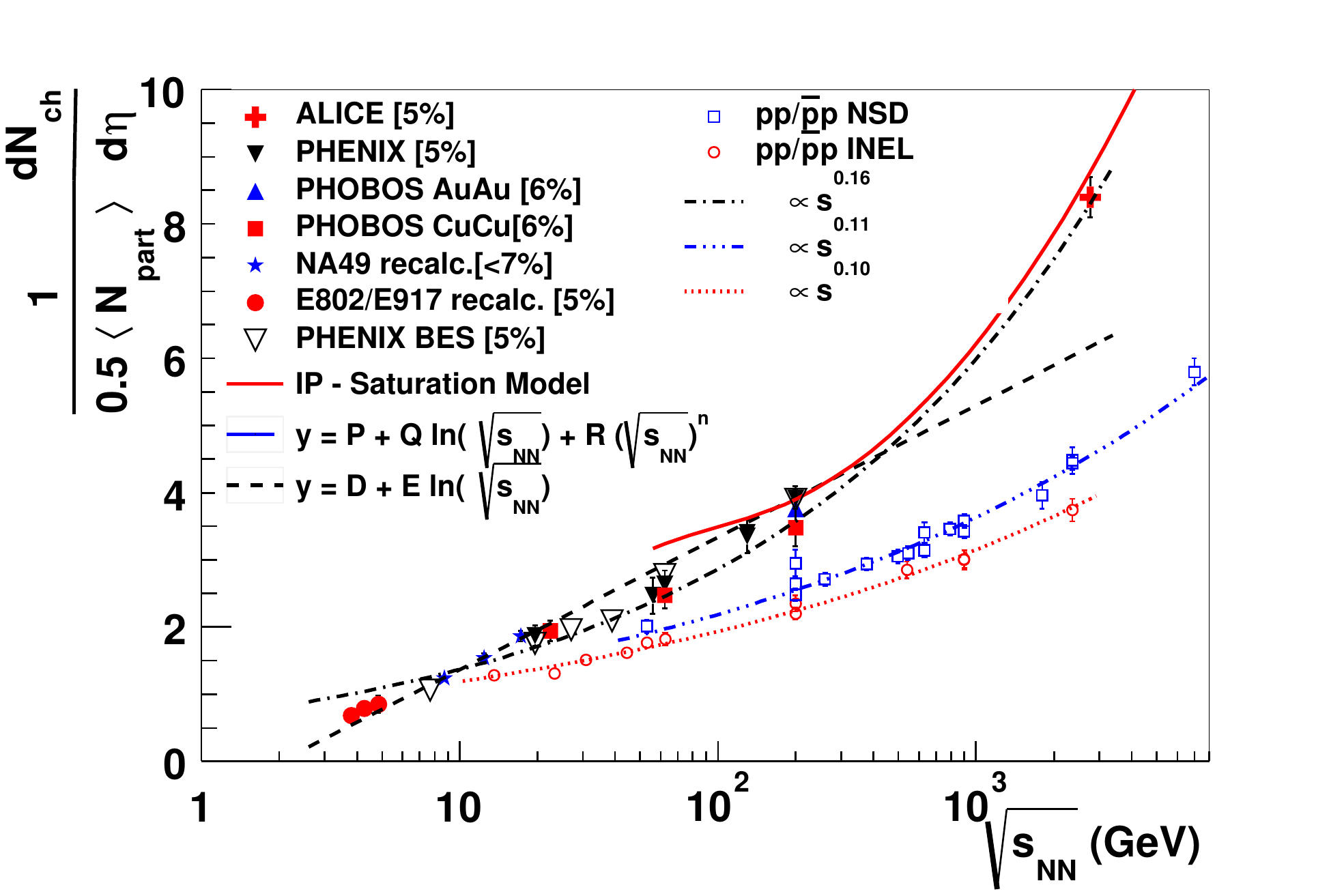}
\caption{ Energy dependence of charged particle multiplicity density
  distribution per participant pair for most central collisions at midrapidity. Compared are the corresponding measurements in $pp/p{\bar{p}}$ NSD and INEL collisions.}
\label{EngDep}
\end{center}
\end{figure}
The energy dependence of charged particle multiplicity density
  distribution per participant pair for most central collisions in heavy-ion collisions at
 midrapidity and for nucleon-nucleon non-single diffractive (NSD) and inelastic (INEL) collisions,
 as a function of collision energy is shown in Figure \ref{EngDep}. The data points are from different energy and different collision species. To explain the normalized particle distribution in the midrapidity, different phenomenological functions are fitted.  Upto top RHIC energy $\frac{dN_{\rm{ch}}}{d\eta}$ for heavy-ion collisions is well described by a logarithmic function. However, the LHC data is underestimated by logarithmic function upto $26\%$. On the other hand, a power law fit seems to overestimate the low energy data for nucleus-nucleus collisions while explaining the high energy data upto LHC energies. 
Looking at the low-energy and high-energy behaviours of charge particle production being well-described by a logarithmic function and power-law functions, respectively, we have tried to fit a hybrid function (a combination of both) and find a very good agreement with the nucleus-nucleus data at all energies upto LHC 2.76 TeV. The physics motivation of the hybrid function can be explained by considering the result by Wolschin {\it et al.} which states that at high energy, charged particle multiplicity can be explained by a combination of midrapidity gluonic source predicted by the power law function and a fragmentation source predicted by logarithmic function \cite{wolschin}. The predictions from IP-saturation model for the top RHIC energy and higher are also shown for a comparison with the corresponding nucleus-nucleus experimental data.
For a direct comparison with A+A data, we have put together the $p+p(\bar{p})$ NSD and INEL data. Both the data seem to fit to a power-law behaviour with the power decreasing while going from A+A to  $p+p(\bar{p})$ collisions.

\subsection{Scaling of $N_{\rm{ch}}^{\rm{total}}$ with $N_{\rm{part}}$}
\begin{figure}
\begin{center}
\includegraphics[width=3.6in]{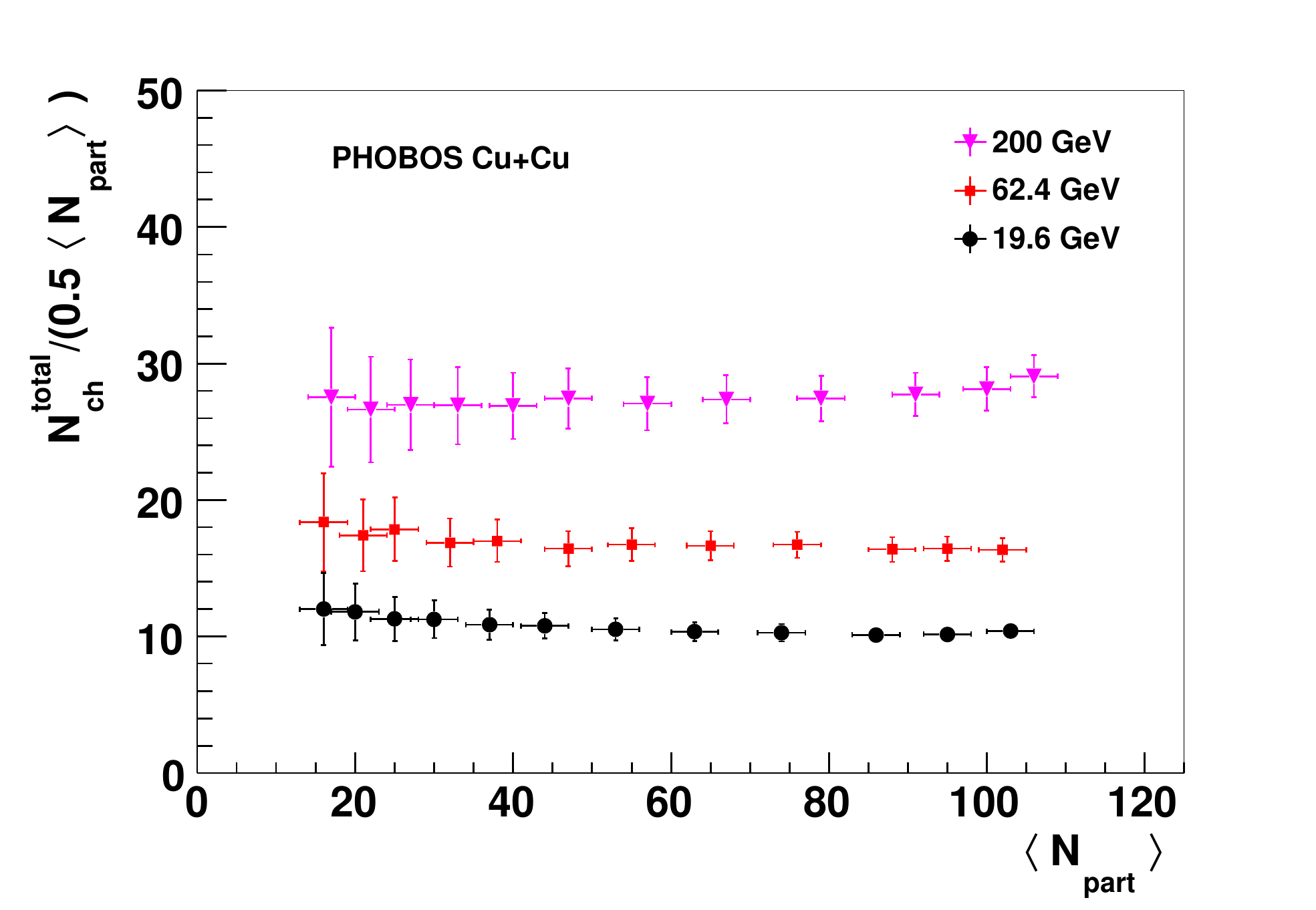}
\caption{$N_{\rm{ch}}^{\rm{total}}$ normalized to participant pair as a function of
$N_{\rm{part}}$ for Cu+Cu collisions at different collision energies.}
\label{NtotCu}
\end{center}
\end{figure}

\begin{figure}
\begin{center}
\includegraphics[width=3.6in]{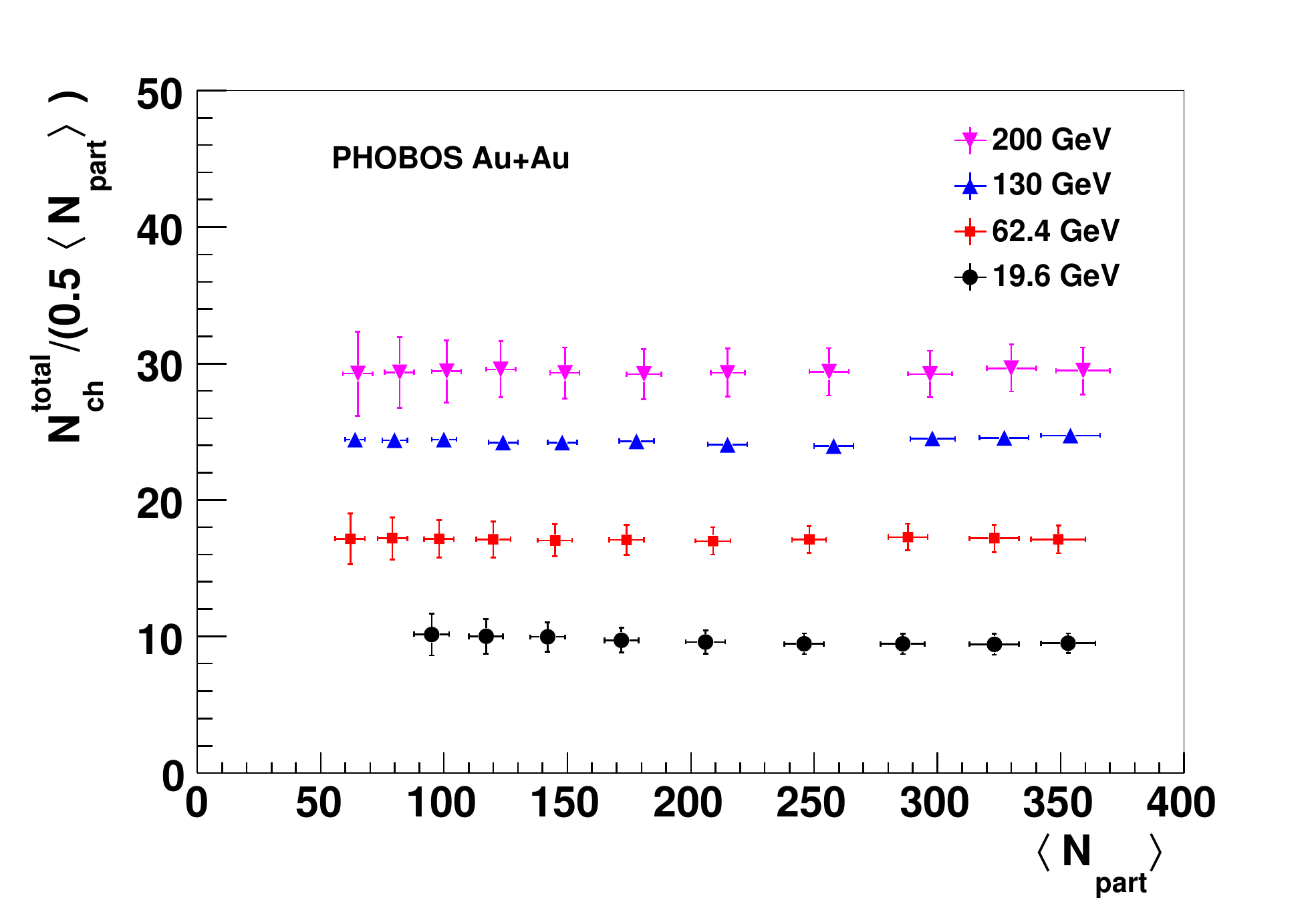}
\caption{$N_{\rm{ch}}^{\rm{total}}$ normalized to participant pair as a function of
$N_{\rm{part}}$ for Au+Au collisions at different collision energies.}
\label{NtotAu}
\end{center}
\end{figure}

It is observed that the particle multiplicity at midrapidity does not
scale with the number of participant nucleons, i.e. $N_{\rm{part}}$. It is
observed from Ref \cite{a3, c4}
that the total charged particle measured over a wide range of
pseudorapidity, when normalized per participant pair, scales
with $N_{\rm{part}}$. We considered different collision energies and
collision systems to see the scaling behaviour of total charged
particles. The normalized $N_{\rm{ch}}^{\rm{total}}$ per participant pair as a
function of $N_{\rm{part}}$ are shown in Figure \ref{NtotCu} and Figure
\ref{NtotAu} for Cu+Cu and Au+Au collisions, respectively. The error
bars shown in the figures are statistical only.
 
It is observed from Figure \ref{NtotCu} and Figure \ref{NtotAu} that
the participant pair normalized $N_{\rm{ch}}^{\rm{total}}$ scales perfectly with
$N_{\rm{part}}$ within the statistical uncertainties. Both for Cu+Cu and Au+Au systems, the normalized value of $N_{\rm{ch}}^{\rm{total}}$ with respect to $N_{\rm{part}}$ is constant as a function
of $N_{\rm{part}}$ and increases with increase of collision energy. It
implies that modifications to particle production at forward
rapidities are strongly correlated with compensating changes at
midrapidity.

\subsection{Energy dependence of $N_{ch}^{total}$}

\begin{figure}
\begin{center}
\includegraphics[width=3.5in]{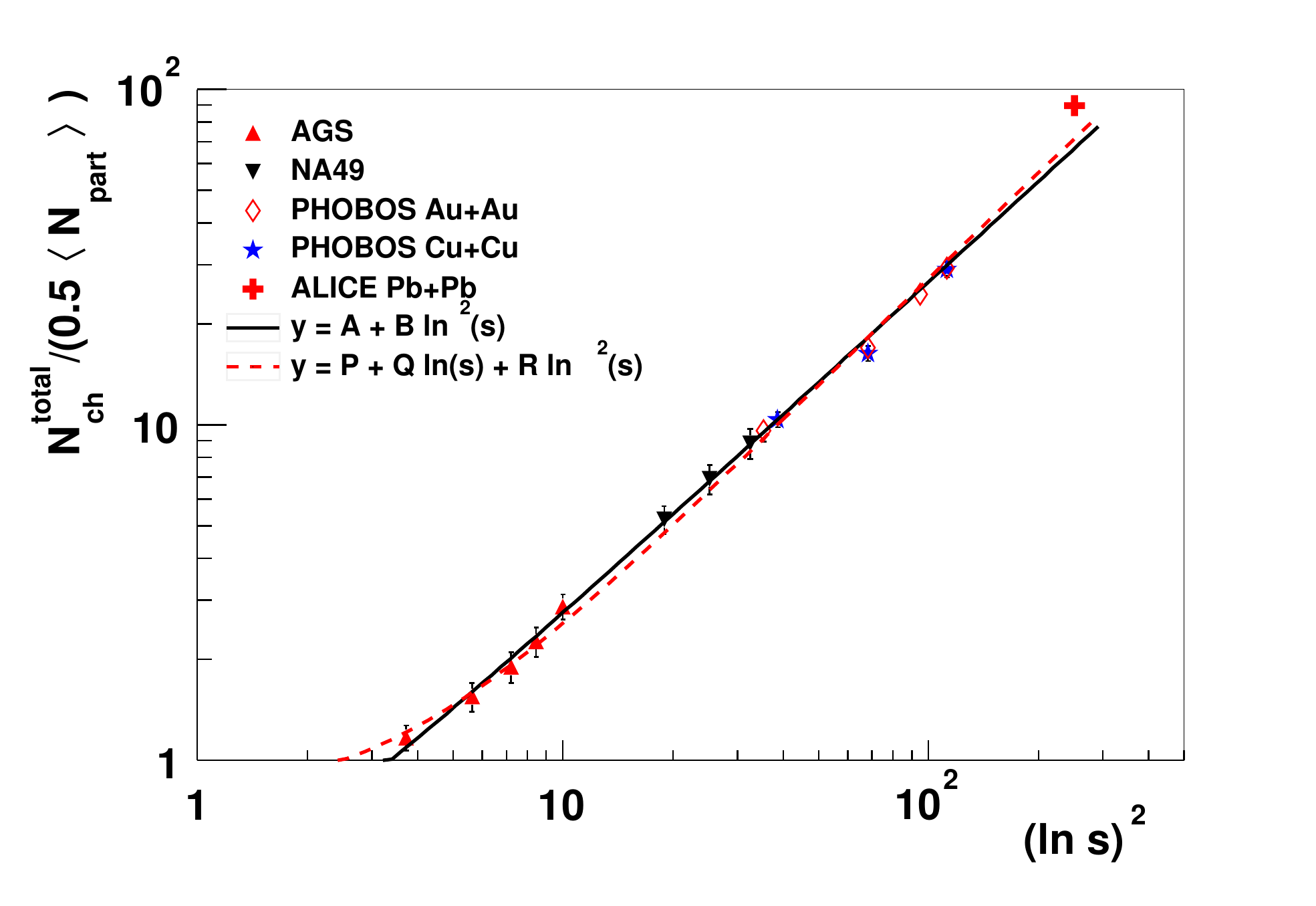}
\caption{ Energy dependence of total charged particle multiplicity per
  participant pair for most central collisions.}
\label{engDepNch}
\end{center}
\end{figure}

As discussed earlier, the total charged particles normalized par
participant pair ($N_{\rm{ch}}^{\rm{total}}/0.5~\langle N_{\rm{part}} \rangle$ ) for Cu+Cu, Au+Au
systems at different collision energies are independent of
centrality. In addition to this, the $N_{\rm{ch}}^{\rm{total}}$ value increases
with increase of energy for all centralities. The energy dependence of
$N_{\rm{ch}}^{\rm{total}}$ from AGS to LHC is shown in Figure \ref{engDepNch}.
It is observed that the $dN_{\rm{ch}}/d\eta$ distribution in the
midrapidity is almost flat \cite{c4} and the width of the distribution
decreases with decrease of collision energy. The fragmentation
region can be explained by $dN_{\rm{ch}}/d\eta = \alpha ( y_{\rm{beam}} +
\eta_{0} - \eta)$. So the overall $dN_{\rm{ch}}/d\eta$ distribution now can
be thought of a trapezoid and hence, the total charged particles can be
given by the trapezoidal rule as follows \cite{c4}.
\begin{equation}
N_{ch}^{tpz} = \frac{dN_{ch}|_{0}}{d\eta} \left( 2\eta_{0} + 2y_{beam}
  - \frac{\langle N_{part} \rangle}{2\alpha}
  \frac{dN_{ch}|_{0}}{d\eta}  \right)
\label{eqnTrap}
\end{equation}
As $y_{\rm{beam}} \simeq \frac{1}{2} \text{ln} s_{\rm{NN}} -
\text{ln}(m_{0}c^{2})$ for $\sqrt{s_{\rm{NN}}} \gg m_{0}$, $m_{0}$ is the
mass of the nucleon, Eq. \ref{eqnTrap} reduces to

\begin{equation}
\frac{N_{ch}^{tpz}}{0.5~\langle N_{part} \rangle} \simeq  \text{A} + \text{B}~\text{ln}
~s_{NN} +\text{C} {(\text{ln}~s_{NN})}^{2} .
\label{eqnTrap1}
\end{equation}
To explain the evolution of $N_{\rm{ch}}^{\rm{total}}/(0.5~\langle N_{\rm{part}} \rangle)$ with respect
to $\sqrt{s_{\rm{NN}}}$, parametrized form of Eq. \ref{eqnTrap1} is used
and fitted with the collision data which is shown by a dashed line in
Figure \ref{engDepNch}. It is found that this equation explains the PHOBOS Cu+Cu and Au+Au
data at RHIC. However, it fails to explain the data at lower energies. Only after
considering the leading term (${\text{ln}~s_{\rm{NN}}})^{2}$ in Eq. \ref{eqnTrap1}, it explains the whole spectrum of energy dependence of total  charged particles very nicely starting from $\sqrt{s_{\rm{NN}}}$ = 2.4 GeV to $\sqrt{s_{\rm{NN}}}$ = 200 GeV. The general form is,
\begin{equation}
\frac{N_{ch}^{tpz}}{0.5~\langle N_{part} \rangle} =\text{A} + \text{C}~ {(\text{ln}~s_{NN})}^{2} .
\label{eqnTrap2}
\end{equation}
The fitting of Eq. \ref{eqnTrap2} to the data point is shown in Figure \ref{engDepNch}
by the solid line. It can be seen from Figure  \ref{engDepNch} that
derived form of trapezoidal rule given by Eq. \ref{eqnTrap1} and
Eq. \ref{eqnTrap2} underestimate the $N_{\rm{ch}}^{\rm{total}}$ of Pb+Pb
collisions at $\sqrt{s_{\rm{NN}}}$ = 2.76 TeV measured by the ALICE experiment. This is
because the $dN_{\rm{ch}}/d\eta$ distribution of Pb+Pb data has a dip which
in principle deviate from a trapezoidal shape. 
\par 
Measuring $N_{\rm{ch}}^{\rm{total}}$ as a function of ${(\sqrt{s_{\rm{NN}}})}^{1/2}$ is
important in terms of Landau hydrodynamics. According to Landau
hydrodynamics, the ratio of entropy density to the number density for
a thermally equilibrated system is constant. In other words, the number
density is proportional to the entropy density and hence the total
number of particles is proportional to the total entropy. To be noted
that during the hydrodynamic expansion of the system, the total
entropy remains constant. So by measuring the total observed particles,
the initial entropy can be determined and vice versa. For a system
which is in local thermal equilibrium, the entropy density is
proportional to the energy density and under this assumption, we can
arrive at this relationship of $N_{\rm{ch}}^{\rm{total}}$ with respect to the
center of mass energy $\sqrt{s_{\rm{NN}}}$ as follows \cite{d4}.
\begin{equation}
\frac{N_{ch}^{total}}{0.5~\langle N_{part} \rangle} = K {(\sqrt{s_{NN}}/GeV)}^{1/2}
\label{landau}
\end{equation}

\begin{figure}
\begin{center}
\includegraphics[width=3.5in]{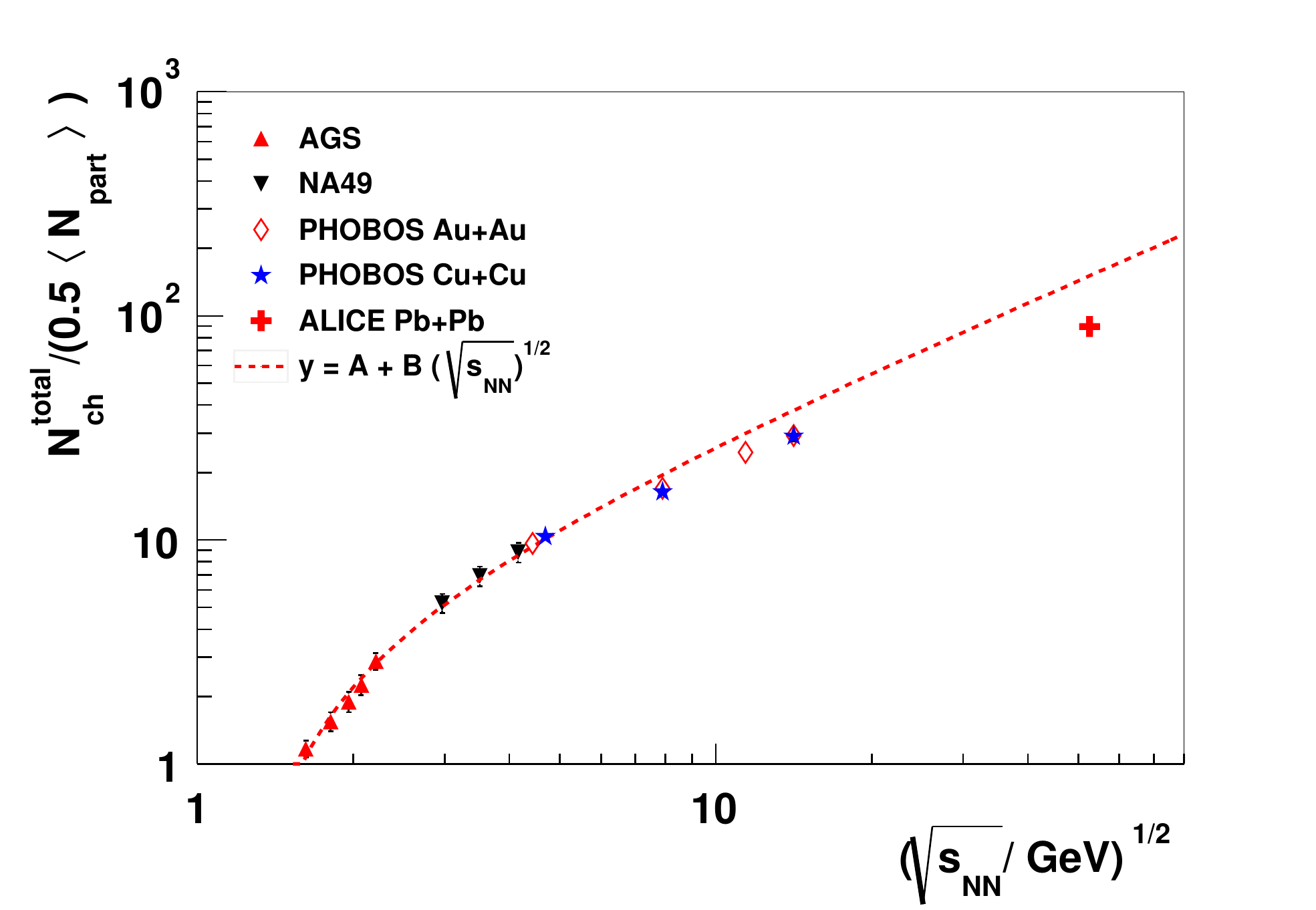}
\caption{Total charged particle multiplicity per
  participant pair as a function of ${(\sqrt{s_{\rm{NN}}}/{\rm GeV})}^{1/2}$. The
data points are fitted with the parametrized form of Eq. \ref{landau},
which is shown by the dotted line.}
\label{engDepNchLandau}
\end{center}
\end{figure}

\begin{figure}
\begin{center}
\includegraphics[width=3.5in]{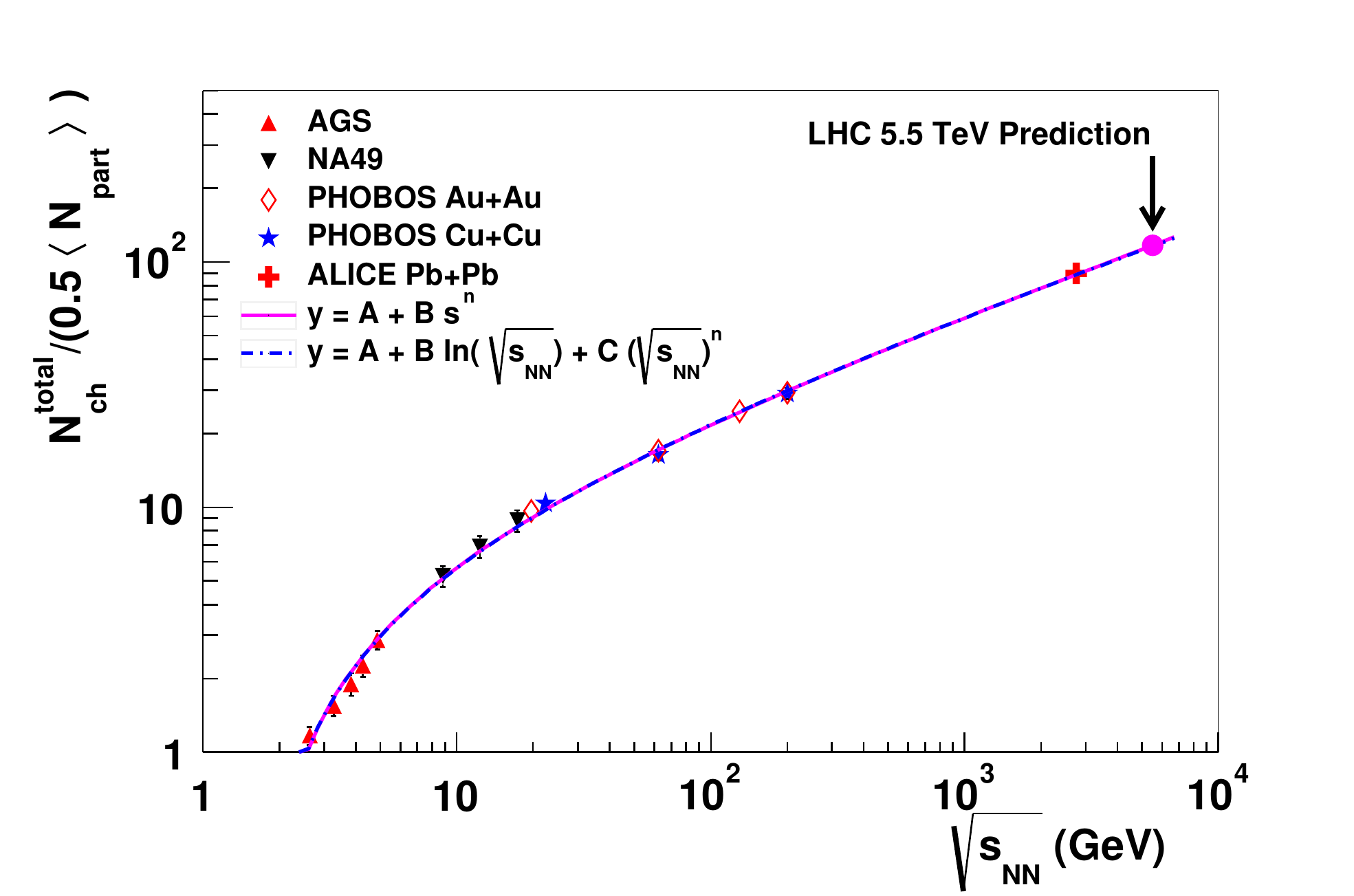}
\caption{Total charged particle multiplicity per
  participant pair as a function of $\sqrt{s_{\rm{NN}}}$. The
data points are fitted with the parametrized form of Eq. \ref{NtotHy},
which is shown by the dotted line. The continuous line shows a comparison with a power law form of energy dependence.}
\label{NtotPred}
\end{center}
\end{figure}

The parametrized form of Eq. \ref{landau} is obtained for PHOBOS 
Au+Au data, which is given by \cite{d4}
\begin{equation}
\frac{N_{ch}^{total}}{0.5~\langle N_{part} \rangle} = 1.135 + 2.019 {(\sqrt{s_{NN}}/GeV)}^{1/2}
\end{equation}

and in general can be written as,
\begin{equation}
\frac{N_{ch}^{total}}{0.5~\langle N_{part} \rangle} = \text{A} + \text{B}
{(\sqrt{s_{NN}}/GeV)}^{1/2}
\label{landau1}
\end{equation}
We have tried to fit Eq. \ref{landau1} to the $N_{\rm{ch}}^{\rm{total}}/(0.5~\langle N_{\rm{part}} \rangle)$ data as a function of
$({\sqrt{s_{\rm{NN}}}/\rm{GeV}})^{1/2}$ obtained from AGS to LHC
experiments which is shown by the dotted line in Figure \ref{engDepNchLandau}.
It is observed that Eq. \ref{landau1} fails to explain the LHC data, as it over predicts. This observation goes inline with
the measurement as shown in Figure \ref{WidthRatio}, {\it i.e.} the width of $dN_{\rm{ch}}/d\eta$ of Pb+Pb data at $\sqrt{s_{\rm{NN}}}$ = 2.76 TeV is more than the expectation of Landau hydrodynamics. It
is seen that the hybrid function nicely describes the whole
$dN_{\rm{ch}}/d\eta$ distribution as a function of $\sqrt{s_{\rm{NN}}}$ and
  Landau hydrodynamics can't explain the LHC data. With
  this motivation, we tried to fit a hybrid form as given in
  Eq. \ref{NtotHy} to fit the $N_{\rm{ch}}^{\rm{total}}/(0.5~\langle N_{\rm{part}} \rangle)$ as a function of
  $\sqrt{s_{\rm{NN}}}$ \cite{rnsAditya}.
\begin{equation}
\frac{N_{ch}^{total}}{0.5~\langle N_{part} \rangle}= \text{A} + \text{B} ~\text{ln}~(\sqrt{s_{NN}}) +
  \text{C} {(\sqrt{s_{NN}})}^{n}
\label{NtotHy}
\end{equation}
It is found that this hybrid function can explain the whole range of
the data upto the LHC energy as shown in Figure \ref{NtotPred}. The extrapolation of this function
for the upcoming LHC Pb+Pb collisions at $\sqrt{s_{\rm{NN}}}$ = 5.5 TeV is shown by the filled circle in Figure \ref{NtotPred}. 


\section{PSEUDORAPIDITY DENSITY DISTRIBUTION OF PHOTONS}
Photons are produced from every phase of the fireball expansion, like
from hard scattering to the decay of hadrons in heavy-ion collision experiment. Photons hardly interact
with the medium. So when photons get
thermalized, their mean free paths become same as the system size and
they leave the system unaffected. Thus it is believed that the photons carry the
information of the thermalized system at all stages of the evolution of the produced fireball. Direct photons
created from the QCD process are treated as
golden probe to measure the thermodynamic parameters like
initial temperature of the fireball. The inclusive photon spectra contain all photons
including the photons produced from particle's decay, e.g.
$\pi^0$ and $\eta^0$. So photons can be used to
estimate the degree of thermalization of the system. It
is also proposed that as majority of photons are produced from $\pi^0$
decay, so they can be used as a complementary measurement to the charged pion
measurements. Photons can be used to study the anisotropic flow of the
system. Photons can be used as a precursor for the
measurement of pseudorapidity density distribution of charged
particles. It is proposed that simultaneous measurement of charged
particles with photons can be used in the search for Dis-oriented Chiral
Condensate (DCC) \cite{y1}. Keeping the importance of measurement of photons in mind
as a probe for QGP, we will be discussing the pseudorapidity
distribution of photons.
\par
In this review, the pseudorapidity density
of photons for different collision systems and at different energies are
discussed. Then the expansion hydrodynamics of photons are discussed by invoking Landau
hydrodynamics along with its advanced forms. In the forward rapidity,
longitudinal scaling of photons are discussed. At the end, the
scaling of total measured photons as a function of $\langle N_{\rm{part}}
\rangle$ is discussed for two collision systems.

\subsection{System size and energy dependence of photon distributions
  ($dN_{\gamma}/d\eta$) }
\begin{figure}
\begin{center}
\includegraphics[width=3.6in]{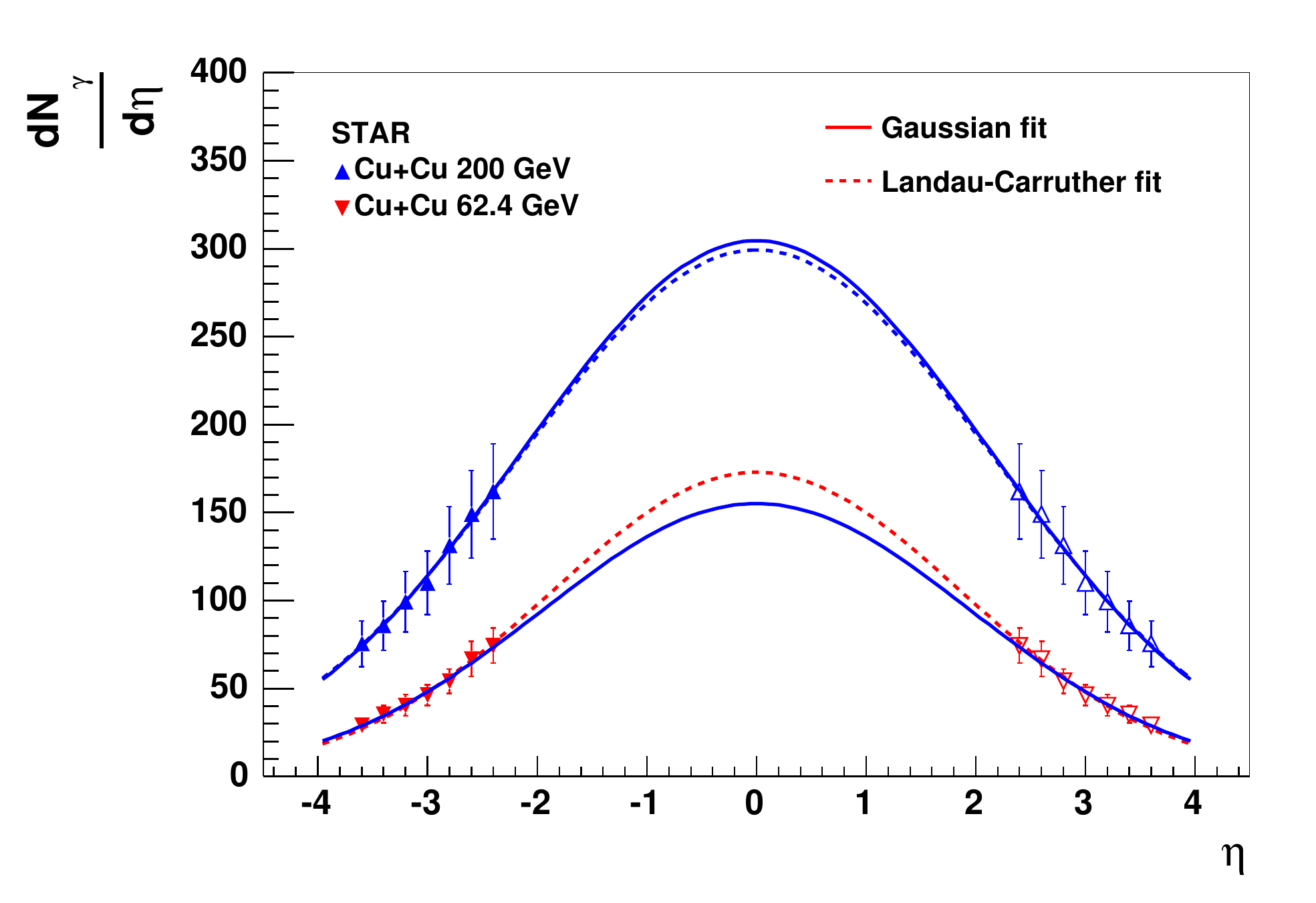}
\caption{Photon multiplicity distributions of Cu+Cu
  collision system as a function pseudorapidity for the most central events for different collision energies.}
\label{dndetaCuGama}
\end{center}
\end{figure}

\begin{figure}
\begin{center}
\includegraphics[width=3.6in]{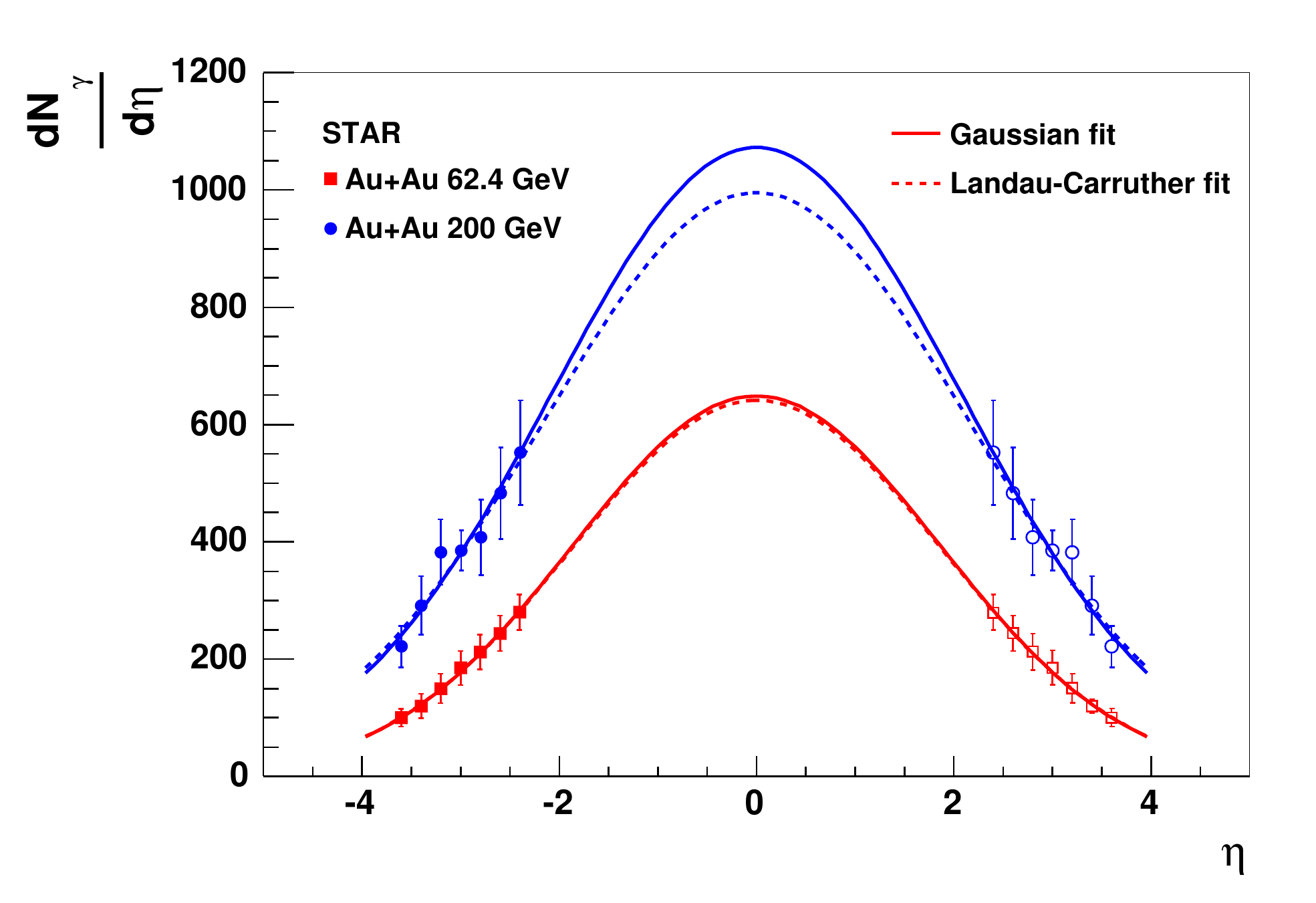}
\caption{Photon multiplicity distributions of Au+Au
  collision system as a function pseudorapidity for the most central events for different collision energies.}
\label{dndetaAuGama}
\end{center}
\end{figure}

\begin{figure}
\begin{center}
\includegraphics[width=3.6in]{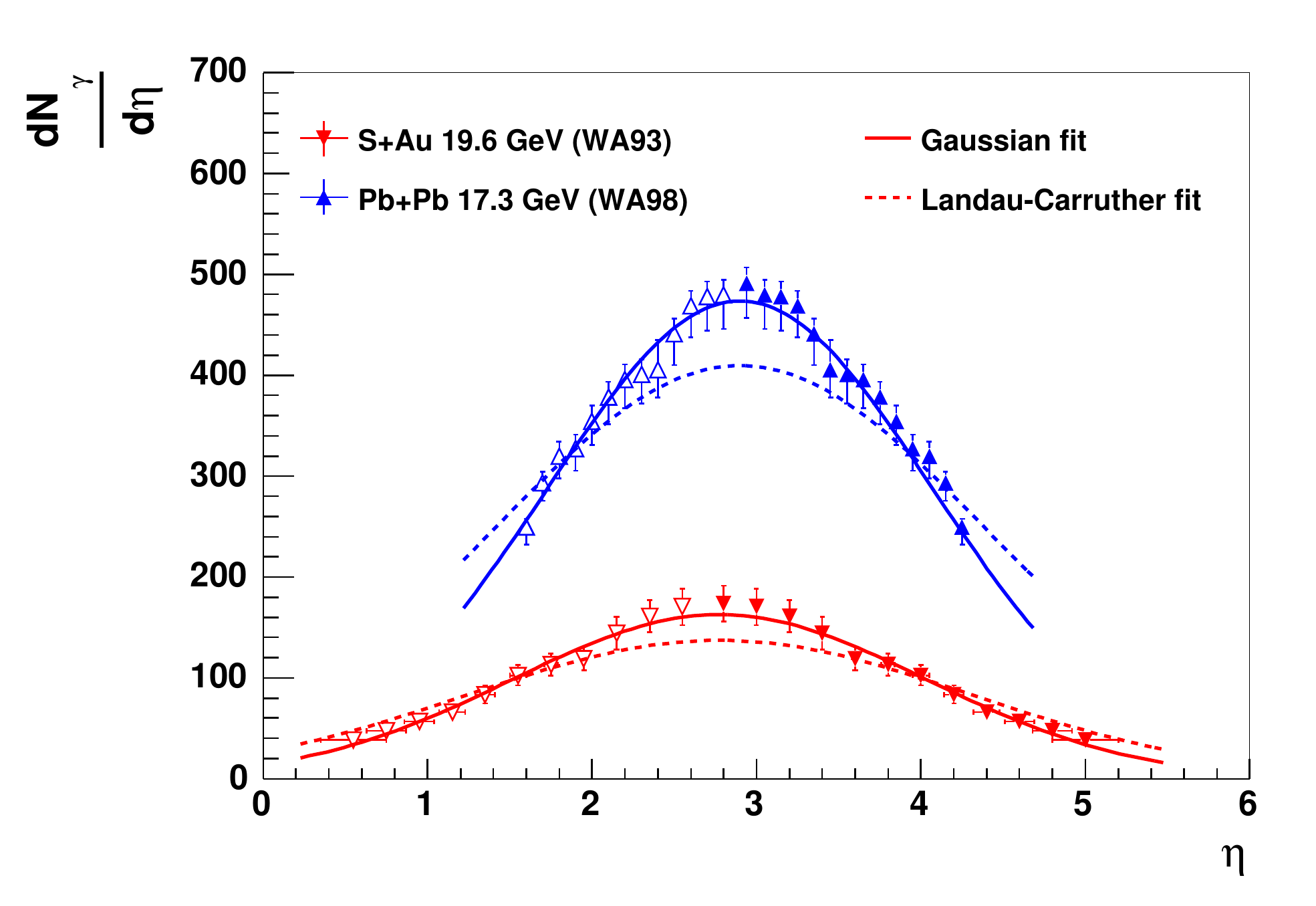}
\caption{Photon multiplicity distributions of Pb+Pb
  collision system as a function pseudorapidity for the most central events for different collision energies.}
\label{dndetaPbGama}
\end{center}
\end{figure}

\begin{figure}
\begin{center}
\includegraphics[width=3.6in]{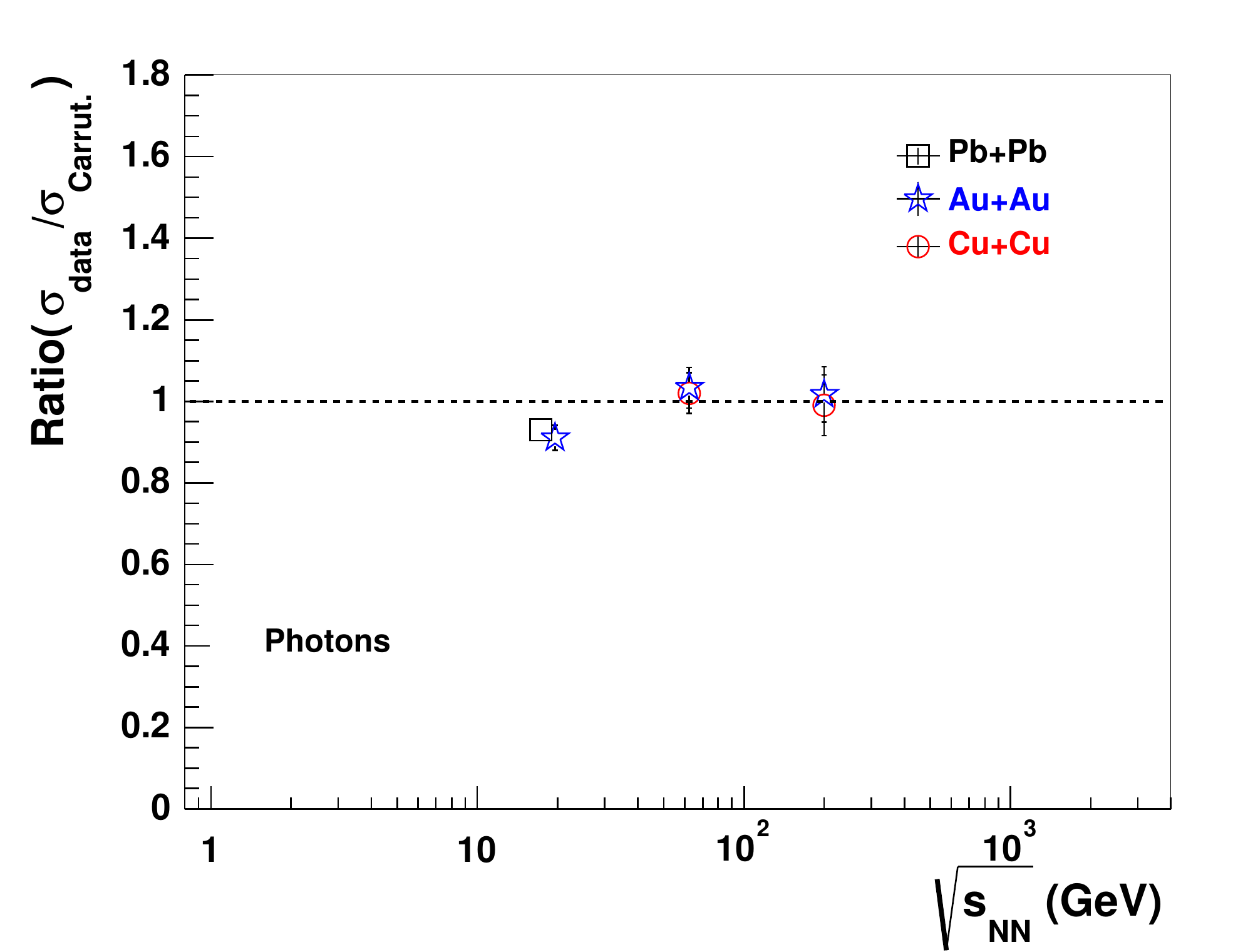}
\caption{Ratio of the widths of the data and that obtained from the fitting
  of Photon multiplicity distributions of different
  collision systems as a function of collision energy.}
\label{RatioGama}
\end{center}
\end{figure}

The energy dependence of pseudorapidity distributions of photons are shown for Cu+Cu, Au+Au systems in Figure
\ref{dndetaCuGama} and \ref{dndetaAuGama}. In Figure
\ref{dndetaCuGama}, the pseudorapidity distribution of photons for
Cu+Cu collisions at $\sqrt{s_{NN}}$= 62.4 and 200 GeV are shown. In
Figure \ref{dndetaAuGama}, $dN_{\gamma}/d\eta$ for Au+Au collisions at
$\sqrt{s_{NN}}$= 62.4 and 200 GeV are shown. The Cu+Cu and Au+Au
collision data are taken from STAR experiment at RHIC \cite{y2}. The
pseudorapidity distribution of photons of S+Au collisions data at 19.3
GeV and Pb+Pb collisions at 17.6 GeV are shown in Figure
\ref{dndetaPbGama}. The S+Au collision data and Pb+Pb collision data
are taken from Ref \cite{y3} and \cite{y4}, respectively. Data
collected are at the forward rapidity. However, to get the photon
distribution in the backward rapidities, a reflection of the data
about the midrapidity is done assuming that the
$dN_{\gamma}/d\eta$ is symmetric about $\eta$ = 0 for collider
experiments, e.g. Cu+Cu and Au+Au collisions. For fixed target
experiments, like S+Au and Pb+Pb, the reflection is carried out with respect to the $\eta_{\rm{peak}}$. The closed
markers represent the mirror reflection of the data recorded by the
detectors. The $dN_{\gamma}/d\eta$ distributions are fitted with
a double Gaussian and Landau-Carruthers functions to
understand the expansion dynamics of the system. To see the extent of the Landau
hydrodynamics is applicable to the system, the ratio of width of the
$dN_{\gamma}/d\eta$ of data and the width obtained from
Landau-Carruthers fitting are shown as a function of collision energies in Figure \ref{RatioGama}. It can be
observed that at lower energy, it deviates from 1, but at RHIC
energy, it agrees with Landau-Carruthers hydrodynamical model. It would be interesting to have corresponding LHC data to look into the validity of Landau hydrodynamics for photons at forward rapidities.

\subsection{Longitudinal Scaling of Photon}
\begin{figure}
\begin{center}
\includegraphics[width=3.6in]{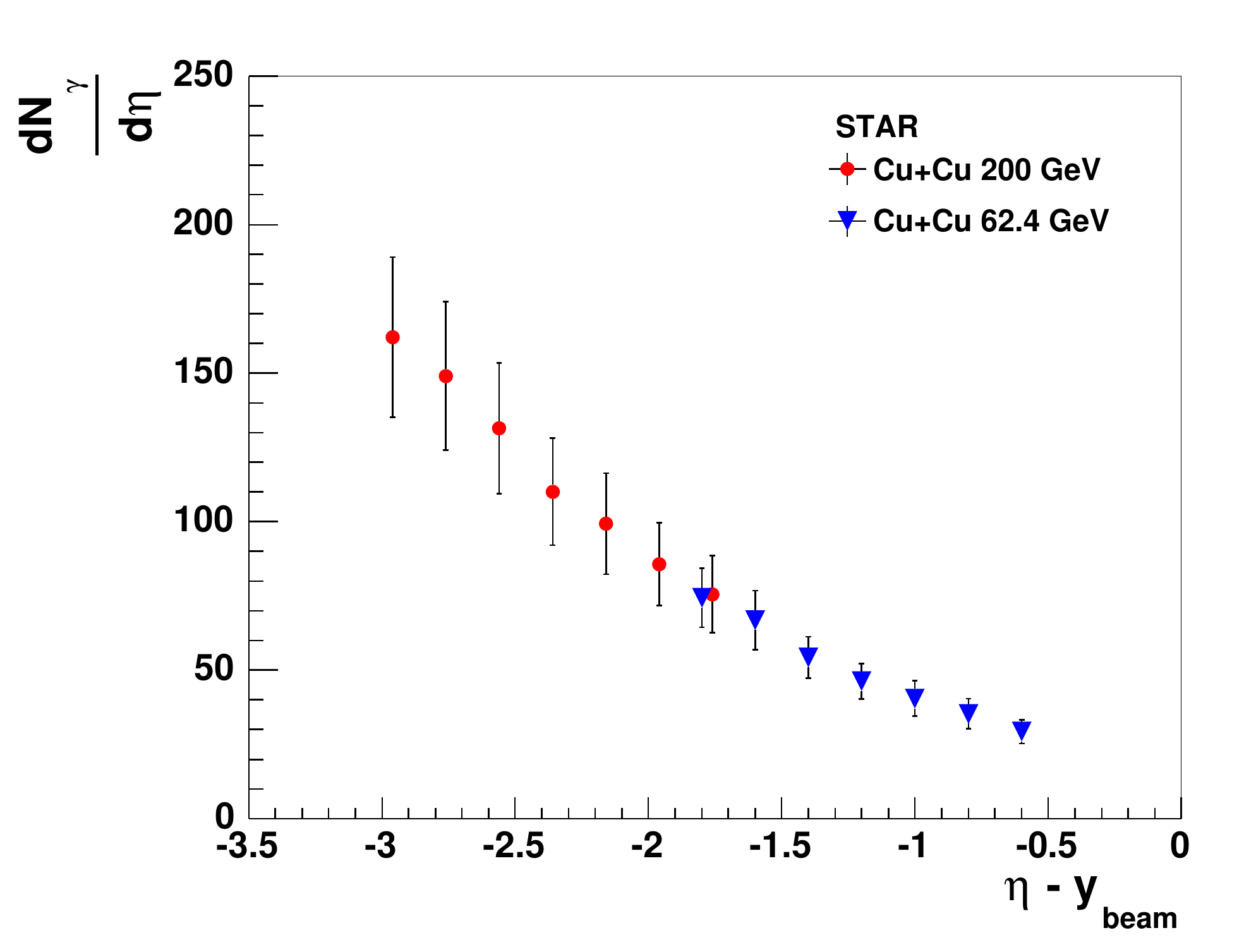}
\caption{ Photon multiplicity density normalized per
  participant pair for different energies shown in the frame one of
  the rest frame of projectile.}
\label{dGamadEtaCu}
\end{center}
\end{figure}

\begin{figure}
\begin{center}
\includegraphics[width=3.6in]{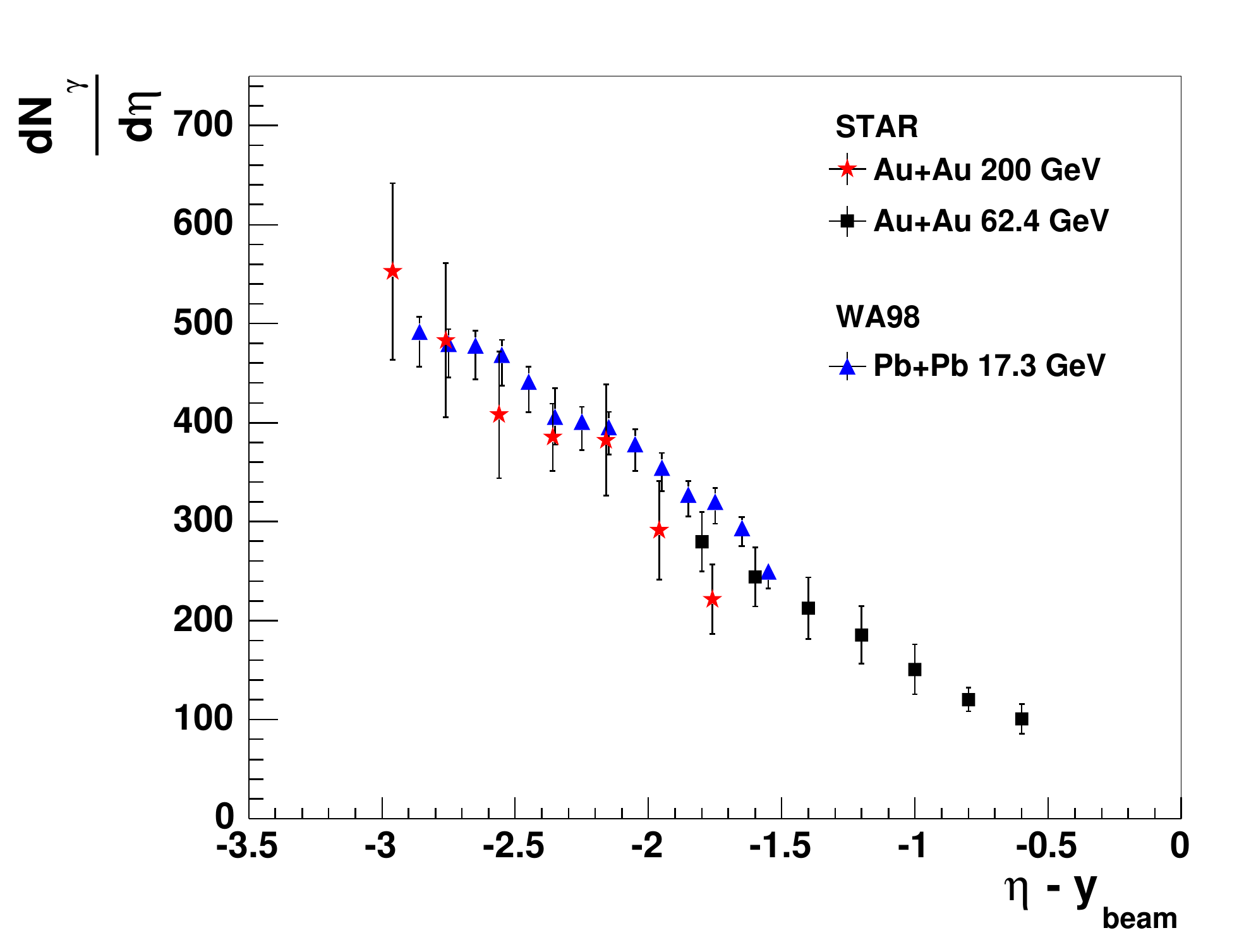}
\caption{ Photon multiplicity density normalized per
  participant pair for different energies shown in the projectile rest frame.}
\label{dGamadEtaAu}
\end{center}
\end{figure}

In the previous section, the longitudinal scaling of charged particles
in the forward rapidities are discussed. Is this longitudinal scaling
a global phenomena of the heavy-ion collision or only specific to
charged particle productions? To confirm this phenomena, the
longitudinal scaling of photons is studied separately for two different collision
species. In Figure \ref{dGamadEtaCu}, the
$dN_{\gamma}/d\eta$ as a function of $\eta^{\prime}$ for Cu+Cu collision data at $\sqrt{s_{NN}}$= 62.4
and 200 GeV are shown. In Figure \ref{dGamadEtaAu}, the $dN_{\gamma}/d\eta$ for Au+Au collisions at
$\sqrt{s_{NN}}$ = 62.4, 200 GeV and Pb+Pb collision data at beam
energy 158 AGeV as a function of $\eta^{\prime}$ are shown. The
$dN_{\gamma}/d\eta$ data are available
for only small pseudorapidity coverage. Still, the nature of the
$dN_{\gamma}/d\eta$ distribution as a function of $\eta^{\prime}$ show the longitudinal
scaling behaviour as a consequences of limiting
fragmentation. It is observed from the
Figure \ref{dGamadEtaCu} and \ref{dGamadEtaAu} that photon also
shows the energy independent limiting fragmentation behaviour.  It is
seen that the limiting fragmentation of pions are same as the photon
and independent of centrality unlike charged hadrons. It is also
  reported in Ref \cite{y5} that the limiting fragmentation behaviour
  of photons for $p+\bar{p}$ collisions at $\sqrt{s}$ = 540 GeV are in close
  agreement with the measured photon yield in Au+Au collisions at
  $\sqrt{s_{NN}}$ = 62.4 GeV unlike the charged particle results. Study
  from HIJING event generator indicates that about 93-96$\%$ of
  measured photons are from $\pi^{0}$ decays. Hence, the centrality
  independent behaviour of photons is interpreted as indirect measure
  of meson limiting fragmentation. This contrasting behaviour of photon results of the
  limiting fragmentation with respect to charged hadrons may be due to
  nuclear remnants and baryon stopping. It indicates that mesons are not
affected by baryon transport at forward rapidities \cite{y5}. The
study of identified charged particles with photon results done in Ref
\cite{y5} clearly indicates that net-proton results violate the
energy independent behaviour of limiting fragmentation: a clear
indication of baryon-meson anomaly. The centrality and energy
independence behaviour of mesons contrary to inclusive charged hadrons
and identified baryons implies that baryon transport plays an
important role in particle production at forward rapidities. It
is argued that although the baryon stopping is different for different
collision energies, the mesons are not affected by it. In the context of
baryon junction picture, baryons would have shown the energy
independent limiting fragmentation behaviour at forward rapidities, if they carry the valance
quarks like the mesons produced from the valance quarks. This
suggests that baryon number resides in a nonperturbative configuration
of gluon fields rather in the valence quarks \cite{y5}.
\par
The longitudinal scaling behaviour observed for charged particles and photons ensure the
universality of hypothesis of limiting fragmentation and put forward
many deeper questions about the actual processes behind it.

\begin{figure}
\begin{center}
\includegraphics[width=3.6in]{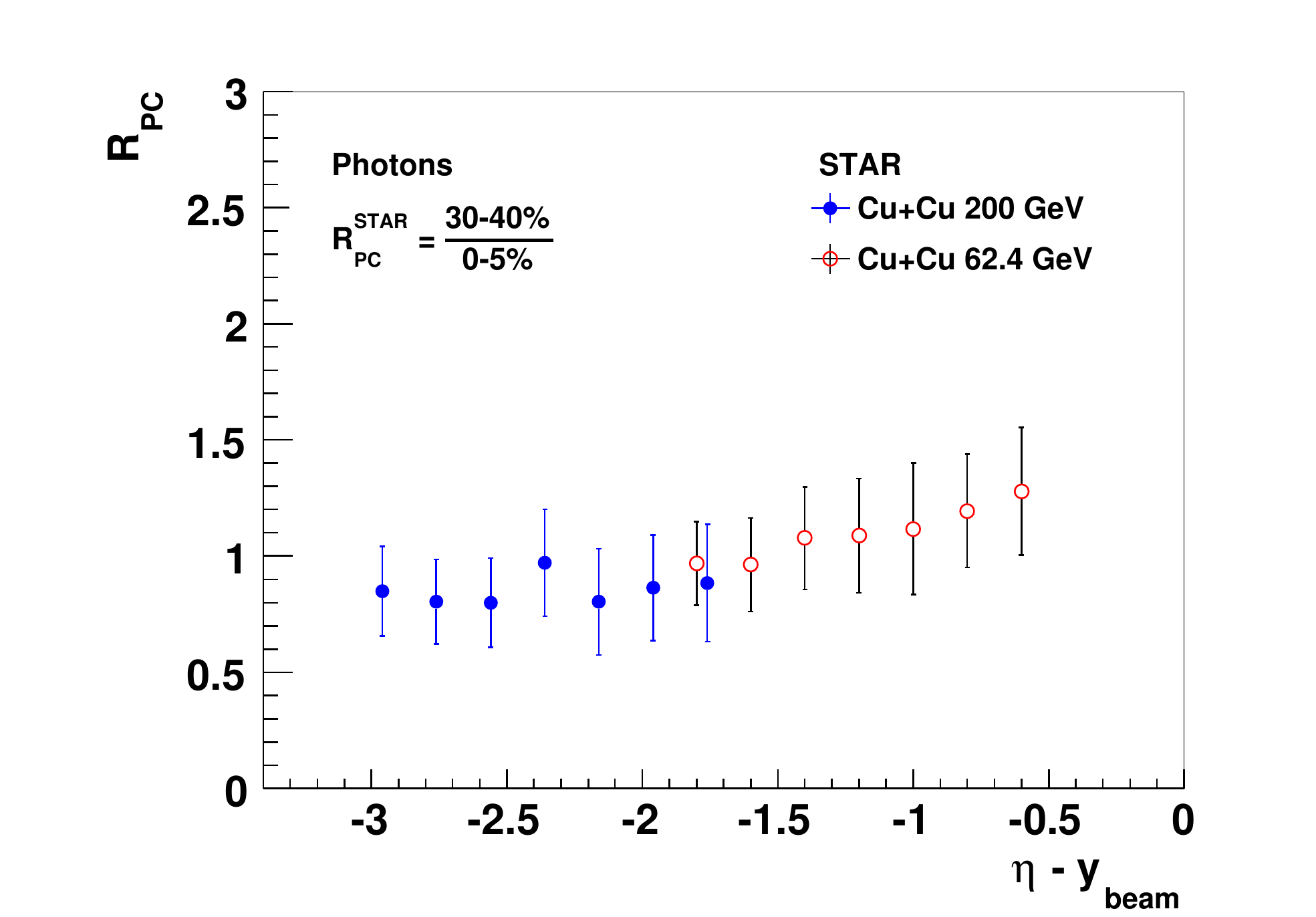}
\caption{$R_{\rm{PC}}$ of photons as a function of $\eta^{\prime}= \eta - y_{\rm{beam}}$ for Cu+Cu
  collisions for different energies.}
\label{RcpCuGamma}
\end{center}
\end{figure}

\begin{figure}
\begin{center}
\includegraphics[width=3.6in]{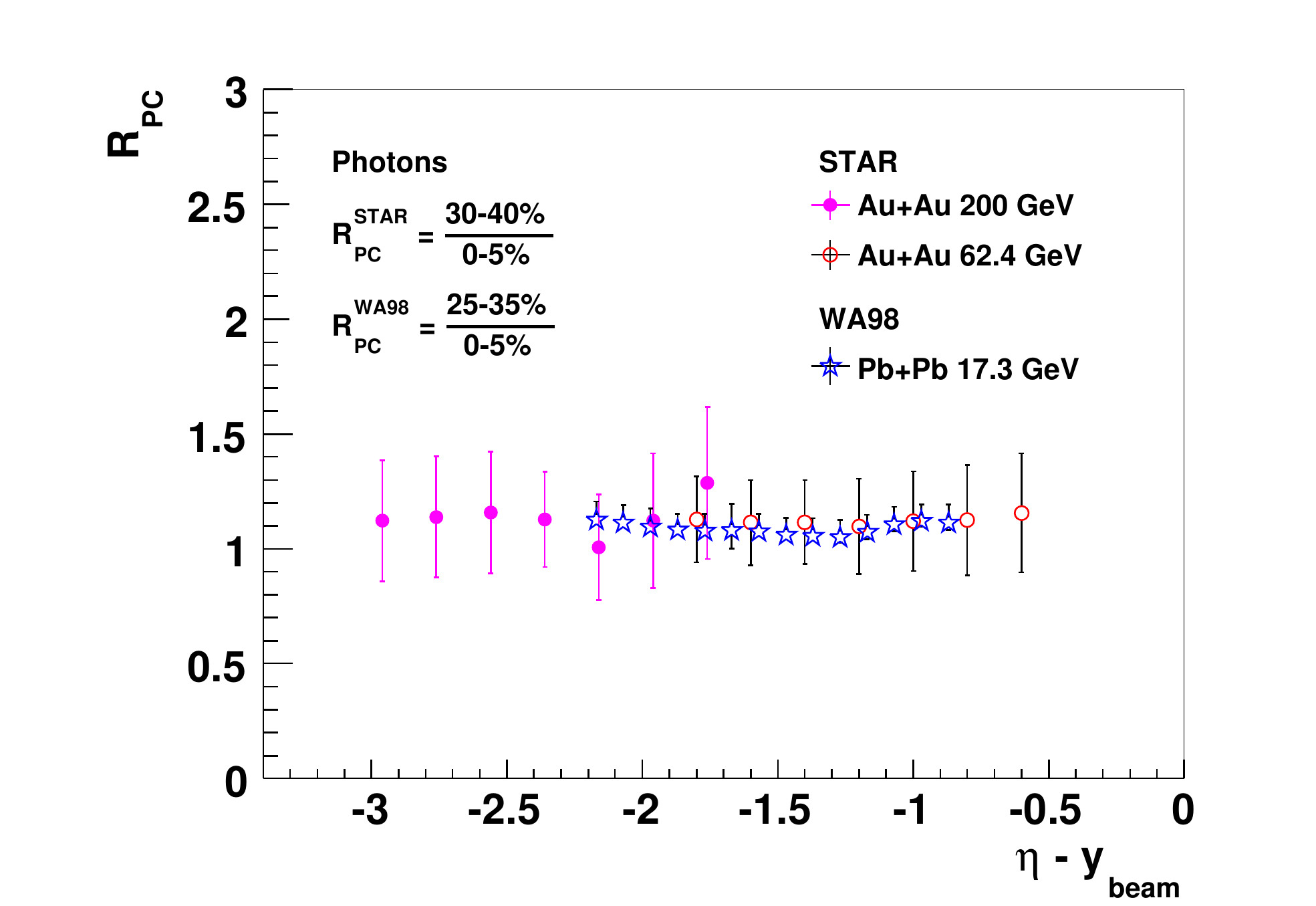}
\caption{$R_{\rm{PC}}$ of photons as a function of $\eta^{\prime}= \eta - y_{\rm{beam}}$ for Au+Au
  collisions for different energies.}
\label{RcpAuGamma}
\end{center}
\end{figure}

During the discussion of extended longitudinal scaling of charged
particles, we have encountered that this is independent of energy but
shows some dependence of collision geometry, i.e. centrality. Then
$R_{\rm{PC}}$ variable was adopted to deal with this issue. But in the
limiting fragmentation of photons, it is found to be centrality independent
\cite{y5}. But to see the consistency, we tried to do the same
exercise for photons by evaluating the
$R_{\rm{PC}}$ for different collision systems at different energies. The
$R_{\rm{PC}}$ is defined as given in Eq. (4). For the RHIC energies, the
peripheral events correspond to 30-40$\%$ centrality and central events 
correspond to 0-5$\%$ centrality. For WA98 experiment, 25-35$\%$ centrality and 
0-5$\%$ events are considered as peripheral events and central events,
respectively. In Figure \ref{RcpCuGamma}, $R_{\rm{PC}}$ for Cu+Cu collision
data at $\sqrt{s_{\rm{NN}}}$ = 62.4 and 200 GeV are shown as a function of
$\eta^{\prime}$. Similarly, in Figure \ref{RcpAuGamma}, $R_{\rm{PC}}$ of
Au+Au collision data at $\sqrt{s_{\rm{NN}}}$ = 62.4 and 200 GeV superimposed
with Pb+Pb data at beam energy 158 AGeV are shown as a function of
$\eta^{\prime}$. The error bars shown in Figure \ref{RcpCuGamma}
and \ref{RcpAuGamma} are of statistical only. We observe from Figure \ref{RcpCuGamma}
and \ref{RcpAuGamma} that within error bars, the $R_{\rm{PC}}$ is constant
and equal to one as a function of $\eta^{\prime}$ irrespective of
collision energies. This observation strengthen our argument that
extended longitudinal scaling is a global phenomena for charged
particles as well as for photons produced in the heavy-ion collision
experiments. 
  
\subsection{Scaling of $N_{\gamma}^{\rm{total}}$ with $N_{\rm{part}}$}
\begin{figure}
\begin{center}
\includegraphics[width=3.6in]{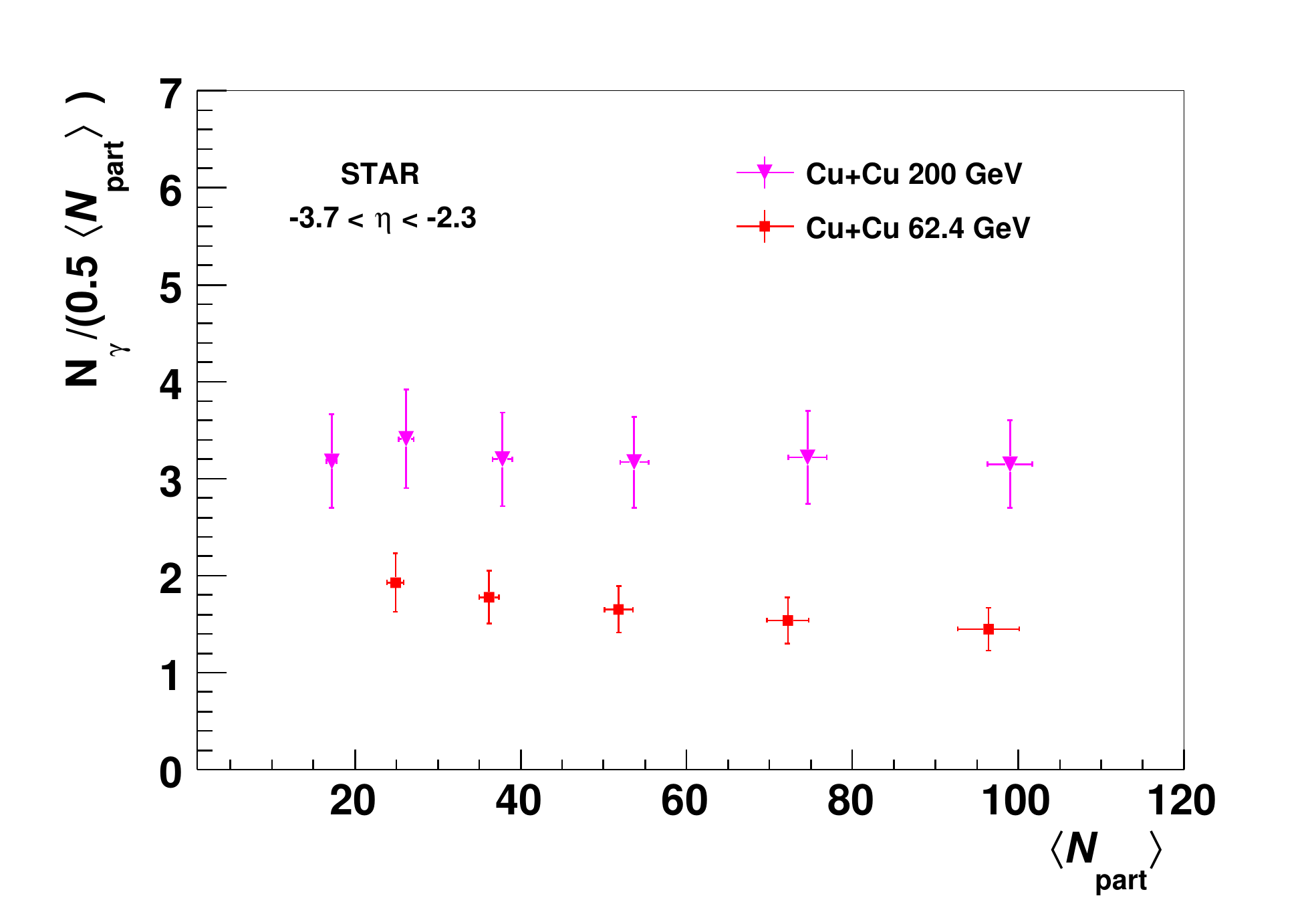}
\caption{$N_{\gamma}$ normalized per participant pair as a function of
$N_{\rm{part}}$ for Cu+Cu collisions.}
\label{NtotGamaCu}
\end{center}
\end{figure}

\begin{figure}
\begin{center}
\includegraphics[width=3.6in]{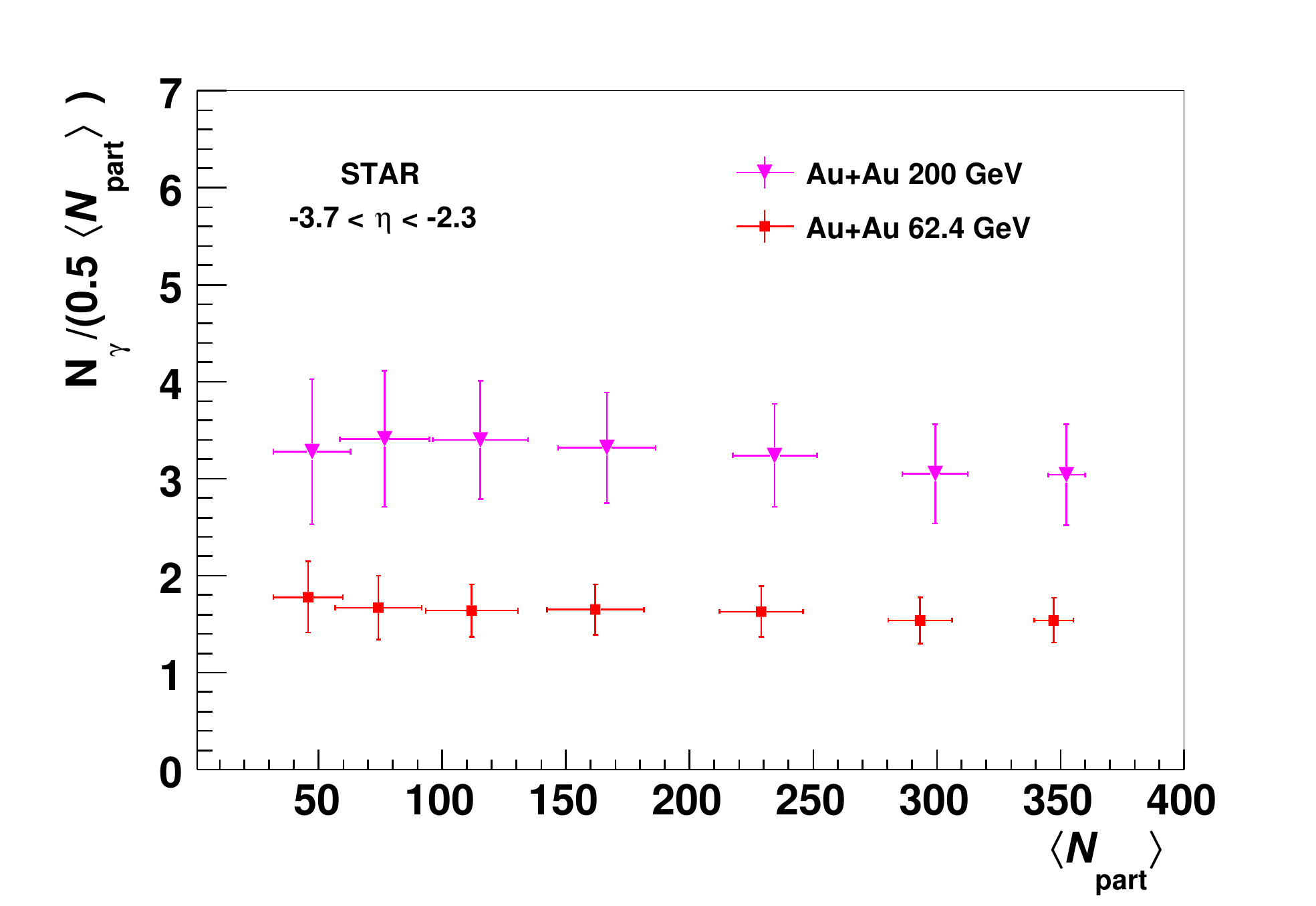}
\caption{$N_{\gamma}$ normalized per participant pair as a function of
$N_{\rm{part}}$ for Au+Au collisions.}
\label{NtotGamaAu}
\end{center}
\end{figure}

Like the scaling of total charged particles with $N_{\rm{part}}$, the total
photons normalized per participant pairs as a function of average
participant pairs are shown for Cu+Cu and Au+Au collision systems at
62.4, 200 GeV in Figure \ref{NtotGamaCu} and \ref{NtotGamaAu},
respectively. Both the data scale nicely and the normalized 
$N_{\gamma}$ values increase with increase of collision
energy. Note that $N_{\gamma}$ is the value of total number of photons measured within the detector acceptance ($-3.7 < \eta < -2.3$) \cite{starPhoton}.

From Figure \ref{NtotGamaCu} and \ref{NtotGamaAu}, we observed that 
$N_{\gamma}$ scales with the collision centrality like charged particles. 

\begin{figure}[htbp22]
\begin{center}
\includegraphics[width=3.6in]{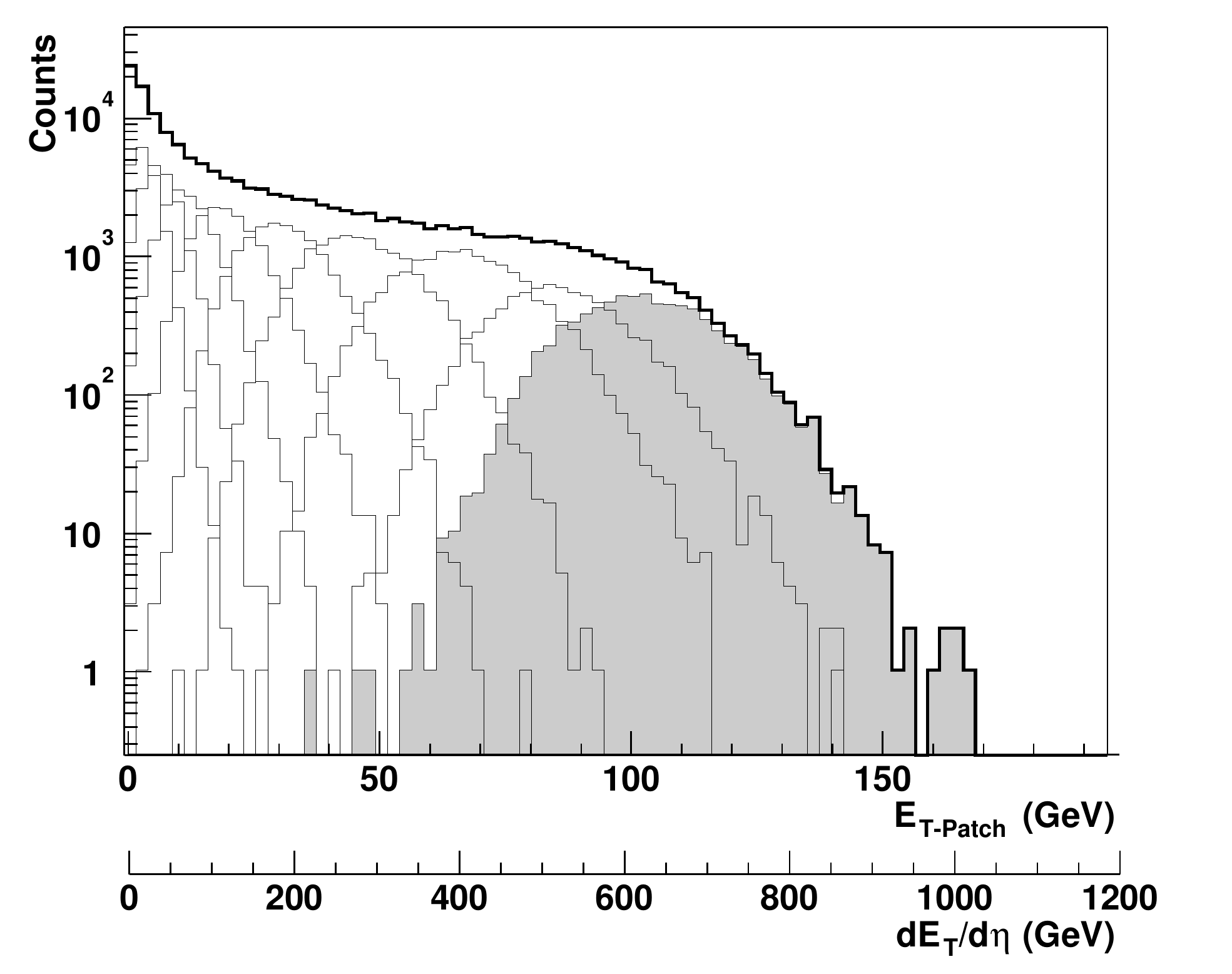}
\caption{The midrapidity ($0 < \eta < 1$) minimum bias distribution of total transverse energy along with distributions for different centrality bins for $\sqrt{s_{\rm{NN}}} = 200$ GeV, as is measured by the STAR experiment at RHIC. The shaded area corresponds to the $5\%$ most central bin. The main axis scale corresponds to the $E_{\rm T}$ measured in the detector acceptance and the bottom axis is corrected to represent the extrapolation to full azimuthal acceptance. The Figure demonstrates the use of $E_{\rm T}$ distribution for estimation of collision centrality. Figure taken from Reference \cite{star200GeV}.} 
\label{ET-min}
\end{center}
\end{figure}

\section{Transverse energy and Collision Cross section}
The transverse energy is one of the important global observables used to characterize the system formed in heavy-ion collisions at extreme conditions of temperature and energy density, where the formation of Quark-Gluon Plasma (QGP) is expected. The transverse energy ($E_{\rm T}$) is the energy produced transverse to the beam
direction and is closely related to the collision geometry. $E_{\rm T}$ is an event-by-event variable defined as
\begin{equation}
{E_T = \sum_i E_i sin\theta_i ~~\rm{and}~~ \frac{dE_T(\eta)}{d\eta} 
  = sin\theta(\eta)\frac{dE(\eta)}{d\eta}.}
\end{equation}
The sum is taken over all particles produced in an event within the detector acceptance. $E_i$ and $\theta_i$ are the energy and polar angle of the final state particles. The energy of the individual particle can be determined by knowing their momenta and particle identification using tracking detectors and/or the total energy deposited in a calorimeter. The source of transverse energy production could be ``soft" multi-particle production and/or the ``hard" scattering jets, depending on the collision energy.  The transverse energy distribution is related to the multiplicity distribution by

\begin{equation}
\frac{dE_T}{d\eta} \sim \left\langle p_T\right\rangle \times \frac{dN}{d\eta}
\label{EtPt}
\end{equation}

To probe the early stages of the produced fireball, it is ideal to take transverse
observables like $E_{\rm T}$, $p_{\rm T}$ etc. This is because, before the collision of two
nuclei, the longitudinal phase space is filled by the beam particles whereas
the transverse phase space is empty. The $E_{\rm T}$ is produced due to the initial
scattering of the partonic constituents of the incoming nuclei and also by
the re-scattering among the produced partons and hadrons \cite{etGen1,etGen2}.
The $E_{\rm T}$ production tells about the explosiveness of the interaction.  Additionally,
in the framework of boost-invariant hydrodynamics, the measurement of $E_{\rm T}$ helps in the quantitative estimation of the initial energy density produced in the interaction \cite{bjorken}. A comparison of this initial energy density with that of estimated by the lattice QCD (lQCD) calculations, gives indication of a possible formation of QGP in the corresponding heavy-ion interactions \cite{lQCD}.
However, there are several competing processes to make a difference between the initially generated and finally observed $E_{\rm T}$.
In an ideal case, if the fireball of the produced quanta, namely the partons or the hadrons depending on the case, break apart instantaneously without significant interactions, the observed transverse energy per unit rapidity $dE_{\rm T}/dy$ will be the same as that was generated initially. On the other hand, if the system interacts very strongly achieving an early thermal equilibrium, which is maintained though out the system expansion, $dE_{\rm T}/dy$ would decrease significantly due to the longitudinal work done by the hydrodynamic pressure \cite{gyulassy,ekrt}. This decrease may however, be moderated by the build up of transverse hydrodynamic flow, which increases $E_{\rm T}$ \cite{kolb}. At higher collision energies, the difference between initially generated and finally observed $E_{\rm T}$ may be reduced because of the gluon saturation in the wave function of the colliding heavy nuclei. This delays the onset of hydrodynamic flow and hence reduces the effective pressure, which decides the above difference \cite{dumitruMG}.

The collision centralities can be estimated by using the minimum bias $E_{\rm T}$ distribution in a way it is done using the charged particle minimum bias distribution. This is shown in Figure \ref{ET-min}. The shaded area in the figure corresponds to the most central ($0-5\%$) collisions having the highest transverse energy. This corresponds to the $5\%$ of the total cross section.
Different centralities are defined by the percentages of total cross sections and are shown in the same figure. Each centrality class follows a Gaussian type of distribution with different mean and variance following the central limit theorem.  The lower edge of the minimum bias distribution shows a peak, which corresponds to the most peripheral collisions. For the most central collisions corresponding to largest values of $E_{\rm T}$, the shape of the distribution is mainly governed by the statistical fluctuations and the experimental acceptance. For larger acceptances, the fall off with increasing $E_{\rm T}$ is very sharp \cite{ET-accept}.

Ref. \cite{ET-accept} gives a very interesting account of addressing a fundamental question like  if $E_{\rm T}$ or the  multiplicity is primary. In other words, whether $E_{\rm T}$ production is primary, followed by fragmentation to final state particles, or whether $E_{\rm T}$ is a random product of the particle multiplicity and the $p_{\rm T}$ distribution. The method as discussed in the above report is as follows. If one assumes that the $E_{\rm T}$ production is a result of the creation of the particles according to the semi-inclusive multiplicity distribution followed by the random assignment of transverse momentum to each particle in accordance with the single-particle semi-inclusive $p_{\rm T}$ distribution, the process could be described by the equation,

\begin{equation}
\frac{d\sigma}{dE_T} = \sigma \sum_{n=1}^{n_{max}} f_{\rm{NBD}}(n,1/k,\mu) ~ f_{\Gamma}(E_T,np,b),
\label{Et-NBD}
\end{equation}

where the multiplicity distribution in A+A collisions is represented by a Negative Binomial Distribution (NBD), $f_{\rm{NBD}}(n,1/k,\mu)$ \cite{E802-95,prem}. The $E_{\rm T}$ distribution for $n$ particles in the final state is represented by a Gamma function, $f_{\Gamma}(E_T, np, b)$, where $p$ and $b$ are the parameters of the $E_{\rm T}$ distribution for a single particle \cite{etGamma}. The details of the NBD and Gamma distributions with their properties are given in the Appendix.  
If we assume that the $E_{\rm T}$ spectra for individual particles are independent of each other and in addition, it is also independent of the multiplicity $n$, then the $E_{\rm T}$ spectrum for $n$ particles is the $n^{\rm{th}}$ convolution of the single particle spectrum. As one finds difficulty in the convergence of fits to Eq. \ref{Et-NBD}, NBD was restricted to Poisson by fixing $1/k =0$, which in turn makes the convergence easier. If one assumes a simpler proportionality between $E_{\rm T}$ and $n$, so that the number of particles in an event, $n$ with transverse energy $E_{\rm T}$ are related by $n=E_T/\langle p_T \rangle$ (the nearest integer). The plot of $\langle E_T \rangle d\sigma/dE_T$ in barn as a function of $E_T/\langle E_T \rangle$ is fitted by the function given by Eq. \ref{EtPt} \cite{ET-accept} and to the NBD given by

\begin{equation}
\frac{d\sigma}{dE_T} = \sigma f_{\rm{NBD}}(E_{\rm T}/\langle p_{\rm T} \rangle,1/k,\mu).
\label{Et-NBD-mod}
\end{equation}

Note that the above NBD is now modified because of the simple relationship of $E_{\rm T}$ and the multiplicity $n$, given by the Eq. \ref{EtPt}. It is observed that the trend of the data leads to a better fit of single-Gamma distribution at higher values of  $E_{\rm T}$ compared to NBD and the reverse at lower values of $E_{\rm T}$. Usually fitting functions with more number of parameters give flexibility to the fitting leading to a better fit. However, in this case, complicated functions like Eq. \ref{Et-NBD} with more number of parameters give worse fit compared to simpler functions like Eq. \ref{Et-NBD-mod}. Single-Gamma distribution fitting to the above distribution is better than the other two functions. If multiplicity were the primary quantity compared to transverse energy, which leads to the form of Eqs. \ref{Et-NBD} and \ref{Et-NBD-mod}, then one would expect these equations to fit better than the single-Gamma distribution. It is interesting and compelling to speculate on the implications of these results for a detailed relationship of multiplicity with transverse energy and the effect of hadronization. However, it remains as an open question to be addressed by more controlled experiments.


\section{Collision energy dependence of Transverse energy}
\begin{figure}
\begin{center}
\includegraphics[width=3.6in]{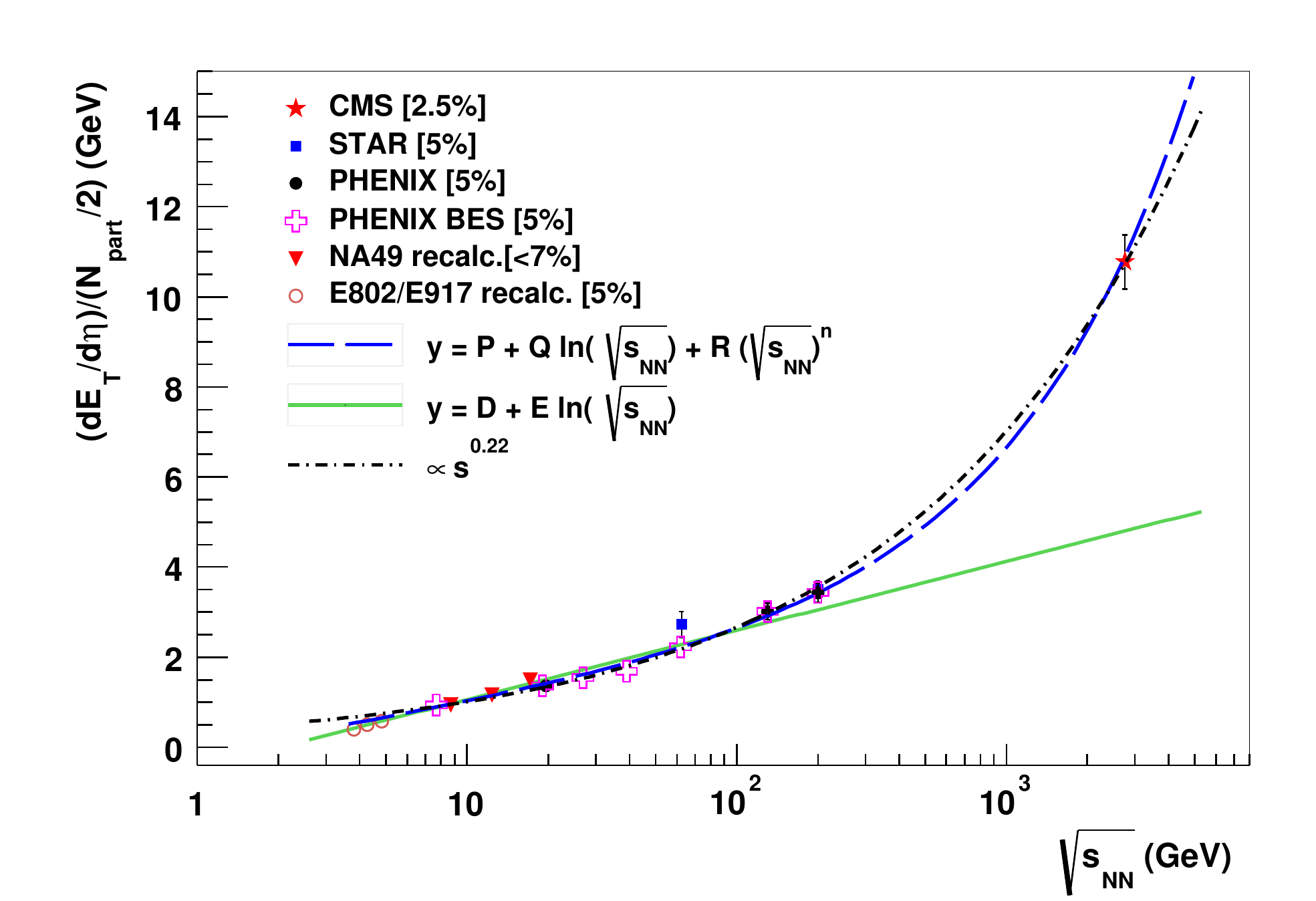}
\caption{\label{EtVsCOM} Collision energy dependence of midrapidity $\frac{1}{N_{\rm {part}}/2}\frac{dE_{\rm{T}}}{d\eta}$. Shown are different phenomenological fitting functions to explain the transverse energy production.}
\end{center}
\end{figure}

Figure \ref{EtVsCOM} shows the collision energy dependence of $\frac{1}{N_{\rm {part}}/2}\frac{dE_{\rm{T}}}{d\eta}$ for central collisions at midrapidity. A logarithmic growth of transverse energy upto the top RHIC energy underestimates the LHC measurement, which is better described by a power-law fit. However, the later overestimates the low energy measurements. A hybrid function, which is a combination of logarithmic and power law, motivated by midrapidity gluonic source and a fragmentation source seems to explain the data for wide range of energies starting from few GeV to TeV \cite{rnsAditya,wolschin}. $\frac{1}{N_{\rm {part}}/2}\frac{dE_{\rm{T}}}{d\eta}$ increases by a factor of $3.07$ from $\sqrt{s_{\rm NN}} = 200$ GeV to 2.76 TeV. The CMS experiment estimate of $\frac{dE_{\rm{T}}}{d\eta} = 2007 \pm 100$ GeV and $\frac{dN_{\rm{ch}}}{d\eta} = 1612 \pm 55$ for top $5\%$ central Pb+Pb collisions at 2.76 TeV \cite{cmsEt,cmsNch}. Division of both leads to transverse energy per charged particle of $1.25 \pm 0.08$ GeV at $\sqrt{s_{\rm NN}} = 2.76$ TeV, for top $5\%$ central Pb+Pb collisions, which is almost $42\%$ higher than its corresponding value at top RHIC energy ($0.88 \pm 0.07$ GeV).

\section{Centrality dependence of Transverse energy}
\begin{figure}
\begin{center}
\includegraphics[width=3.6in]{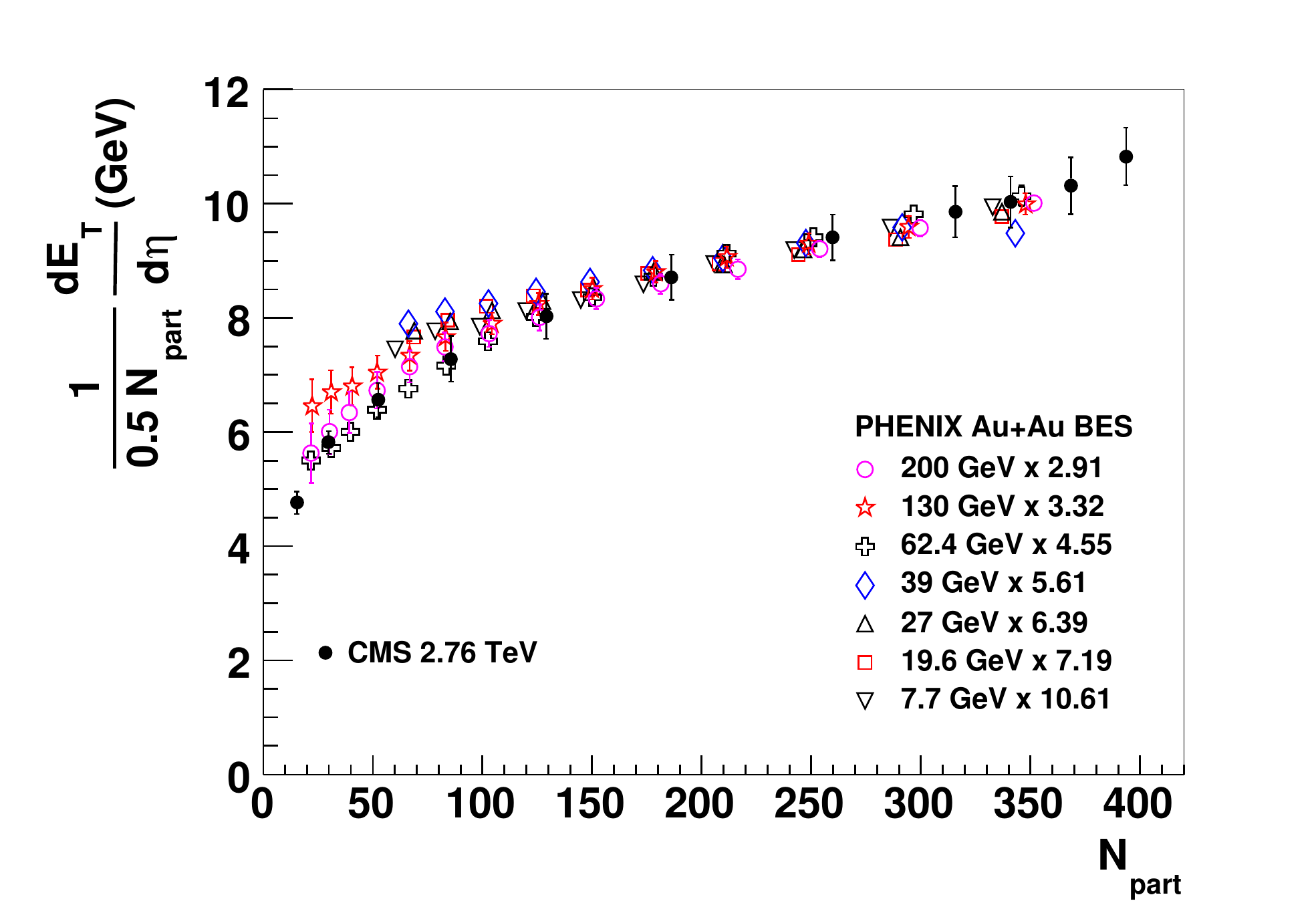}
\caption{\label{EtVsCent} Centrality dependence of midrapidity $\frac{1}{N_{\rm {part}}/2}\frac{dE_{\rm{T}}}{d\eta}$. }
\end{center}
\end{figure}
Figure \ref{EtVsCent} shows the centrality dependence of $\frac{1}{N_{\rm {part}}/2}\frac{dE_{\rm{T}}}{d\eta}$ at midrapidity. Various lower energy measurements are multiplied with some numbers to look into the similarity in the shape at higher energies. Except extreme peripheral events, within experimental uncertainties the centrality shows a universal shape for all energies.
The value of $\frac{1}{N_{\rm {part}}/2}\frac{dE_{\rm{T}}}{d\eta}$ shows a monotonic increase with collision centrality. 
\par

\begin{figure}[htbp26]
\begin{center}
\includegraphics[width=3.6in]{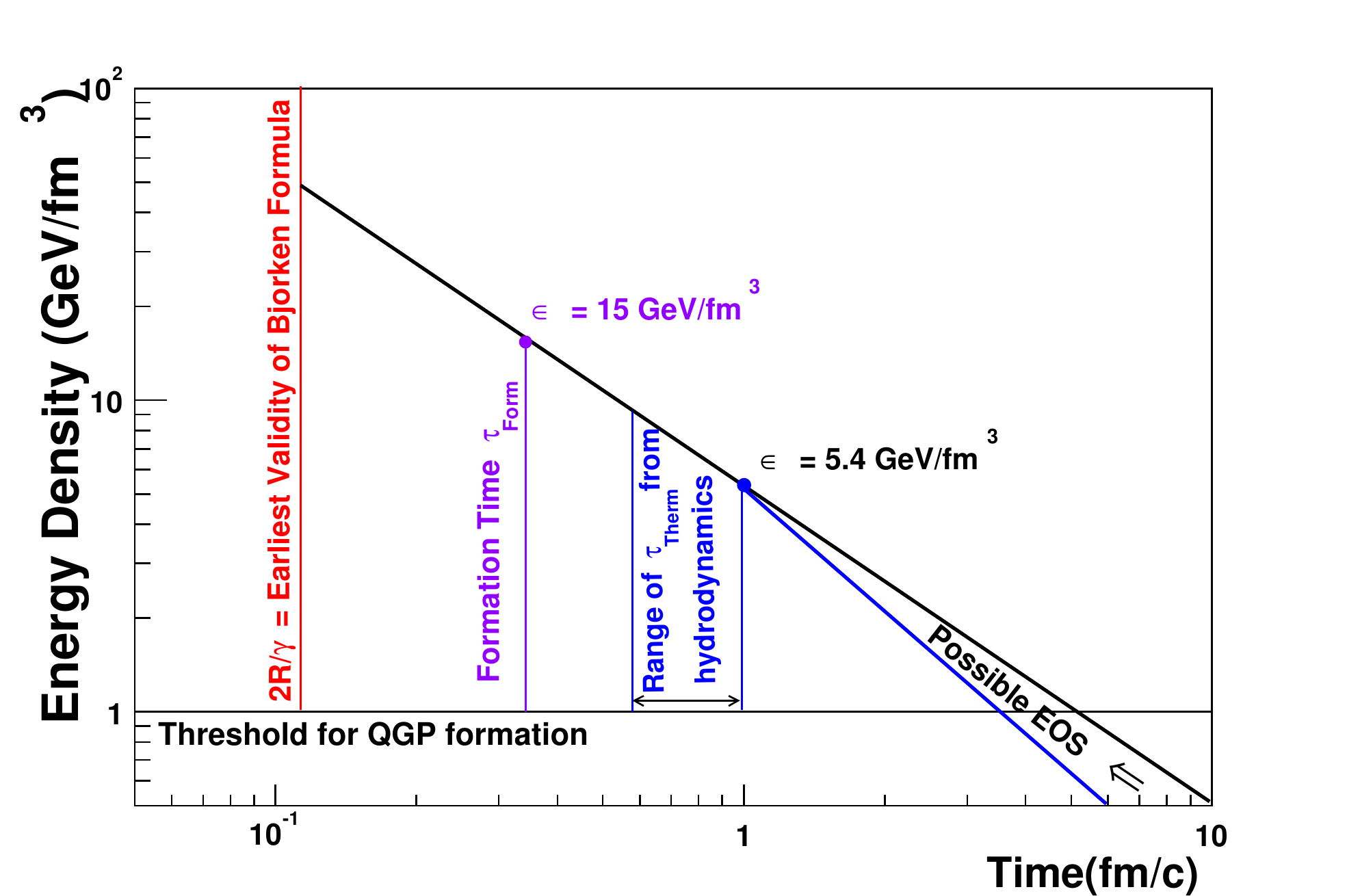}
\caption{Schematic diagram of the time and energy density scales derived through the Bjorken picture \cite{phenixW}.} 
\label{EngDensityVsTime}
\end{center}
\end{figure}

One of the goals of heavy-ion collision experiments is to create QGP in the laboratory and a pre-requisite of this is to ensure that sufficiently large energy density has been produced in the heavy-ion collisions. To ensure this, the estimation of the initial energy density through the measurement of the final state multiplicity and transverse energy is done through Bjorken hydrodynamic model.  Numerical simulations on lattice \cite{lattice, lQCD} give a lower bound for the initial energy density for the formation of a Quark Gluon Plasma, which is of the order of $1~ {\rm GeV/fm^3}$ \cite{lQCD}. A comparison of the estimated energy density from Bjorken model may establish the possible formation of QGP in heavy-ion collisions at a given collision energy. A schematic diagram of energy density as a function of fireball evolution time is given in Figure \ref{EngDensityVsTime}. In general one can think of three different energy density estimates and two different time scales: 
\begin{enumerate}
\item The peak {\it general energy density}, which is achieved when the incoming nuclei overlap with each other.

\item The peak {\it formed energy density}, which involves the produced particles at a proper time $\tau_{\rm form}$.

\item The peak {\it thermalized energy density},  present at proper time $\tau_{\rm therm}$, when local thermal equilibrium is first achieved, assuming that this occurs. 
\end{enumerate}
In this review we shall restrict ourselves to the discussion on the formed energy density estimated through Bjorken boost invariant hydrodynamics. However, for a detailed discussions one can refer to Ref. \cite{phenixW}. 


\section{Bjorken Hydrodynamics and Initial Energy Density}
\begin{figure}
\begin{center}
\includegraphics[width=3.6in]{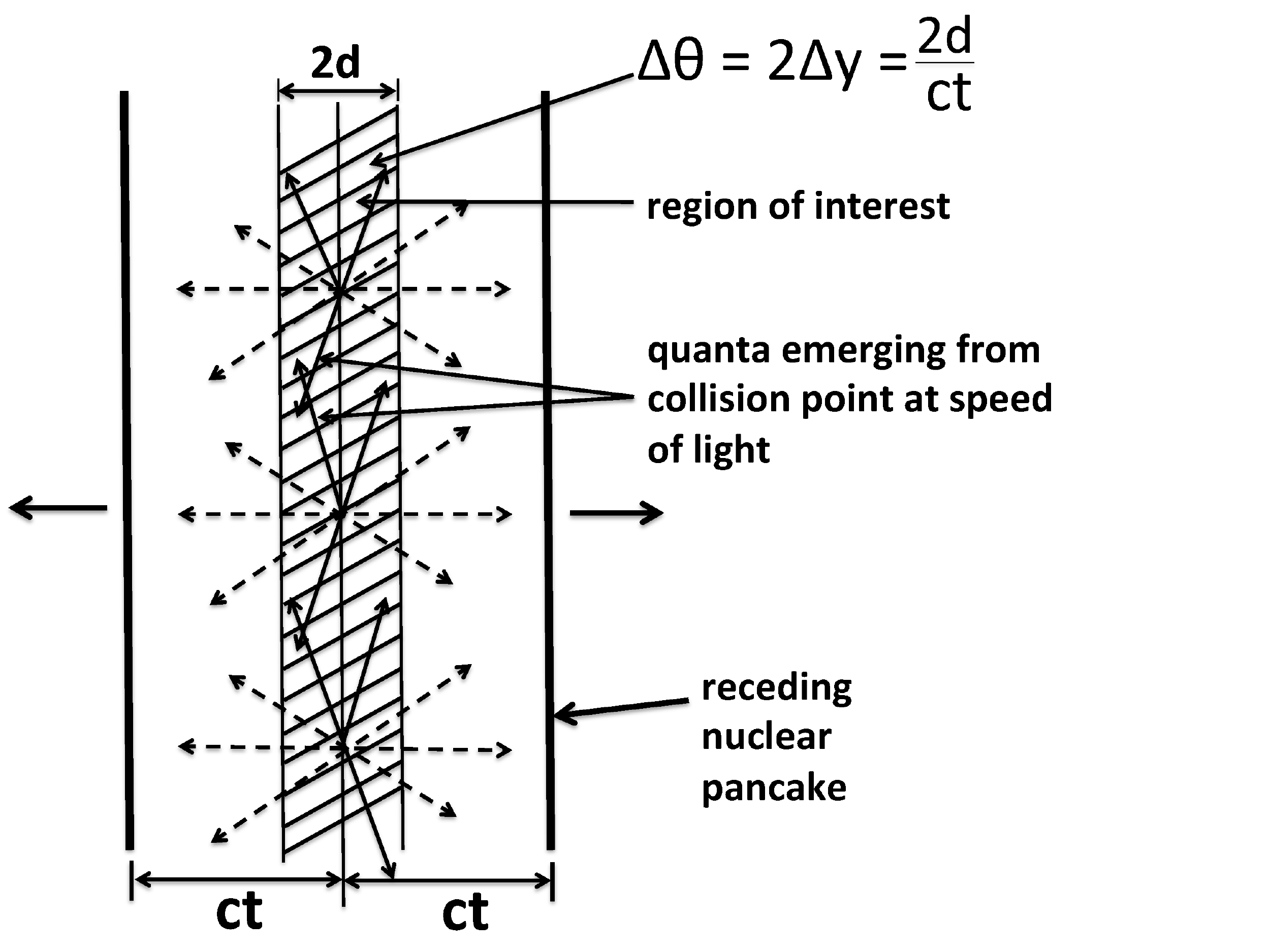}
\caption{\label{bjorken}Geometry for the initial state of centrally produced
plasma in nucleus-nucleus collisions. This picture is valid in any frame in
which the incoming nuclei have very high energies and so are Lorentz contracted. Figure from Reference \cite{bjorken}}.
\end{center}
\end{figure}
The energy density, in general is defined as the ratio of total mass-energy within some region of space and the volume of that region, as seen at some instant of time in some Lorentz frame. As discussed in Ref \cite{phenixW}, this definition is not satisfactory, as one can easily raise any energy density  by viewing the system from a different frame of reference. For example, a gold or lead nucleus with constant energy density $\rho_0$, when viewed in a boosted frame will appear to have energy density $\gamma^2 \rho_0$, where $\gamma$ is the value of the Lorentz boost factor. In a region having total momentum zero, one can meaningfully calculate the energy density as ratio of mass-energy and volume. Considering symmetric heavy-ion collisions (A+A) in collider experiments, with an overlapping of two nuclei, viewed in the center of momentum frame, the total energy density in the overlapping region is given by $\langle \epsilon \rangle = 2\rho_0 \gamma^2$. If we take the normal nuclear matter density, $\rho_0 = 0.14 ~{\rm {GeV/fm^3}}$, for a nucleus at rest and $\gamma = 106$ at $\sqrt{s_{\rm {NN}}} = 200$ GeV, then the general energy density is $\langle \epsilon \rangle = 3150 ~{\rm {GeV/fm^3}}$ at RHIC. For LHC $\sqrt{s_{\rm {NN}}} = 2.76$ TeV Pb+Pb collisions, $\gamma = 1471.2$, which leads to $\langle \epsilon \rangle = 606053 ~{\rm {GeV/fm^3}}$. As these numbers are spectacularly high, when compared to the lQCD predicted value of $1 ~ {\rm {GeV/fm^3}}$ energy density as a condition for the formation of a QGP phase, seem to be absurd. Hence, our interest would be to consider the energy density of the produced particles in order to infer about the possible formation of a QGP phase. This is done through the measurement of transverse energy at midrapidity and further the estimation of initial energy density in Bjorken hydrodynamic model.  

In the framework of Bjorken boost invariant hydrodynamic model, in any frame where the two incoming nuclei have very high energies, the region when/where the nuclei overlap will be very thin in the longitudinal direction
and very short in duration. In this scenario, it is fair to describe that all produced particles are created at the same time and radiated out from a thin disk. This is the Bjorken hydrodynamic picture of nucleus-nucleus collision \cite{bjorken}.

Once the Lorentz contracted beam ``pancakes'' recede after their initial overlap, the region
between them is occupied by secondaries at intermediate rapidities. We can
calculate the local energy densities of these created particles, if we 
assume the secondaries are formed at some proper time $\tau_{\rm {form}}$.

Our region of interest, in any frame, will be a slab perpendicular to the beam
direction, with longitudinal thickness $dz$, with one face of the ``source''
plane in this frame, and the transverse overlap area $A$. The region described 
here corresponds to half the shaded region shown in Figure \ref{bjorken}. 
Since $\beta_{\parallel}\simeq 0$ for particles near the source location, 
this is an appropriate region over which we can calculate a meaningful energy density. At time $t = \tau_{\rm {form}}$, this volume will contain all the (now-formed) 
particles with longitudinal velocities $0 \leq \beta_{\parallel} \leq dz/\tau_{\rm{form}}$ 
(since we assume particles can't scatter before they are formed!). Then we can 
write this number of particles as $dN = (dz/\tau_{\rm{form}})\frac{dN}{d\beta_{\parallel}}$, 
or equivalently $dN = (dz/\tau_{\rm{form}})\frac{dN}{dy}$, where $y$ is longitudinal
rapidity, since $dy = d\beta_{\parallel}$ at $y = \beta_{\parallel} = 0$.
If these particles have an average total energy $\langle m_{\rm T} \rangle$ in this frame 
($E = m_T$ for particles with no longitudinal velocity), then the total energy
divided by the total volume of the slab at $t = \tau_{\rm{form}}$ is 
\begin{eqnarray}
\langle \epsilon(\tau_{form})\rangle &=& \frac{dN \langle m_T \rangle}{dz A} \nonumber \\
&=& \frac{dN(\tau_{form})} {dy} \frac{\langle m_T \rangle}{\tau_{form} A} \nonumber \\
&=& \frac{1}{\tau_{form} A} \frac{dE_T(\tau_{form})}{dy},
\label{bj}
\end{eqnarray}
where, we have equated $\frac{dE_T}{dy} = \langle m_{\rm T} \rangle \frac{dN}{dy} \approx 
\langle m_{\rm T} \rangle \frac{3}{2}\frac{dN_{ch}}{dy}$ and emphasized that Eq.~(\ref{bj}) is true for 
the transverse energy density present at time $t = \tau_{\rm{form}}$. The factor 3/2 
compensates for the neutral particles.

Eq.~(\ref{bj}) is referred as $Bjorken ~energy ~density, ~\epsilon_{B_j}$. It is a valid
measure of peak energy density in created particles, on very general grounds and
in all frames, as long as two conditions are satisfied: (1) A finite formation time
$\tau_{\rm{form}}$ can meaningfully be defined for the created secondaries; and (2) The
thickness/``crossing time'' of the source disk is small compared to $\tau_{\rm{form}}$,
that is, $\tau_{\rm{form}} >> 2R/\gamma$. Here $R$ is the rest-frame radius of the nucleus 
and $\gamma$ is the Lorentz factor. In particular, the validity of Eq.~(\ref{bj})
is completely independent of the shape of the $dE_T(\tau_{\rm{form}})/dy$ distribution
to the extent that $\beta_{\parallel}$ is infinitesimally small in a co-moving frame;
a plateau in $dE_T/dy$ is not required. For practical purposes at RHIC, we will 
consider condition (2) above to be satisfied as long as $\tau_{\rm{form}} > 2R/\gamma$
is true.

Historically, $\epsilon_{B_j}$ has been calculated using the final state $dE_T/dy$
and simply inserting a nominal value of 1 fm/{\it c} for $\tau_{\rm{form}}$. In addition, fixed 
target experiments have been using $dE_T/d\eta$ as an estimate for $dE_T/dy$, which
is a good approximation for these experiments. For collider experiments,
a correction is made for the Jacobian $dy/d\eta$: ($\sqrt{1-m^2/\langle m_{\rm T} \rangle ^2}\frac{dN}{dy}
= J\frac{dN}{dy}  = \frac{dN}{d\eta}$). However, we can't take 
$\epsilon_{B_j}$ as an exact estimate of energy density without some justification
for the value of 1 fm/{\it c} taken for $\tau_{\rm{form}}$. Hence, we term it as 
$\epsilon_{B_j}^{Nominal}$.  An indication of potential problems
with this choice arises immediately when considering AGS Au+Au and SPS Pb+Pb collisions,
where the center of mass ``crossing times'' $2R/\gamma$ are 5.3
fm/{\it c} and 1.6 fm/{\it c}
respectively, which implies that this choice for $\tau_{\rm{form}} =
1$ fm/{\it c} actually
violates the validity condition $\tau_{\rm{form}} > 2R/\gamma$ we set for the use of 
Eq.(\ref{bj}). So we will deprecate the use of $\epsilon_{B_j}^{Nominal}$ as 
a quantitative estimate of actual produced energy density and instead 
treat it only as a compact way of comparing $dE_T/d\eta$ measurements across 
different systems, centralities and beam energies.

The Bjorken
energy density obtained in this framework is given by 
\begin{eqnarray}
\epsilon_{Bj} &=& \frac{dE_{T}}{dy} ~\frac{1}{\tau_0 \pi R^2} \\
&=& \frac{dE_{T}}{d\eta} J(y,\eta)~\frac{1}{\tau_0 \pi R^2} \\
&\simeq & \langle m_{\rm T} \rangle \frac{3}{2} \frac{dN_{ch}}{dy} ~\frac{1}{\tau_0 \pi R^2}
\label{bjEqn}
\end{eqnarray}
where, $\tau_0$ is the formation time, usually assumed to be 1 fm/{\it
  c} and $\pi R^2$ is
the transverse overlap area of the colliding nuclei. The formation time is usually 
estimated from model calculations and has been a matter of debate. There are 
different ways to estimate the transverse overlap area. It goes like $N_{\rm{part}}^{2/3}$
in an approach which accounts for only the common area of colliding nucleons but not 
the nuclei (chosen by STAR). In this approach, the transverse overlap area 
$F = \pi R^2$, where $R =R_0 A^{1/3}$. When we replace $A$ with the number of 
participants by, $A=N_{\rm{part}}/2$ \cite{kharzeevNpart}, $F$ becomes,
\begin{equation}
{F = \pi R_0^2 ~(\frac{N_{part}}{2})^{2/3} }
\label{FEqn}
\end{equation}

\begin{figure}[htbp26]
\begin{center}
\includegraphics[width=3.6in]{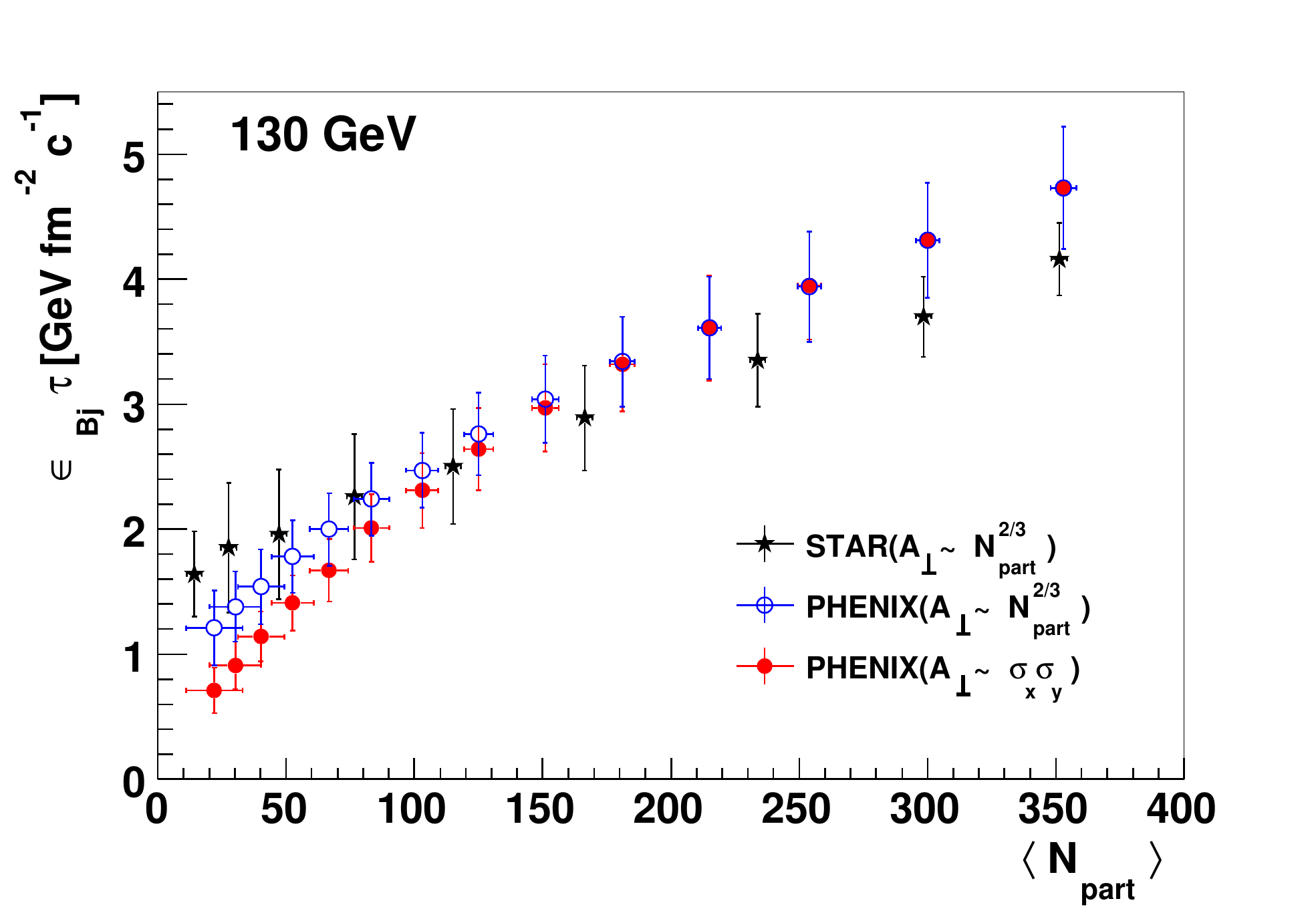}
\caption{The Bjorken energy density vs $N_{\rm{part}}$ using different estimates of
the transverse overlap area at $\sqrt{s_{NN}} = 130$ GeV. Figure taken from Reference.~\cite{phenixSyst}.} 
\label{phenixTransA}
\end{center}
\end{figure}
In the other approach (adopted by PHENIX) \cite{phenixSyst}, the transverse overlap 
area of the colliding species, $F$, is estimated in the following way.
The Woods-Saxon parametrization for the nuclear density profile is given by
\begin{equation}
{\rho(r) = \frac{1}{(1+e^{(r-r_n)/d})}},
\label{woods}
\end{equation}
where $\rho(r)$ is the nuclear density profile, $r_n$ is the nuclear radius and
$d$ is a diffuseness parameter. Based on the measurements of electron scattering 
from Au nuclei \cite{eAuScatt}, $r_n$ is set to $(6.38 \pm 0.27)$ fm and $d$ to 
$(0.54 \pm 0.01)$ fm. A Monte Carlo-Glauber model with $F\sim \sigma_x \sigma_y$, 
(where $\sigma_x$ and $\sigma_y$ are the widths of $x$ and $y$ position 
distributions of the participating nucleons in the transverse plane) is used to 
estimate the transverse overlap area of two colliding nuclei. In this approach, 
$F$ is the transverse overlap area of two colliding nuclei, not the 
participating nucleons. The normalization to $\pi R^2$, where $R$ is the sum of 
$r_n$ and $d$ parameters in the Woods-Saxon parametrization (given by 
Eq.~\ref{woods}), is done for most central collisions at the impact parameter $b=0$.

The results obtained in these two methods, as shown in Figure \ref{phenixTransA}, 
are different only in the peripheral bins. The results obtained by STAR agree 
with PHENIX results within systematic errors. However, STAR data show 
a smaller rate of increase of the energy density with $N_{\rm{part}}$. As can be seen from 
the figure, the results agree rather well within uncertainties for central collisions, where we expect
a deconfinement of quarks and gluons to take place.
\par
In the estimation of  $\epsilon_{Bj}.\tau$, one uses the energy and rapidity dependent Jacobian factor, $J(y,\eta)$, for the conversion of pseudorapidity to rapidity phase space. The value of the Jacobian is smaller at higher energies,  as the average transverse momentum of particles increases with beam energy. STAR collaboration uses a factor of 1.18 for $\eta \rightarrow y$-phase space conversion, as compared to 1.25 used by
PHENIX \cite{phenixEt, phenixSyst} for the estimation of Bjorken energy density at 200 GeV. The value of $\epsilon_{Bj}$ for Au+Au collisions at $\sqrt{s_{NN}} =$ 19.6, 130 \cite{phenixEt,phenixSyst} and 200 GeV \cite{star200GeV} are $2.2 \pm 0.2, ~4.7 \pm 0.5$ and $4.9 \pm 0.3~~ {\rm GeV/fm^3}$ 
($5.4 \pm 0.6 ~ {\rm GeV/fm^3}$, PHENIX), respectively. Compared to this, $\epsilon_{Bj}$
at SPS for Pb+Pb collisions at $\sqrt{s_{NN}} =$ 17.2 GeV is found to be 
$3.2 ~{\rm {GeV/fm^3}}$ \cite{Alber}. This value of $\epsilon_{Bj}$ is much higher than the
same for Au+Au collisions at the SPS-like energy i.e $\sqrt{s_{NN}} =$ 19.6 GeV
at RHIC. The CMS collaboration has estimated $\epsilon_{Bj} = 14 ~
{\rm GeV/fm^3}$ with transverse overlap area of $A= \pi \times (7~{\rm
  fm}^2)$ and $J(y,\eta) = 1.09$ for top $5\%$ central Pb+Pb
collisions at $\sqrt{s_{NN}} = 2.76$ TeV \cite{cmsEt}.  As all these
estimations assume the same formation time of 1 fm/{\it c}, there 
is an over estimation of $\epsilon_{Bj}$ at SPS. In any case these energy densities 
are significantly larger than the energy density ($\sim ~ 1~ {\rm {GeV/fm^3}}$) predicted 
by lattice QCD calculations \cite{lQCD} for a transition to a deconfined 
quark gluon plasma phase. Following the deconfinement transition, there is a
hydrodynamic expansion. Subsequently local equilibrium is achieved at 
$\tau_0 \sim 1$ fm/{\it c}. This picture is indeed valid, if we compare the RHIC data
for elliptic flow to the hydrodynamic calculations \cite{hydro1, hydro2, hydro3}.

\begin{figure}[htbp22]
\begin{center}
\includegraphics[width=3.6in]{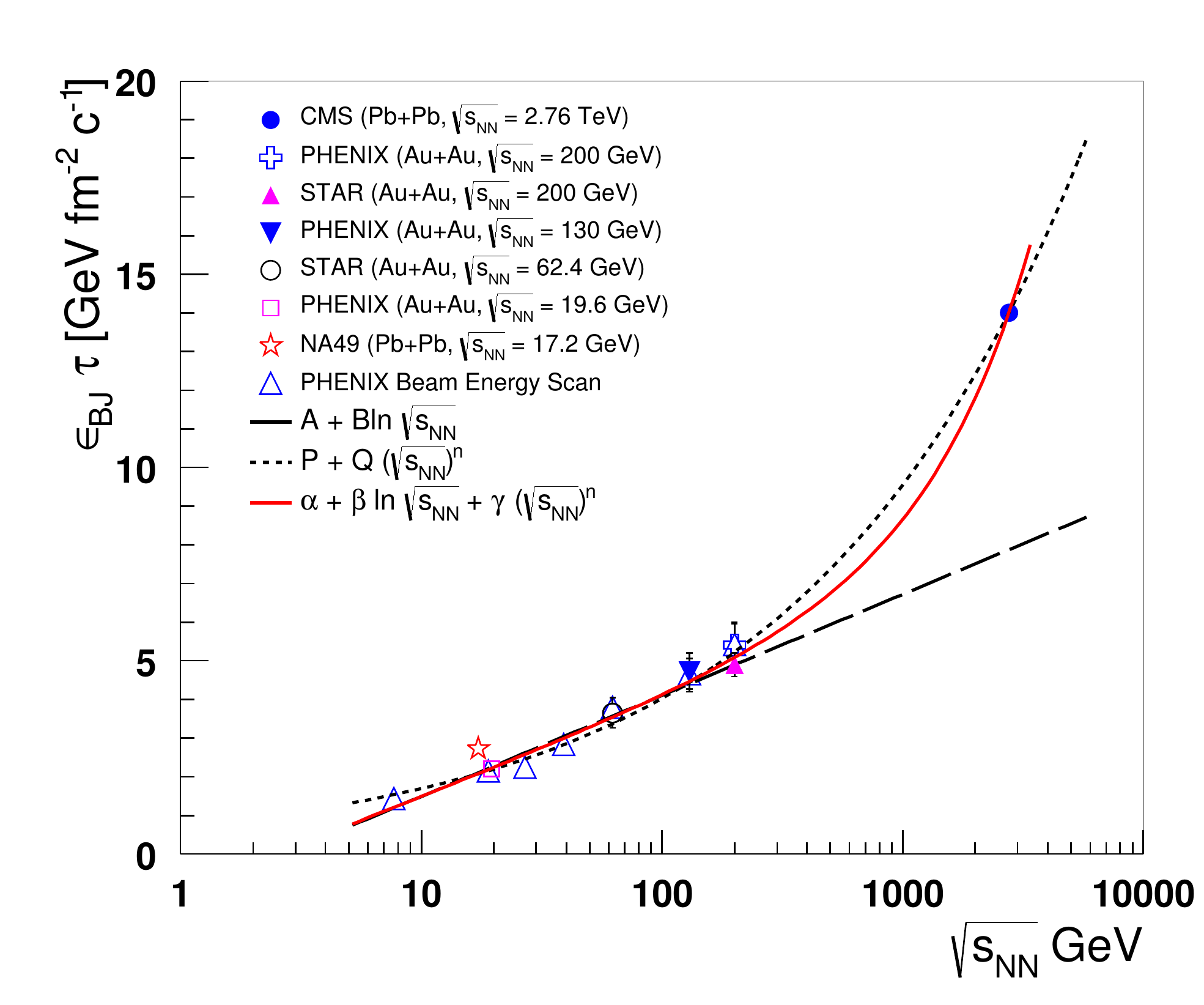}
\caption{The excitation function of $\epsilon_{Bj}. \tau~ [GeV~fm^{-2}~c^{-1}]$.
Logarithmic prediction fails to explain the LHC data. Shown are different phenomenological data driven fitting functions to describe the observable as a function of collision energy.} 
\label{bjLHC}
\end{center}
\end{figure}

\begin{figure}[htbp222]
\begin{center}
\includegraphics[width=3.6in]{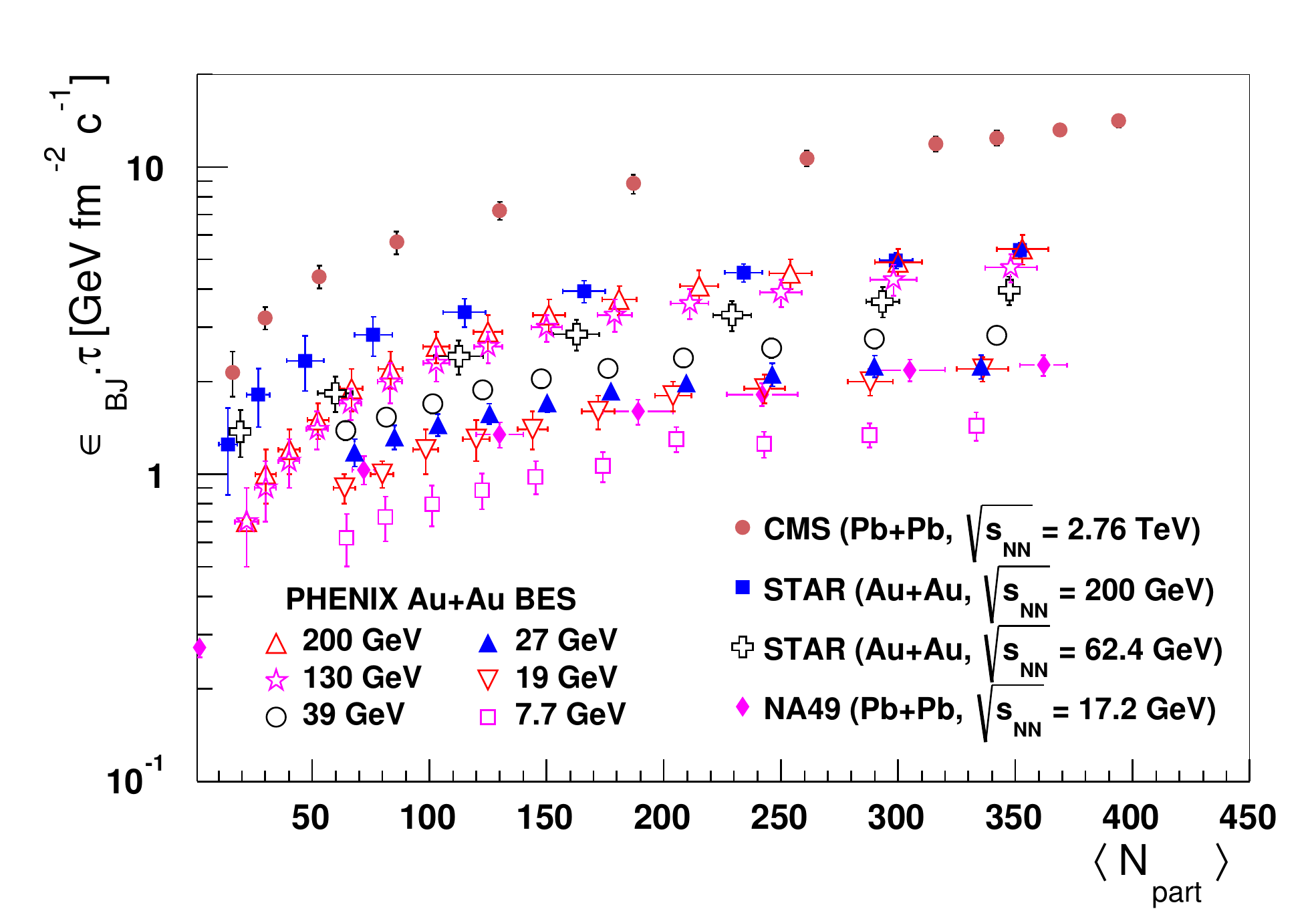}
\caption{The $N_{\rm{part}}$ dependence of the product of the Bjorken energy density
and the formation time ($\epsilon_{Bj}. \tau$) for Au+Au system at different 
energies at RHIC compared to Pb+Pb collisions at LHC.}
\label{bjNpart}
\end{center}
\end{figure}

Taking all $\epsilon_{Bj}$ measured for heavy-ion collisions at different energies and colliding species, we show  $\epsilon_{Bj}.\tau$ as a function of collision energy in Figure \ref{bjLHC}. This is done using the Eq.~\ref{bjEqn}. The dashed line is a logarithmic fit. The logarithmic 
extrapolation of $\epsilon_{Bj}.\tau$  for Pb+Pb collisions at $\sqrt{s_{\rm NN}} = 2.76$ TeV at LHC is around $7.17 ~{\rm GeV~fm^{-2}~c^{-1}}$. However, the experimental estimation gives a value of $\epsilon_{Bj}.\tau = 14 ~ {\rm GeV~fm^{-2}~c^{-1}}$, showing almost $50\%$ underestimation by the logarithmic trend of the data. On the other hand, a hybrid fitting function, which is a combination of logarithmic and power law functions in center of mass energy, describes the data from few GeV to TeV energies. Fitting a power law function overestimates the low energy measurements.
It should be noted that  the formation time at LHC will be much less than 1 fm.
The above value sets a lower bound to the initial energy density formed at LHC. Going from top RHIC energy to LHC 2.76 TeV, $\epsilon_{Bj}.\tau$ increases almost 3 times. 
Figure \ref{bjNpart} shows the estimate of the product of the Bjorken energy
density and the formation time ($\epsilon_{Bj}. \tau$) as a function of the 
centrality of the collision in terms of $N_{\rm{part}}$.  As expected there is a monotonic increase in $\epsilon_{Bj}. \tau$ with increasing centrality of the collision.

While comparing the results from different experiments, related to the initial 
energy density, one needs to take care of the following factors: (i) value of the 
formation time taken into the calculations, (ii) the procedure of estimation of 
the transverse overlap area and (iii) the value of the Jacobian used to transform 
$\eta$ to $y$ phase space.

\subsection{The Formation Time}
Is it possible to justify a better estimate for $\tau_{\rm{form}}$? From general quantum
mechanical arguments, in a frame where it's motion is entirely transverse, a particle
of energy $m_T$ can be considered to have ``formed'' after a time $t = \hbar/m_T$. To estimate the average transverse mass, we can
use the final state $dE_T/d\eta$ to estimate $dE_T(\tau_{\rm{form}})/dy$ and, 
correspondingly, use the final state $dN/d\eta$ as an estimate for $dN(\tau_{\rm{form}})/dy$
to obtain
\begin{equation}
{\langle m_T \rangle = \frac{dE_T(\tau_{form})/dy}{dN(\tau_{form})/dy} \simeq
 \frac{dE_T/d\eta}{dN/d\eta}~~(Final ~state).} 
\label{mt}
\end{equation}
It has been observed experimentally that the ratio of final state transverse energy 
density to charge particle density, each per unit pseudorapidity is constant at
about 0.85 GeV for central Au+Au collisions at top RHIC energy. This value is constant 
for a wide range of centrality and shows a very little change with beam energy, 
decreasing to 0.7 GeV, when $\sqrt{s_{NN}}$ is decreased by a order of magnitude
down to 19.6 GeV. However, at LHC, its observed value is $1.25 \pm 0.08$ GeV, which will be discussed in the next section. If we approximate $dN_{ch}/d\eta = (2/3)dN/d\eta$ in the final 
state, then Eq.(\ref{mt}) would imply $\langle m_{\rm T} \rangle \simeq 0.57$ GeV and corresponding 
$\tau_{\rm{form}} \simeq 0.35$ fm/{\it c}, a value shorter than the ``nominal'' 1 fm/{\it c} but long enough to satisfy the given validity condition $\tau_{\rm{form}} > 2R/\gamma$ at RHIC. With $R= 7$ fm for Au+Au collisions and Lorentz factor $\gamma = \frac{\sqrt{s_{\rm {NN}}}}{2m_{\rm p}} = 106.6$, at $\sqrt{s_{\rm {NN}}} = 200$ GeV, $2R/\gamma = 0.13$ fm/{\it c}. For LHC, at $\sqrt{s_{\rm {NN}}} = 2.76$ TeV, the observed $\langle p_{\rm T} \rangle \sim$ 0.678 GeV for Pb+Pb collisions \cite{LHC-pt}. Taking pion mass, one gets $\langle m_{\rm T} \rangle \sim$ 0.81 GeV, which leads to $\tau_{\rm{form}} \simeq 0.25$ fm/{\it c}. For Pb+Pb collisions at $\sqrt{s_{\rm {NN}}} = 2.76$ TeV, taking $R= 7.1$ fm, Lorentz factor, $\gamma =1471.22$,  we get $2R/\gamma = 0.01$ fm/{\it c}. Hence, the condition of $\tau_{\rm{form}} > 2R/\gamma$ is also satisfied at LHC.
\par
It's worth noting that the value of energy density obtained by Eq.~(\ref{bj}) 
represents a conservative lower limit on the actual $\langle \epsilon(\tau_{\rm{form}})\rangle$
achieved at RHIC. This follows from two observations: (1) The final state measured
$dE_T/d\eta$ is a solid lower limit on the $dE_T(\tau_{\rm{form}})/dy$ present at formation
time; and (2) The final state ratio $(dE_T/d\eta)/(dN/d\eta)$ is a good lower limit
on $\langle m_{\rm T} \rangle$ at formation time, and so yields a good upper limit on $\tau_{\rm{form}}$. The justification of these statements could be realized as follows.
\par
There are several known mechanisms that will decrease $dE_T/dy$ as the collision
system evolves after the initial particle formation, while no mechanism is known
that can cause it to increase (for $y = 0$, at least). Therefore, its final state
value should be a solid lower limit on its value at any earlier time. A list of
mechanisms through which $dE_T/dy$ will decrease after $t = \tau_{\rm{form}}$ includes:
(i) The initially formed secondaries in any local transverse ``slab'' will, in a 
co-moving frame, have all their energy in transverse motion and none in longitudinal
motion; if they start to collide and thermalize, at least some of their $E_T$ will
be converted to longitudinal modes in the local frame; (ii) Should near local
thermal equilibrium be obtained while the system's expansion is still primarily
longitudinal, then each local fluid element will lose internal energy through $pdV$
work and so its $E_T$ will decrease; (iii) If there are pressure gradients during
a longitudinal hydrodynamic expansion then some fluid elements may be accelerated
to higher or lower rapidities; these effects are complicated to predict, but we can 
state generally that they will always tend to $decrease~ dE_T/dy$ where it has its
maximum, namely at $y = 0$. Given that we have strong evidence of thermalization
and hydrodynamic evolution at RHIC collisions, it's likely that all these
effects are present to some degree, and so we should suspect that final state
$dE_T/d\eta$ is substantially lower than $dE_T(\tau_{\rm{form}})/dy$ at midrapidity.

Coming to the estimate of $\tau_{\rm{form}}$, the assumption that $\tau_{\rm{form}} = \hbar/\langle m_{\rm T} \rangle$
can't be taken as exact, even if the produced particles' $m_T$'s are all identical,
since ``formed'' is not an exact concept. However, if we accept the basic validity
of this uncertainty principle argument, then we can see that the approximation in
Eq.~(\ref{mt}) provides a lower limit on $\langle m_{\rm T} \rangle$. First, the numerator $dE_T/d\eta$
is a lower limit on $dE_T(\tau_{\rm{form}})/dy$, as above. Second, the argument is often
made on grounds of entropy conservation that the local number density of particles
can never decrease~\cite{entropy}, which would make the final state denominator in 
Eq.~(\ref{mt}) an upper limit on its early-time value.

\section{Transverse energy per Charged particle ($E_{\rm{T}}/N_{\rm{ch}}$) and Freeze-out Criteria}
\begin{table}{ht}
\caption{$E_{\rm{T}}/N_{\rm{ch}}$ (GeV) as a function of $\sqrt{s_{\rm{NN}}}$, plotted in Figure \ref{etEch}.}
\centering
\begin{tabular}{c|c| c| c}
\hline \hline
$\sqrt{s_{\rm{NN}}}$ (GeV) &  Coll. Species & $E_{\rm{T}}/N_{\rm{ch}}$ (GeV) & Reference \\ [0.5ex]
\hline
2.05 & Au+Au  & 0.13 $\pm$ 0.03 &   \cite{fopi} \\
3.81 & Au+Au   & 0.598 $\pm$ 0.060 & \cite{phenixSyst} \\
4.27 & Au+Au   & 0.634 $\pm$ 0.063 & \cite{phenixSyst} \\
4.84 & Au+Au   & 0.680 $\pm$ 0.068 & \cite{phenixSyst} \\
8.7   & Pb+Pb  & 0.760 $\pm$ 0.060 & \cite{SPS-1,SPS-2} \\
12.4 & Pb+Pb  & 0.780 $\pm$ 0.060 & \cite{SPS-1,SPS-2} \\
17.2 & Pb+Pb  & 0.810 $\pm$ 0.060 & \cite{SPS-3} \\
19.6 & Au+Au  & 0.738 $\pm$ 0.070 & \cite{phenixSyst} \\
62.4 & Au+Au & 0.867 $\pm$ 0.121 & \cite{RHIC-62} \\
130 &  Au+Au & 0.869 $\pm$ 0.066 & \cite{phenixEt} \\
200 &  Au+Au & 0.881 $\pm$ 0.071 & \cite{phenixSyst} \\
2760 & Pb+Pb & 1.283 $\pm$ 0.085 & \cite{cmsEt} \\ [1ex]

\hline
\end{tabular}
\label{table-excitation}
\end{table}

\begin{center}
\begin{table*}[ht]
\caption{$E_{\rm{T}}/N_{\rm{ch}}$ (GeV) for different center of mass energies as a function of $N_{\rm {part}}$, the measure of collision centrality (shown in Figure \ref{etEchCent})}    
\centering                           
\begin{tabular}{c c c c c c c c c c c c c c c }             
\hline\hline   
$\sqrt{s_{\rm{NN}}}$ (GeV) & 0-5 & 5-10 & 10-15 & 15-20 & 20-25 & 25-30 & 30-35 & 35-40 & 40-45 & 45-50 & 50-55 & 55-60 & 60-65 & 65-70 \\[0.5ex]             
\hline
7.7 & 0.70& 0.73 & 0.74 & 0.77 & 0.78 & 0.81 & 0.84 & 0.87 & 0.90 & 0.94 \\
19.6 & 1.07 & 1.11 & 1.14 & 1.17 & 1.18 & 1.22 & 1.24 & 1.27 & 1.30 & 1.36 \\
27 & 1.22 & 1.24 & 1.27 & 1.30 & 1.34 & 1.38 & 1.40 & 1.44 & 1.47 & 1.54 \\
39 & 1.41 & 1.45 & 1.47 & 1.51 & 1.54 & 1.58 & 1.61 & 1.66 & 1.71 & 1.69 \\
62.4 & 1.21 & 1.26 & 1.32 & 1.40 & 1.49 & 1.57 & 1.67 & 1.76 & 1.84 & 1.92 & 2.01 & 2.07 & 2.17 & 2.23 \\
130 & 1.95 & 2.02 & 2.05 & 2.12 & 2.21 & 2.31 & 2.38 & 2.48 & 2.56 & 2.65 & 2.73 & 2.80 & 2.89 & 3.01 \\
200 & 1.93 & 2.06 &  2.18 & 2.31 & 2.46 & 2.57 & 2.66 & 2.76 & 2.87 & 2.95 & 3.04 & 3.16 & 3.29 & 3.44 \\
\hline
$\sqrt{s_{\rm{NN}}}$ (TeV) & 0-2.5 & 2.5-5 & 5-7.5 & 7.5-10 & 10-20 & 20-30 & 30-40 & 40-50 & 50-60 & 60-70 & 70-80 &  &  &  \\[0.5ex]   
\hline
2.76 & 1.26 & 1.25 & 1.24 & 1.23 & 1.22 & 1.21 & 1.19 & 1.17 & 1.16 & 1.15 & 1.05\\
\hline
\end{tabular}
\label{table-etCent}
\end{table*}
\end{center}

\begin{figure}[htbp00]
\begin{center}
\includegraphics[width=3.6in]{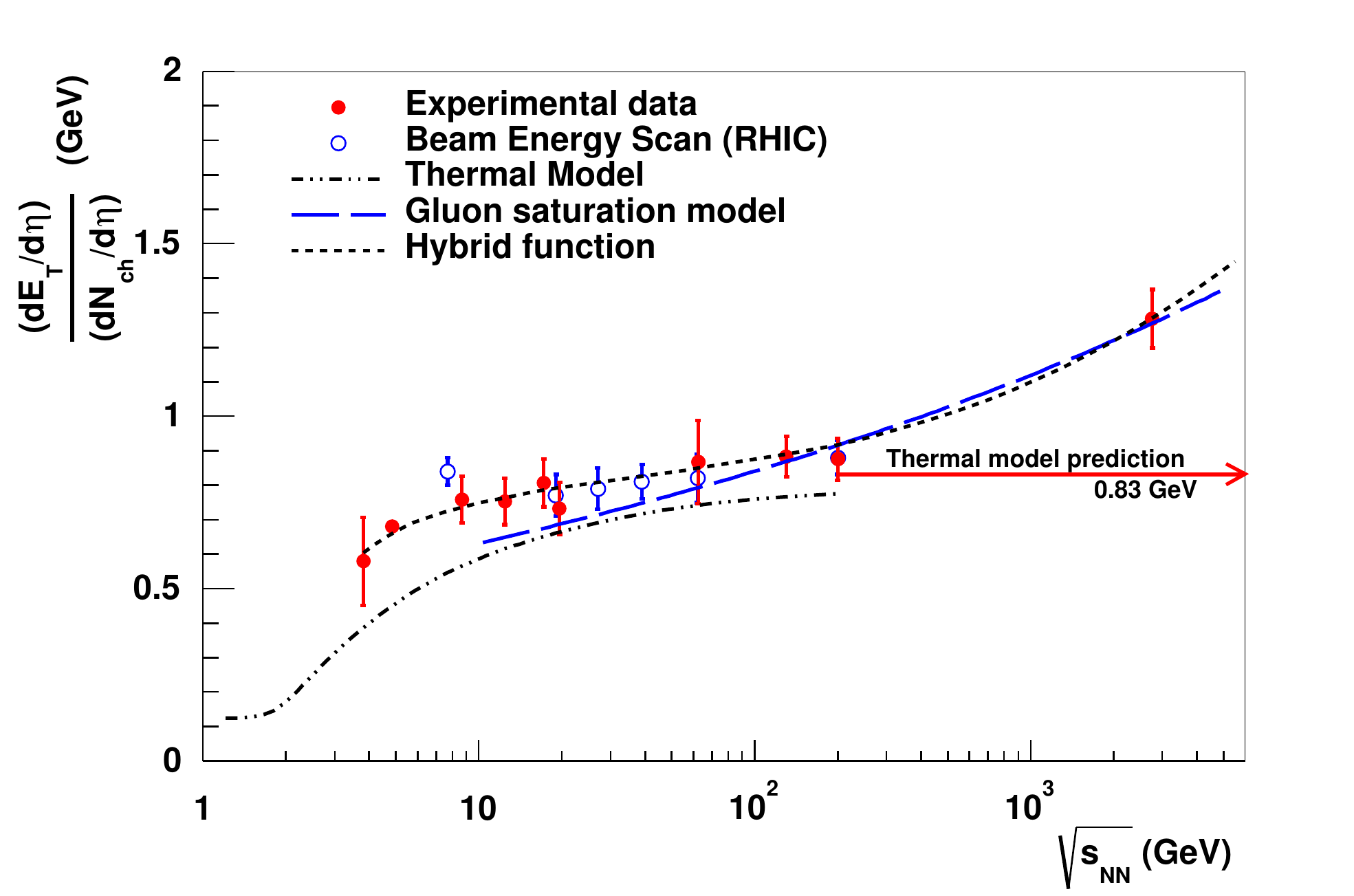}
\caption{The ratio of $\frac{dE_T}{d\eta}$ and $\frac{dN_{ch}}{d\eta}$ at midrapidity, as a function of center of mass energy. Experimental data are compared to the predictions from thermal model, gluon saturation model and the estimations obtained in the framework of the hybrid model fitting to transverse energy and charged particle data.\protect\label{etEch}}
\end{center}
\end{figure}

The ratio of pseudorapidity densities of transverse energy and number of charged particles at midrapidity, i.e. $\frac{dE_{\rm T}}{d\eta}/\frac{dN_{\rm {ch}}}{d\eta}(\equiv E_{\rm{T}}/N_{\rm{ch}}$) has been studied both experimentally \cite{phenixSyst,cmsEt,star200GeV} and phenomenologically \cite{rns,rnsJPG,prorok} to understand the underlying particle production mechanism. This observable is known as global barometric measure of the internal pressure in the ultra-dense matter produced in heavy-ion collisions. This quantity depends on the initial state of the collision and the viscosity of the matter as it expands to its final state, when it is observed by the detectors. This observable when studied as a function of collision energy (as shown in Figure \ref{etEch} and the values are given in Table \ref{table-excitation}),  shows three regions of interest. The first one from the lower SIS energies to SPS energies shows a steep increase of $E_{\rm{T}}/N_{\rm{ch}}$ values, thereby indicating that the mean energy of the system increases (at midrapidity, $\langle E \rangle \sim  \langle m_{\rm T} \rangle$). In the second region, from SPS to top RHIC energy, $E_{\rm{T}}/N_{\rm{ch}}$ shows a very weak collision energy dependence, i.e. like a saturation behaviour. In this region the mean energy doesn't increase, whereas the collision energy increases. This may indicate that the increase in collision energy results in new particle production in this energy domain, which is consistent with higher particle multiplicity observed at these energies. This behaviour has been well described in the context of a statistical hadron gas model (SHGM) \cite{rns,rnsJPG}. In the framework of SHGM, it has been predicted that $E_{\rm{T}}/N_{\rm{ch}}$  would saturate at energies higher to that of top RHIC energy with a limiting value of 0.83 GeV \cite{rns,rnsJPG}.  Here a static fireball is assumed at the freeze-out. However, a value of $1.25 \pm 0.08$ GeV is observed at the LHC 2.76 TeV center of mass energy recently, by the CMS collaboration \cite{cmsEt}. This creates a third region in the excitation function of $E_{\rm{T}}/N_{\rm{ch}}$, showing a jump from top RHIC to LHC energies. In this region, along with particle multiplicity, the mean energy per particle also increases, which needs to be understood from theoretical models taking the dynamics of the time evolution of the created fireball. It is however, observed that models based on final state gluon saturation (CGC like) seems to explain this behaviour in the excitation function of $E_{\rm{T}}/N_{\rm{ch}}$ \cite{rnsAditya}. The RHIC Beam Energy Scan (BES) data seem to follow the overall trend of the collision energy dependence of  $E_{\rm{T}}/N_{\rm{ch}}$. It has been seen in one of the previous works of one of us (RS) \cite {rnsJPG} that various freeze-out criteria like constant energy per particle ($\langle E \rangle/\langle N \rangle = 1.08$ GeV) \cite{CleymansRed}, fixed baryon+anti-baryon density ($n_B + n_{\bar{B}} \simeq 0.12~fm^{-3}$) \cite{PBM}, fixed entropy density per $T^3$ ($\frac{s}{T^3} \simeq 7$) \cite{tawfik,JCHO} seem to describe the qualitative energy dependent behaviour of $E_{\rm{T}}/N_{\rm{ch}}$ quite consistently upto RHIC energies. As shown in the figure, a hybrid function which is a combination of logarithmic and power law in center of mass energy, seems to describe the data quite well. At very high energies, the creation and annihilation of gluons balances out leading to gluon saturation. In the framework of gluon saturation models,  the high energy behaviour of this observable is well described \cite{rnsAditya}.

\begin{figure}[htbp22]
\begin{center}
\includegraphics[width=3.6in]{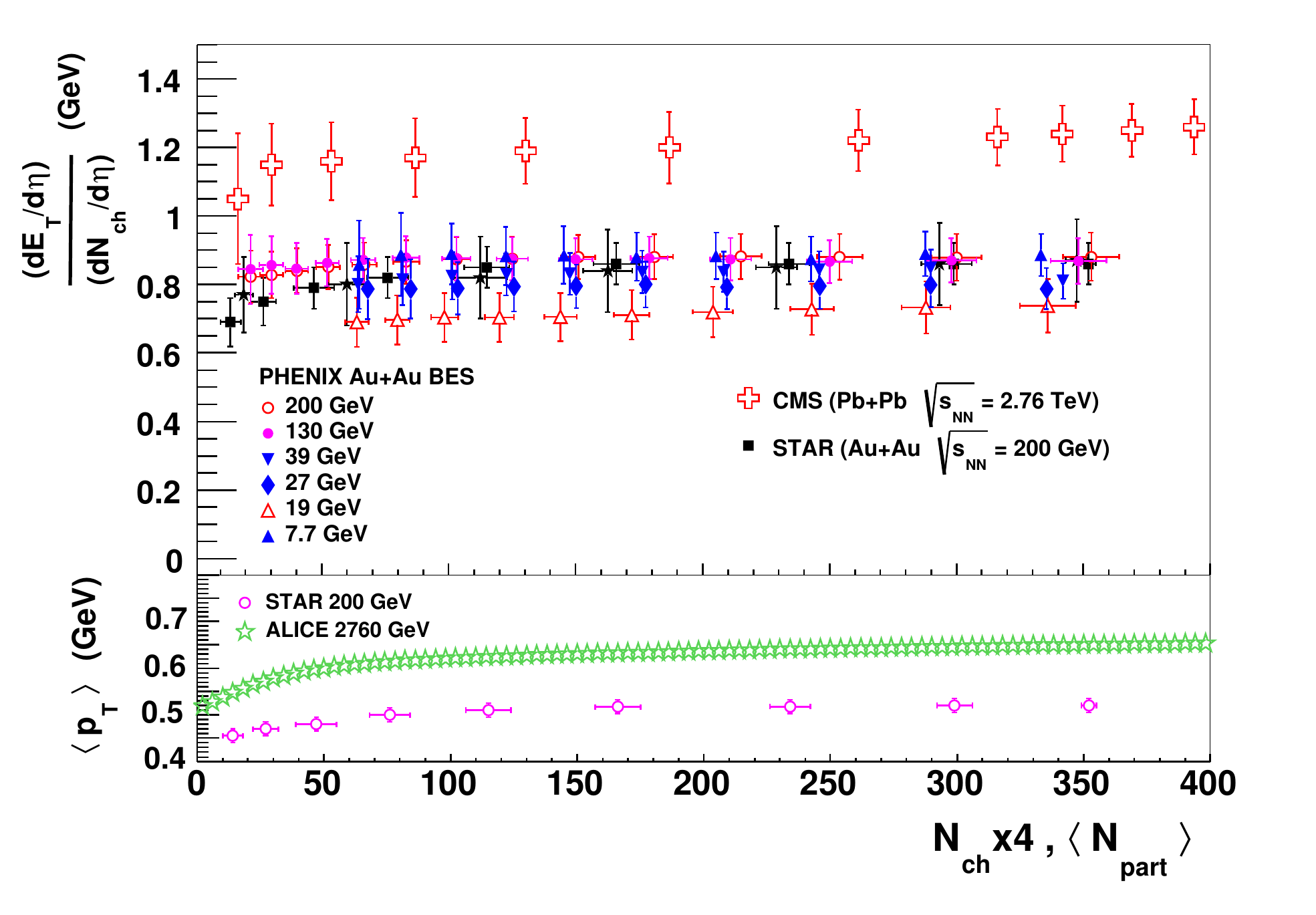}
\caption{Upper panel: The midrapidity ($0 < \eta < 1$) $E_{\rm{T}}/N_{\rm{ch}}$ as a function of collision centrality for a wide range of energies spanning RHIC beam energy scan to LHC. Lower panel: the mean transverse momentum as a function of collision centrality, both for top RHIC energy and LHC 2.76 TeV center of mass energy. A similar spectral behaviour is observed for $\langle p_{\rm T} \rangle$ and the barometric observable, $E_{\rm{T}}/N_{\rm{ch}}$.} 
\label{etEchCent}
\end{center}
\end{figure}

Figure \ref{etEchCent} (upper panel) shows the centrality dependence of $E_{\rm{T}}/N_{\rm{ch}}$ from $\sqrt{s_{\rm {NN}}}$ = 7.7 GeV to 2.76 TeV energy. These data are enlisted in Table \ref{table-etCent}. Since the centrality definitions by the CMS experiment for $\frac{dN_{\rm {ch}}}{d\eta}$ and $\frac{dE_{\rm T}}{d\eta}$ are different, fitting the centrality dependent $\frac{dN_{\rm {ch}}}{d\eta}$
by a function, $\frac{1}{0.5 N_{\rm{part}}}\frac{dN_{\rm {ch}}}{d\eta} = A ~ N_{\rm{part}}^{\alpha}$, with $A= 2.63 \pm 0.24$ and $\alpha = 0.19 \pm 0.02$, we have evaluated the $\frac{dN_{\rm {ch}}}{d\eta}$ values corresponding to the $N_{\rm{part}}$ values used to define the centrality classes for $\frac{dE_{\rm T}}{d\eta}$. Then we have estimated the LHC values of $E_{\rm{T}}/N_{\rm{ch}}$ at different centralities, which are given in Table \ref{table-etCent} and are shown in Figure \ref{etEchCent}.  Within the systematic errors, $E_{\rm{T}}/N_{\rm{ch}}$ for all energies upto top RHIC energy, show a weak centrality dependence with a modest increase from most peripheral collisions to $N_{\rm {part}} = 100$, reaching a roughly constant value of around $0.8$ GeV towards central collisions. The LHC data also show a similar behaviour but the constant value of $E_{\rm{T}}/N_{\rm{ch}}$ is around 1.25 GeV.
This centrality dependence of $E_{\rm{T}}/N_{\rm{ch}}$ is shown to be equivalent to the behaviour of $\langle p_{\rm T}\rangle$ as a function of centrality for top RHIC energy \cite{star200GeV} and for $\sqrt{s_{\rm{NN}}} = 2.76$ TeV \cite{LHCpT} at LHC. This is shown in the lower panel of Figure \ref{etEchCent}.
The value of $\langle p_{\rm T}\rangle = 0.678 \pm 0.007$ GeV/c at $\sqrt{s_{\rm{NN}}} = 2.76$ TeV, which is almost 36\% increase when compared with its value ($\sim 0.5$ GeV/c) at top RHIC energy \cite{star200GeV,LHCpT}. The value of $E_{\rm{T}}/N_{\rm{ch}}$ increases almost $45\%$ from top RHIC to LHC energy. This shows that not only particle multiplicity increases while going from top RHIC to LHC energy, the $\langle p_{\rm T}\rangle$ also increases, making a third region in the excitation function of $E_{\rm{T}}/N_{\rm{ch}}$. The near centrality independent behaviour of  $E_{\rm{T}}/N_{\rm{ch}}$ is explained by statistical hadron gas model (SHGM) with a static fireball approximation at freeze-out \cite{rns}. However, to explain the energy dependent behaviour of $E_{\rm{T}}/N_{\rm{ch}}$ in the whole range of energies upto LHC, one needs to consider the dynamical effects during the time evolution of fireball. Irrespective of the collisions species, the center of mass energies and the collision centrality, starting from the lower energies to top RHIC energy, the system evolves to the same final state, which could be characterized by a constancy in chemical freeze-out temperature. On the other hand, LHC data shows a different trend of $E_{\rm{T}}/N_{\rm{ch}}$, while the chemical freeze-out temperature doesn't change that much from RHIC to LHC. This needs to be understood from the thermodynamics point of view \cite{rhicTemp,lhcTemp}.

A theoretical description of the time evolution of the produced fireball in heavy-ion collisions is difficult, as it involves different degrees of freedoms at different points. The SHGM uses hadronic degrees of freedom at later times, when the chemical composition of the matter is frozen (known as chemical freeze-out). Then the particles mean free path becomes higher than the system size, which forbids the elastic collision of the constituents and the system is said to be kinetically frozen (known as thermal or kinetic freeze-out). In general, freeze-out could be a complicated process involving duration in time and a hierarchy where different types of particles and reactions switch-off at different times. This gives to the concept of {\it ``differential freeze-out" }. From kinetic arguments, it is expected that reactions with lower cross-sections switch-off at higher densities/temperature compared to reactions with higher cross-sections. Hence, the chemical freeze-out, which corresponds to inelastic reactions occurs earlier in time compared to the kinetic freeze-out, which corresponds to elastic reactions. In accordance with the above discussions, one can think of strange or charmed particles decoupling from the system earlier than the lighter hadrons.  A series of freeze-outs could be thought of corresponding to particular reaction channels \cite{baran}. However, in general one focuses on chemical and kinetic freeze-outs, considering the freeze-out to be an instantaneous process. At higher energies, when $\mu_B \sim 0$, the transverse energy production is mainly due to the meson content of the system. The experimental observations go in line with the above fact, when we observe the ratio of $\bar{p}/p \sim 1$ at higher energies \cite{lhcTemp}. The intersection points of lines of constant  $E_{\rm{T}}/N_{\rm{ch}}$ and the freeze-out line give the values of $E_{\rm{T}}/N_{\rm{ch}}$ at the chemical freeze-out \cite{rns}. 

\section{SUMMARY AND CONCLUSIONS}
 Pseudorapidity distribution of charged particles is proposed to be one of the important  global observables to characterize the hot and dense medium produced in the heavy-ion collisions. We reviewed the charged particle and photon multiplicity density distribution results obtained by various heavy-ion collision experiments starting from AGS energies to top RHIC and LHC energies. Before going to the results, a brief introduction on determination of centrality is given. Centrality determination is important in terms of relating theoretical observables, like impact parameters and numbers of nucleon participant ($N_{\rm{part}}$), to the collision geometry and observed particle multiplicity. To correlate them, for example, two-component model with NBD is fitted with the V0 amplitude and the centrality percentile is evaluated to classify the events into different centralities. In the mean time the respective $N_{\rm{part}}$, $N_{\rm{coll}}$ and impact parameters are estimated by Monte Carlo Glauber model. The $dN_{\rm{ch}}/d\eta$ spectra are discussed for Cu+Cu, Au+Au and Pb+Pb collisions at different energies. It is observed that the width and amplitude of the distribution increases with increase of collision energy. A double-Gaussian function is fitted with the distribution and it is found that the ratio of amplitudes, the widths are similar from one centrality to other in their respective collision energy. More interestingly that more dip in the mid rapidity is observed in Pb+Pb collisions at $\sqrt{s_{\rm{NN}}}$= 2.76 TeV. This is an indication of different hadro-chemistry at LHC energy than RHIC. Still this needs to be understood in details. Similarly, the energy dependence of $dN_{\gamma}/d\eta$ of Cu+Cu, Au+Au, S+Au and Pb+Pb collisions are discussed. 
\par
Then the limiting fragmentation behaviour of charged particles as well as photons are discussed for Cu+Cu, Au+Au and Pb+Pb collisions at different energies. The compilation of various experimental data go in line with the hypothesis of limiting fragmentation. Moreover, after observing the centrality dependence of longitudinal scaling of charged particles, $R_{\rm{PC}}$ is used to confirm the scaling behaviour and the scaling seems to be valid for a wide range of energies. In contrast to charged particles, photons do not show any centrality dependence. It is interpreted as majority of photons in the forward rapidities are coming form $\pi^{0}$ decays. Hence, mesons are not affected by baryon stopping as they are originated from valence quarks. CGC model has successfully explained the limiting fragmentation upto some extent. However, it needs more development and complete understanding of the final state effect and inclusion of quark distribution. This longitudinal scaling of hadrons still needs more insight to understand the physics process and its predicted violation at LHC energies in the frame work of SHGM and the validity from experimental data are to be understood from theoretical considerations. 
\par 
During the discussion of factorization, it is also observed that the centrality dependence of $dN_{\rm{ch}}/d\eta$ can be factorized to beam energy and collision centrality. By taking the ratio of $dN_{\rm{ch}}/d\eta$ of Pb+Pb collision at $\sqrt{s_{\rm{NN}}}$= 2.76 TeV to other collision energies show a scaling behaviour as a function of $N_{\rm{part}}$. To understand the expansion dynamics of the system, the $dN_{\rm{ch}}/d\eta$ of charged particles and photons are fitted with Landau-Carruther and Gaussian functions. By taking the ratio of widths of data to the Landau-Carruther's function, it is found that the system is expanding more or less like a  Landau hydrodynamic fluid  upto the RHIC energy. But the LHC data deviates from the Landau hydrodynamic model. Similarly, photons at RHIC energies also obey the Landau hydrodynamics. It is observed that the $N_{\rm{ch}}^{\rm{total}}$ normalized to $N_{\rm{part}}$ scales with centrality. It is to be noted here that in the midrapidity, $dN_{\rm{ch}}/d\eta$ normalized to $N_{\rm{part}}$ doesn't scale with centrality, whereas the total charged particles do. This is because of the modification of charged particles at forward rapidities are strongly correlated with compensating changes at midrapidity. $dN_{\gamma}/d\eta$ also shows similar scaling. 
It is found that trapezoidal rule can be used to explain the $N_{\rm{ch}}^{\rm{total}}$ normalized to participant pair from AGS energies to RHIC energies. However, it fails at LHC energy. When a hybrid function constructed by adding the power law and logarithmic of $\sqrt{s_{\rm{NN}}}$, it explains  the whole range of data indicating that the charged particle production is a combined process of midrapidity gluonic source (power law) and fragmentation process (logarithmic).
\par
The transverse energy measurement and the estimation of initial energy density in the framework of Bjorken boost invariant hydrodynamics are presented for collision energies ranging from few GeV to TeV. In this energy domain, the centrality and energy dependence of $\frac{dE_{\rm T}}{d\eta}$ and Bjorken energy density multiplied with formation time, $\epsilon_{\rm{Bj}}.\tau$ have been studied. A comparison of $\epsilon_{\rm{Bj}}$ with that of lQCD value, indicates the formation of a QGP phase both at RHIC and LHC energies.
The barometric observable, {\it i.e.} transverse energy per charged particle, is related to the chemical freeze-out.
Various freeze-out criteria seem to describe the energy dependent behaviour of $E_{\rm{T}}/N_{\rm{ch}}$ starting from
few GeV to top RHIC energies. A static fireball approximation at freeze-out, however, fails to reproduce
the corresponding data at LHC and necessitates the inclusion of fireball evolution dynamics in space and time
in order to describe the behaviour for the whole range of energies. The similarity in the centrality dependence
upto the top RHIC energy indicates that irrespective of the collision species and center of mass energies, the system
evolves to a similar final state at freeze-out.  \\
\par
{\bf Note-}
{\it In this review, we have made an attempt to give the developments in
heavy-ion collisions towards the measurements of charged particle and
photon multiplicities alongwith transevrse energy production from few
GeV to TeV energies. Although we have tried to cover in some details,
it is not an easy task and we can never assume the task to be
complete. However, we believe that the references
mentioned in this review shall guide the readers in the related fields. We apologize to
those authors whose valuable contributions in this area have not been
mentioned properly.}
\appendix*
\section{The Gamma and Negative Binomial Distributions}
The Gamma distribution represents the probability density for a continuous variable $x$, and has two parameters $b$ and $p$. This is given by 

\begin{equation}
f(x) ~=~f_{\Gamma}(x,p,b) ~=~ \frac{b}{\Gamma(p)}(bx)^{p-1}~e^{-bx},
\label{gammaFn}
\end{equation}
where\\

$p > 0$, ~~$b > 0$, ~~~~$0 \leq x \leq \infty$, \\

$\Gamma(p) ~=~ (p-1)! $ is the Gamma function if $p$ is an integer, and $f(x)$ is normalized, $\int_0^{\infty} f(x) ~dx ~=~1$. The first few moments of the distribution are\\

 \begin{equation}
\mu \equiv \langle x \rangle ~=~ \frac{p}{b},  ~~~~~ \sigma \equiv \sqrt{\langle x^2 \rangle -\langle x \rangle ^2}~=~ \frac{\sqrt{p}}{b}, ~~~~ \frac{\sigma^2}{\mu^2} ~=~ \frac{1}{p}. 
\label{gammaMoments}
\end{equation}

The Negative Binomial Distribution (NBD) of an integer $m$ is defined as \\
\begin{equation}
P(m) = \frac{(m+k-1)!}{m! (k-1)!}~~ \frac{\left( \frac{\mu}{k}\right)^m}{\left( 1+\frac{\mu}{k}\right)^{m+k}},
\label{NBDFn}
\end{equation}
where $P(m)$ is normalized for $0 \leq m \leq \infty$, $\mu \equiv \langle m \rangle$, and some of the higher moments are\\

\begin{equation}
\sigma ~=~ \sqrt{\mu \left(1+\frac{\mu}{k}\right)}, ~~ \frac{\sigma^2}{\mu^2} = \frac{1}{\mu}+\frac{1}{k}.
\label{NBDMoments}
\end{equation}

The NBD is having an additional parameter $k$ compared to a Poisson distribution. In the limit $k \rightarrow \infty$ NBD becomes a Poissonian distribution.  With $k$ equals to a negative integer (hence the name) becomes NBD. The NBD is strongly correlated with Gamma distribution, and hence becomes Gamma distribution in the limit $\mu \gg k > 1$. Usually Gamma distributions are replaced with NBD to prove various theorems \cite{GammaTheo}. One important difference between NBD and Gamma distributions is in the limit $m$ or $x \rightarrow 0$: for $p > 1$ the limit is always zero for a Gamma distribution, whereas for the NBD it is always finite.

 The Gamma Distribution has got potential applications as under {\it convolution} it shows an important property. Define the $n-$fold convolution of a distribution with itself as

\begin{equation}
f_n(x) ~=~ \int_0^x ~ dy f(y)~ f_{n-1}(x-y);
\label{convolute}
\end{equation}

then for a Gamma distribution given by Eq. \ref{gammaFn}, the $n-$fold convolution is simply given by the function

\begin{equation}
f_n(x) ~=~ \frac{b}{\Gamma (np)}(bx)^{np-1}e^{-bx} ~=~ f_{\Gamma}(x, np, b), 
\label{GammaConvolute}
\end{equation}

i.e., $p \rightarrow np$ and $b$ remains unchanged. Note that the mean $\mu_n$ and the standard deviation $\sigma_n$ of the $n-$fold convolution obey the familiar rule

\begin{equation}
\mu_n  ~=~ n\mu ~=~ \frac{np}{b}, ~~~~  \sigma_n ~=~ \sigma \sqrt{n} ~=~ \frac{\sqrt{np}}{b}, ~~ \frac{\sigma_n}{\mu_n} ~=~ \frac{1}{\sqrt{np}}. 
\label{GammaConvoluteRule}
\end{equation}
The convolution property of the Gamma distribution also holds good for the NBD, with $\mu_n  \rightarrow n\mu, ~ k \rightarrow nk$, so that $\mu/k$ remains constant \cite{E802-95}. Note that the charged particle multiplicity distribution in proton-proton collisions obey a NBD, whereas the Gamma distribution fits to $E_{\rm T}$ distributions.


\vskip-0.75cm

\end{document}